\documentclass[acmsmall,nonacm,authorversion,screen]{acmart}

\AtBeginDocument{%
	\providecommand\BibTeX{{%
			\normalfont B\kern-0.5em{\scshape i\kern-0.25em b}\kern-0.8em\TeX}}}

\setcopyright{rightsretained}
\acmJournal{TOCHI}
\acmYear{2022} \acmVolume{1} \acmNumber{1} \acmArticle{1} \acmMonth{1} \acmPrice{}\acmDOI{10.1145/3524122}

\acmJournal{TOCHI}

\usepackage[T1]{fontenc}
\usepackage{balance}
\usepackage{multirow}
\usepackage{microtype}
\usepackage{ccicons}
\definecolor{darkpastelgreen}{rgb}{0.01, 0.75, 0.24}
\definecolor{orange-red}{rgb}{1.0, 0.27, 0.0}

\DeclareMathAlphabet{\mathcal}{OMS}{cmsy}{m}{n}

\newcommand{\R}{\mathbb{R}}
\newcommand{\N}{\mathbb{N}}

\newcommand{\INFO}[1]{}%
\usepackage{nicefrac}
\usepackage[normalem]{ulem}
\usepackage[caption=false]{subfig}
\usepackage{anyfontsize}
\usepackage{tikz}
\usepackage{pgfplots}
\pgfplotsset{compat=1.7}

\definecolor{ForestGreen}{RGB}{34,139,34}

\makeatletter
\newcommand\footnoteref[1]{\protected@xdef\@thefnmark{\ref{#1}}\@footnotemark}
\makeatother

\makeatletter
\newcommand\approxsim{\mathchoice
	{\@approxsim {\displaystyle}      {1ex} }
	{\@approxsim {\textstyle}         {1ex} }
	{\@approxsim {\scriptstyle}       {.7ex}}
	{\@approxsim {\scriptscriptstyle} {.5ex}}}
\newcommand\@approxsim[2]{%
	\mathrel{%
		\ooalign{%
			$\m@th#1\sim$\cr
			\hidewidth$\m@th#1.$\hidewidth\cr
			\hidewidth\raise #2 \hbox{$\m@th#1.$}\hidewidth\cr
		}%
	}%
}
\makeatother

\makeatletter
\newenvironment{customlegend}[1][]{%
	\begingroup
	\pgfplots@init@cleared@structures
	\pgfplotsset{#1}%
}{%
	\pgfplots@createlegend
	\endgroup
}%

\def\addlegendimage{\pgfplots@addlegendimage}
\makeatother

\usetikzlibrary{shapes,arrows,shapes.geometric, calc, matrix}
\tikzset{%
	block/.style    = {draw, thick, rectangle}, %
	sum/.style      = {draw, circle}, %
	input/.style    = {coordinate}, %
	output/.style   = {coordinate} %
}
\tikzset{gain/.style={draw=black,text=black,inner sep=2pt, shape = isosceles triangle,shape border rotate=180},
	split/.style={draw, circle, fill=black,inner sep=0.4mm},
	field/.style={outer sep=0pt, draw, minimum height=8mm,
		minimum width=#1\linewidth,anchor=center}}

\newcommand{\inte}{$\int$}

\newcommand{\deleted}[1]{}

\begin{document}

\title{Optimal Feedback Control for Modeling Human-Computer Interaction}

\author{Florian Fischer}
\affiliation{%
  \institution{University of Bayreuth}
  \city{Bayreuth}
  \country{Germany}
}
\email{florian.j.fischer@uni-bayreuth.de}

\author{Arthur Fleig}
\affiliation{%
	\institution{University of Bayreuth}
	\city{Bayreuth}
	\country{Germany}
}
\email{arthur.fleig@uni-bayreuth.de}

\author{Markus Klar}
\affiliation{%
	\institution{University of Bayreuth}
	\city{Bayreuth}
	\country{Germany}
}
\email{markus.klar@uni-bayreuth.de}

\author{Jörg Müller}
\affiliation{%
	\institution{University of Bayreuth}
	\city{Bayreuth}
	\country{Germany}
}
\email{joerg.mueller@uni-bayreuth.de}

\renewcommand{\shortauthors}{Fischer, et al.}

\begin{abstract}

Optimal feedback control (OFC) is a theory from the motor control literature that explains how humans move their body to achieve a certain goal, e.g., pointing with the finger.
OFC is based on the assumption that humans aim to control their body optimally, within the constraints imposed by body, environment, and task.
In this paper, we explain how this theory can be applied to understanding Human-Computer Interaction (HCI) in the case of pointing.
We propose that the human body and computer dynamics can be interpreted as a single dynamical system.
The system state is controlled by the user via muscle control signals, and estimated from observations. Between-trial variability arises from signal-dependent control noise and observation noise.
We compare four different models from optimal control theory and evaluate to what degree these models can replicate movements in the case of mouse pointing.
We introduce a procedure to identify parameters that best explain observed user behavior.
To support HCI researchers in simulating, analyzing, and optimizing interaction movements, we provide the Python toolbox \textit{OFC4HCI}.
We conclude that OFC presents a powerful framework for HCI to understand and simulate motion of the human body and of the interface on a moment by moment basis.

\end{abstract}



\maketitle

\section{Introduction}
We address the problem of understanding, and modeling, how users control a virtual end-effector %
when interacting with computers.
Traditionally, the field of Human-Computer Interaction (HCI) has %
concentrated on models such as Fitts' Law~\cite{Fitts54, Fitts64}, predicting summary statistics of the movement such as movement time.
Recently, more attention has been paid to modeling the underlying process by which the end-effector is controlled, predicting not only movement time, but end-effector position, velocity, and acceleration sequences, as well as applied forces (e.g.,~\cite{Mueller17, Fischer21, Martin21, Do21}). %

We argue that, in order to understand how users control user representations (e.g., mouse pointer) \cite{Seinfeld20}, or virtual objects, %
the field of Human-Computer Interaction needs to learn more from human motor control.
While human motor control mainly addresses the question of how humans control the movement of their body, the theories developed there also apply to and can be adapted to the question of how humans control the state of a computer, e.g., movement of the mouse pointer. 

In the field of human motor control, %
modern understanding of human movement is based on the theory of optimal feedback control (OFC)~\cite{Todorov02, Diedrichsen10}.
This theory understands the human body, and possibly the environment the body is interacting with, as a dynamical system that can be controlled, e.g., via muscle control signals. 
Body and environment put constraints on this control, e.g., via the system dynamics and constant and signal-dependent motor noise.
The theory assumes that humans continuously observe the state of their own body and the environment they are interacting with, e.g., by processing visual and proprioceptive signals.
Humans are assumed to control their body optimally with respect to an internalized cost function, while respecting the constraints given by the system dynamics and motor noise.

We believe that the OFC framework enables a better connection between the field of HCI and recent advances in neighboring scientific disciplines, such as the study of human movement in motor control~\cite{Schmidt05, Flash85} and neuroscience~\cite{Shadmehr05}.
However, OFC models are not very well known in the field of HCI, yet.
In particular, it has not yet been shown whether these models, developed to model how humans control their body, can be used to model how users behave during interaction. %

The objective of this work is to examine the applicability of optimal feedback control to HCI, using the example of mouse pointing.
The contribution of this paper is fourfold:

First, we propose %
a unifying optimal control framework
for understanding movement in interaction with computers. %
This framework allows to predict the kinematics and dynamics of the entire movement trajectory, including, e.g., end-effector position, or muscle excitation.

Second, we present the first qualitative and quantitative evaluation to what degree different optimal control models (either open- or closed-loop, deterministic or stochastic) can replicate movements of the mouse pointer.
To the best of our knowledge, these models have not yet been evaluated quantitatively regarding their ability to predict movement trajectories during interaction.
We also discuss the possibilities and limitations of the presented models regarding their suitability for other HCI tasks such as target tracking, path-following, or handwriting.

Third, we propose a generic parameter fitting process, which can be used to identify the components of both the system dynamics and the cost function that best explain observed user behavior, using any desired optimal control model.
For each of the presented models, we systematically analyze the individual effects of the parameters and show how the proposed parameter fitting can be used to explain typical differences between users and/or task conditions, which would remain hidden when using summary statistics only.

Fourth, we provide \textit{OFC4HCI}, an open-source toolbox accessible from our GitHub repository\footnote{\url{https://github.com/fl0fischer/OFC4HCI}} that contains the underlying Python code of this paper. %
This toolbox includes easy-to-use scripts for three main use cases: running simulations of human pointing movements using any of the presented control methods, comparing the resulting trajectories to data from the Pointing Dynamics Dataset, and optimizing the parameters of a given control model.
While the focus of this toolkit currently is on (one- or multidimensional) pointing tasks, using the toolkit, extensions to other HCI tasks such as target tracking, keyboard typing, or gesture-based input methods are possible.

Our results suggest that stochastic OFC models are able to explain average user behavior significantly better than models that only account for simplified movement dynamics (second-order lag) or pure kinematic models (jerk minimization).
In addition, stochastic models such as LQG are able to fit the \textit{distribution} of entire trajectories, given a specific user and task condition.
Moreover, the fitting is significantly better than using the recently proposed Intermittent Control (IC) model~\cite{Martin21} with respect to both KL divergence~\cite{Kullback51} and the $2-$Wasserstein distance~\cite{Olkin82} serving as evaluation metrics.
The considered deterministic OFC model, which does not take into account any noise terms, 
is able to predict average user behavior, given a slightly modified cost function. 

We strongly believe that a proper modeling of the underlying control process %
can provide intuition to interface designers as to why users move the way they do during interaction, and enables a deeper understanding of the impact of parameters of the interface and input device on the process of interaction.
In the long term, such models could be used for automated optimization of the parameters of interaction techniques and input devices.
Models that work in real-time could further be used in predictive interfaces, which anticipate what the user wants to do and respond accordingly, such as pointing target prediction~\cite{Asano05}. %
While we will focus on the example of mouse pointing throughout this paper, the framework we present is generic and suitable for a wide range of pointing devices, %
using, e.g., joysticks, keyboards, pens, touch-based input, mid-air gestures, etc.  %
~ \\
\noindent The paper is structured as follows:

In Section~\ref{sec:rel-work}, we start with a short overview of existing models and methods from the fields of Human-Computer Interaction, Human Motor Control, and Optimal Control Theory. %
The proposed \textit{optimal control framework for Human-Computer Interaction} is
then introduced in Section~\ref{sec:oc-framework}.
The models %
presented in this paper are evaluated against an existing dataset of one-dimensional pointing movements, which is described in Section~\ref{sec:dataset}.
The generic parameter fitting process we use to identify the model parameters that best explain observed user behavior is described in Section~\ref{sec:param-fitting}.

In Sections~\ref{sec:dynamics}-\ref{sec:IC}, different optimal control models %
are presented, analyzed, and adapted to the case of mouse pointing. 
Moreover, the predicted movements are compared against user data.
Since this paper is also supposed to serve as a tutorial to Optimal Feedback Control for HCI researchers and interaction designers, we start with an analysis of the individual components of the OFC framework before combining them into a final model.
In Section~\ref{sec:dynamics}, we start with a basic model of movement dynamics, the second-order lag, which has been used to describe the overall human-computer system dynamics~\cite{Mueller17} and serves as a baseline for the presented optimal control models. %
The idea of (open-loop) optimal control is introduced in Section~\ref{sec:min-jerk}, using the minimum jerk model~\cite{Flash85}.
In Section~\ref{sec:OFC}, 
both movement dynamics and the assumption of optimality are %
integrated into one closed-loop OFC model, the Linear-Quadratic Regulator, which is based on the assumptions of linear dynamics and quadratic costs~\cite{Todorov05}.
From a didactic point of view, it is important to develop a thorough understanding of deterministic optimal feedback control before progressing to stochastic OFC (SOFC) models.
For this reason, %
we first start with the (substantially simpler) deterministic case, which can be used to predict \textit{average} human movement.
In Sections~\ref{sec:SOFC} and~\ref{sec:SOFC-2}, we extend this framework to the general stochastic case by adding different sensory-input models along with Gaussian motor and sensory noise (thus denoted as Linear-Quadratic Gaussian Regulator). %
We compare the stochastic OFC models to a recently proposed Intermittent Control model~\cite{Martin21}, which is briefly described in Section~\ref{sec:IC}.

Finally, both qualitative and quantitative comparisons between all considered models are given in Section~\ref{sec:model-comparison}.
Difficulties and limitations of the proposed framework with regard to its applicability to other HCI tasks are discussed in Section~\ref{sec:discussion-future}, together with some practical advice for HCI researchers, %
and conclusions are drawn in Section~\ref{sec:conclusion}.

\section{Related Work}\label{sec:rel-work}

In the field of \textbf{Human-Computer Interaction}, interaction is most commonly understood as a sequence of discrete actions, which is reflected %
in the classification of tasks, such as \textit{command selection} or \textit{target acquisition}~\cite{Buxton87}. %
In particular, movement, e.g., of the mouse pointer, is often reduced to summary statistics. %
The most prominent example is the dependency of movement time $\operatorname{MT}$ from distance $D$ and width $W$ of targets, which is described by Fitts' Law \cite{Fitts54, Fitts64} as \mbox{$\operatorname{MT}=a+b\operatorname{ID}$}, with Index of Difficulty (ID) usually defined as \mbox{$\operatorname{ID}=\log_2(D/W+1)$}~\cite{Mackenzie92}.
This affine relationship has shown to apply for a variety of tasks, including reciprocal tapping \cite{Fitts54}, mouse pointing and dragging \cite{Card78, Gillan90}, eye-gazing \cite{Ware86, Jagacinski85}, reaching with a joystick \cite{Jagacinski85, Card78}, and ellipse drawing \cite{Mottet94}.
A very good explanation of the information theoretic interpretation of Fitts' Law has been provided by Gori et al.~\cite{Gori18}.

While aggregated metrics of movement trajectories, e.g., \textit{movement variability} or \textit{movement offset}~\cite{MacKenzie01}, have been used to evaluate task accuracy since the early days of HCI~\cite{Jacob94},
predictive models of movement kinematics and dynamics are less common.
Exceptions include the works of Williamson~\cite{Williamson06, Williamson09}, which introduce an information-theoretic model of interaction with a focus on the amount of uncertainty that is apparent in different sensor and control channels, and M\"uller et al.\ \cite{Mueller17}, in which three feedback control models (without optimization) are compared regarding their ability to model mouse pointer movements.
However, the former model is originally designed for the specific needs of brain-computer interfaces, particularly inference of the user's intention based on noisy signal channels, whereas the latter models only describe the biomechanical apparatus, while high-level factors affecting the movement trajectory such as concrete task requirements or intrinsic motivations are neglected.
Ziebart et al.~\cite{Ziebart12} explore the use of inverse optimal control models for pointing target prediction.
They do not make particular a priori assumptions about the structure of the cost function.
Instead, they use an inverse optimal control approach to fit a generic function with a large number of parameters (36) to a dataset of mouse pointer movements.
While Ziebart et al.~\cite{Ziebart12} focus on the application of inverse optimal control to pointing target prediction, in this paper we investigate the ability of optimal (feedback) control models %
to model movement of the mouse pointer more quantitatively. %
From an engineering perspective, several interaction techniques that take into account the underlying end-effector kinematics have been proposed, including cursor jumping~\cite{Asano05, Murata98}, target expansion~\cite{McGuffin05}, and increased cursor activation areas~\cite{Mott14, Chapuis09}. 
These approaches are either based on target likelihood estimates~\cite{Ziebart12, Murata98} or extrapolate sensor data measured during runtime~\cite{Asano05, McGuffin05, Mott14}. %
Other methods compare observed trajectories to a set of pre-defined templates in order to predict the desired end-point (``kinematic template matching'')~\cite{Pasqual14}.
In general, these methods are restricted to the kinematic end-effector level, i.e., they ignore the movement dynamics of the human body (which play a crucial role for non-standard interaction techniques such as gesture-based input) and cannot be used to model interaction with dynamic objects.

In addition to their functional use in HCI,
movement dynamics have been a research focus within the field of \textbf{Motor Control} for a long time. %
Various models %
have been developed, all of which predict complete trajectories, e.g., end-effector position, velocity, and acceleration profiles over the entire movement (e.g., ~\cite{Bootsma04, Bullock88, Flash85, Flash13, Guiard93, Jagacinski03, Lacquaniti83, Mottet99, Plamondon98}).
Biomechanical and neural models, in addition, explain how these trajectories are dynamically generated.
This can be either done on the joint-, muscle-, or neuronal level, incorporating quantities internal to the human body such as joint angles, joint moments, muscle forces and activations, or neural excitation signals~\cite{Rosenbaum95, Uno89, Nakano99, Kawato93, Kawato96, Berret11}. %

Many models of motor control are also capable of modeling the characteristic between-trial variability that is typically observed in human movements.
This variability is mainly attributed to multiple sources of noise within the human biomechanical and neural system, most of which can be modeled as additive or multiplicative Gaussian random variables~\cite{Sutton67, Schmidt79, Slifkin99, Jones02, Todorov98_thesis, Todorov05}.
Signal-dependent noise terms, e.g., Gaussians with zero mean and with a standard deviation that linearly depends on the magnitude of the muscle control signal, are also considered responsible for the well-known \textit{speed-accuracy trade-off} in goal-directed human movements~\cite{Sutton67, Schmidt79, HarrisWolpert98}.
These noise terms have the effect that larger control signals, which might increase the speed of the end-effector, also result in larger deviations from the desired end-effector position.
Users thus face a trade-off between accurate achievement of the desired goal and fast, but noisy movements.

Another well-established finding from human motor control refers to the amount of information that is used when selecting a specific control signal.
Several experiments suggest that information that becomes available to the controller during the movement, e.g., proprioceptive and/or visual signals regarding the end-effector, are utilized to adjust control signals online and to account for unexpected perturbations~\cite{Woodworth99, Meyer88, Todorov02, Wang01, Thoroughman00}.
This is reflected by feedback control models %
of movement, which %
are able to explain how users correct errors and handle disturbances during the movement. 
An early closed-loop model %
has been provided by Crossman and Goodeve~\cite{Crossman83}. 
They assume that users observe hand and target and adjust their velocity as a linear function of the distance, as a first-order lag.
A physically more plausible extension of the first-order lag is the second-order lag~\cite{Crossman83, Langolf76}.
These dynamics can be interpreted as a spring-mass-damper system, where
a constant force is applied to the mass, such that the system moves to and remains at the target equilibrium.
Because of its simplicity and widespread use, we use this model as a baseline, called~\emph{2OL-Eq}. 
Other models of human movement include VITE~\cite{Bullock88} and the models of Plamondon \cite{Plamondon97}.

The \textit{desired trajectory hypothesis}~\cite{Kawato99} assumes that whenever disturbances occur (e.g., due to internal control noise or external perturbations), feedback is used to push the end-effector towards a predetermined, deterministic trajectory that results from a separated planning phase.
In contrast, Todorov and Jordan~\cite{Todorov02} have demonstrated that deviations are corrected only if they interfere with the task performance, i.e., deviations that are irrelevant for achieving the desired goal remain ignored.
This \textit{minimum intervention principle} %
particularly implies that all task-specific requirements (end-point position, movement time, accuracy, etc.) need to be reflected by an internal formulation that the controller has access to.

\textbf{Optimal control models} provide exactly this internal representation by assuming that humans try to behave optimally with respect to a certain internalized cost function.
Flash and Hogan~\cite{Flash85} proposed that humans aim to generate smooth movements by minimizing the jerk of the end-effector.
We call this model \emph{MinJerk} in the following.
Although the hypothesis that people aim to minimize jerk has been questioned, see, e.g., Harris and Wolpert~\cite{HarrisWolpert98}, the minimum jerk model is one of the most established models. For example, it has been successfully used by Quinn and Zhai~\cite{Quinn18} to model the shape of gestures on a word-gesture keyboard. %

Most modern theories of motor control are based on optimal feedback control (OFC), i.e., they combine the assumptions of optimality and continuously perceived feedback for closed-loop control. %
Excellent overviews of recent progress in OFC theory are provided by Crevecoeur et al.~\cite{Crevecoeur14} and Diedrichsen~\cite{Diedrichsen10}.
An early approach that models perturbed reach and grasp movements by using the minimum-jerk trajectory on a moment by moment basis was presented by Hoff and Arbib~\cite{HoffArbib93}.
A more general, more recent, and better known OFC model is the Linear-Quadratic Gaussian Regulator (LQG)~\cite{Hoff92, Loeb90}, which was mainly used by Todorov to model human movement from a sensorimotor perspective~\cite{Todorov98_thesis, Todorov02, Todorov05}. In this work, we will present and discuss the assumptions and limitations of this model, and analyze its applicability to standard HCI tasks such as mouse pointing. %

An important limitation of the LQG model (and many other optimal control models, e.g., \cite{Flash85, Uno89, HarrisWolpert98}) is that the exact movement time needs to be known in advance.
One way to circumvent this issue is to use infinite-horizon OFC~\cite{Jiang11, Qian13, Li18}, i.e., to formulate the optimal control problem on an infinite-time horizon.
With such models, (quadratic) distance and effort costs are usually applied continuously, resulting in an optimal trajectory that consists of both a transient phase (where the end-effector is moved towards the target) and a steady-state equilibrium (where the end-effector is kept at the target).
The movement time thus emerges implicitly from the optimal control problem. %

Another strand of literature that specifically deals with the duration of movement has produced the \emph{Cost of Time} theory \cite{Hoff94, Shadmehr10, Berret16}.
To account for the fact that humans value earlier achievement more than later achievement, this theory assumes that time is explicitly penalized with a certain cost function (usually hyperbolic or sigmoidal).  %

Recently, methods from the field of \textbf{Reinforcement Learning (RL)} have gained increased attention.
These methods are also based on the principles of optimal control theory, however, they do not require the system dynamics to be known in terms of equations and formulas, but solely rely on sampling from an environment that is usually assumed a black-box to the controller.
For this reason, they are generally applicable to arbitrarily complex systems including highly non-linear dynamics and discontinuous cost functions~\cite{Sutton18}.

Cheema et al.~\cite{Cheema20} have applied recent RL methods to predict fatigue during mid-air movements, using a torque-actuated linked-segment model of the upper limb.
Building on this work, it has recently been shown that RL applied to a more realistic upper-limb model allows to synthesize human arm movements that follow both Fitts' Law and the \nicefrac{2}{3} Power Law and can predict human behavior in mid-air pointing and path following tasks~\cite{Fischer21}.
Moreover, an extension to mid-air keyboard typing has been proposed~\cite{Hetzel21}.

In theory, policy-gradient RL methods can also be applied to model interaction on a muscular level, using state-of-the-art biomechanical models of the human body~\cite{Tieck18, Kidzinski18, Lee19, Nakada18}. However, the high complexity of the neuromuscular system 
has so far imposed considerable restrictions to each of these approaches, including the reduction of degrees of freedom~\cite{Tieck18, Nakada18, Kidzinski18} and the omission of muscle activation dynamics~\cite{Nakada18, Lee19}. %
Most importantly, for most RL algorithms no theoretical convergence guarantees exist, which complicates a profound interpretation or replication of the resulting simulation results~\cite{Sutton18}. For this reason, in this paper we focus on the well-understood theory of optimal control, as this allows us to use convergence guarantees more often, which makes us less reliant on intuition and experience. For example, the LQR introduced in Section~\ref{sec:OFC} is guaranteed to converge to the optimal movement trajectory, given a fixed set of parameters. This is a decisive advantage compared to pure RL-based methods, as it allows to compare optimal trajectories for different task conditions, cost functions, and user models.

In summary, the fundamental question of human movement coordination has produced a vast literature and a deep understanding of the nature of human movement.
Given that almost every interaction of humans with computers involves movement of the body, it is surprising that this field is little known, and applied, in HCI.
It is important to bear in mind, however, that most of
these theories intend to model movement of the human body per se.
In HCI, it is also relevant how users control the movement of user representations (e.g., mouse pointers) \cite{Seinfeld20} and virtual objects in the computer. 
Since the control of user representations and virtual objects is mediated by input devices, operating systems, and programs, requires high precision, and is often learnt very well, it is unclear how the theory of human motor control can be applied to the HCI context.
To our knowledge, it has not yet been investigated whether the above optimal motor control models can be applied to HCI tasks such as mouse movements. 
Adapting and validating such models regarding their ability to model HCI tasks such as pointing thus remains an open research question for HCI.

In order to leverage the strengths of recent motor control theory in the field of HCI, we believe that a general optimal control framework for Human-Computer Interaction is necessary, which can explain both \textit{how} and \textit{why} humans behave in interaction with arbitrary interfaces on a continuous level.
Such a framework constitutes a natural extension of the principle of ``designing interaction, not interfaces''~\cite{BeaudouinLafon04} by conceptualizing interaction based on neuroscientific, psychological, and biomechanical insights within one coherent and mathematically profound framework.

In the following section, we will introduce the optimal control framework for Human-Computer Interaction and explain its main constituents using the example of mouse pointing.

\section{Introducing the Framework}\label{sec:oc-framework}

\begin{figure}
	\centering
	
	\subfloat[Open-Loop Optimal Control]{
	\resizebox{0.99\linewidth}{!}{
    \begin{tikzpicture}[auto, thick, node distance=1.5cm, >=triangle 45,inner sep=2mm]
	
	\matrix (inte1) [matrix of nodes, inner sep=1mm, column sep=-\pgflinewidth, outer sep=0pt, minimum height=5mm, anchor=center] {
		\begin{tabular}{c} ~ \\ ~ \end{tabular} \\[0.1cm] |[field=.15, solid]| \begin{tabular}{c} Forward \\ Model \end{tabular} \\
	};
	
	\matrix (inte1a) at ($(inte1-2-1) + (3.5, 0)$) [matrix of nodes, inner sep=1mm, column sep=-\pgflinewidth, outer sep=0pt, minimum height=5mm, anchor=center] {
		|[field=.15, solid]| \begin{tabular}{c} Computation: \\ $\min_{u} J_N(x,u)$ \end{tabular} \\
	};

	\draw node (split-input1) at ([yshift=-0.7cm]inte1a-1-1.south) {};

	\draw node (humancontrollerlabel) at ($(inte1-1-1) + (1.75, 0)$) {\begin{tabular}{c} Human \\ \textbf{Controller} \end{tabular}};

	\node [draw, rectangle, opacity=1, minimum width=7cm, minimum height=2.5cm] (humancontroller) at ($(inte1-1-1)!0.5!(inte1-2-1) + (1.75, -0.1)$) {};

	\draw node [ellipse, draw=black, text=black, text opacity=1] (input1) at ($($(split-input1)+(-10, 0)$)!($(inte1-2-1)+(-3.35, 0)$)!(split-input1)$) {Task};
	
	\draw[solid,-](input1) -- node[pos=0.05] {$J_{N}$} (split-input1.center);
	\draw[solid,->](split-input1.center) -- (inte1a-1-1.south);

	\matrix (inte2) [matrix of nodes, ampersand replacement=\&, right of=inte1a-1-1, node distance=7.35cm, column sep=-\pgflinewidth, outer sep=0pt, minimum height=5mm, anchor=center] {
		|[field=.22]| \begin{tabular}{c} Human \\ \textbf{Body Dynamics} \end{tabular} \&  |[field=.15]| \begin{tabular}{c} \textbf{Interface} \\ \textbf{Dynamics} \end{tabular} \\
	};

	\node [draw, rectangle, opacity=1, minimum width=7.2cm,minimum height=2.5cm] (hcidynamics) at ($(inte2) + (0, 0.4)$) {};
	\draw node [above of=hcidynamics, node distance=0.75cm, xshift=0cm, align=left] () {\textbf{Human-Computer System Dynamics}};

	\draw [solid, ->] (inte1-2-1) -- (inte1a-1-1);
	
	\draw[->] (inte1a-1-1.east) -- node[xshift=-0.2cm] {\begin{tabular}{c} Muscle \\ Control \\ Signals $u^{*}$ \end{tabular}} (inte2-1-1.west);

	\draw node[style={inner sep=0,outer sep=0}] (split2) at ($(inte2-1-2) - (0, 2)$) {\begin{tabular}{c} Optimal State Trajectory $x^{*}$ \end{tabular}};
	\draw[->](inte2-1-2.south) -- node {} (split2.north);

	\begin{scope}[transparency group, opacity=0.2]

		\node [draw, rectangle, fill=blue, dashed, minimum width=13.4cm,minimum height=4cm] (humanmodel) at ($(humancontroller) + (3, 0)$) {};
		
		\node [draw, rectangle, fill=red, dashed, minimum width=3.4cm,minimum height=4cm] (interfacemodel) at ($(humancontroller) + (11.25, 0)$) {};
		
	\end{scope}

	\draw node [above of=humanmodel, node distance=2.5cm, color=blue, opacity=0.7] () {\textbf{\Large Human}};
	\draw node [above of=interfacemodel, node distance=2.5cm, color=red, opacity=0.7] () {\textbf{\Large Computer}};

	\end{tikzpicture}
	}} 
	\\
	
	\subfloat[Closed-Loop Optimal Control (Discrete-Time Formulation)]{
	\resizebox{0.99\linewidth}{!}{
\begin{tikzpicture}[auto, thick, node distance=1.5cm, >=triangle 45,inner sep=2mm]
	
	\matrix (inte1) [matrix of nodes, inner sep=1mm, column sep=-\pgflinewidth, outer sep=0pt, minimum height=5mm, anchor=center] {
		\begin{tabular}{c} ~ \\ ~ \end{tabular} \\[0cm] |[field=.15, solid]| \begin{tabular}{c} Forward \\ Model \end{tabular} \\[0.6cm] |[field=.15, solid]| \begin{tabular}{c} Computation: \\ $\min_{u} J_N(x,u)$ \end{tabular} \\[1.2cm] ~ \\[0.6cm] ~\\
	};
	
	\matrix (inte1a) at ($(inte1-3-1)!0.5!(inte1-5-1) + (3.5, -0.625)$) [matrix of nodes, inner sep=1mm, column sep=-\pgflinewidth, outer sep=0pt, minimum height=5mm, anchor=center] {
		|[field=.15, solid]| \begin{tabular}{c} Control Strategy \\ $u_{n} = \pi(\hat{x}_{n})$ \end{tabular} \\[1.75cm] \begin{tabular}{c} Human \\ \textbf{Observer} \end{tabular} \\[0.75cm] |[field=.15, solid, inner sep=0]| \begin{tabular}{c} Forward \\ Model \end{tabular} \\
	};
	
	\draw node[draw, circle, inner sep=0.2mm] (sensordiff) at ($(inte1a-3-1.north) + (0, 0.5)$) {$-$ \hfill $+$};
	
	\draw node (humancontrollerlabel) at ($(inte1-1-1) + (1.75, 0)$) {\begin{tabular}{c} Human \\ \textbf{Controller} \end{tabular}};

	\node [draw, rectangle, opacity=1, minimum width=7cm, minimum height=4.3cm] (humancontroller) at ($(inte1-1-1)!0.5!(inte1-3-1) + (1.65, -0.1)$) {};
	
	\node [draw, rectangle, opacity=1, minimum width=3cm, minimum height=3cm] (humanobserver) at ($(inte1a-2-1)!0.5!(inte1a-3-1)$) {};
	
	\draw node [ellipse, left of=inte1-3-1, node distance=3.6cm, draw=black, text=black, text opacity=1] (input1) {Task};
	
	\draw[dashed,->](input1) --  node[near start] {$J_{N}$} (inte1-3-1.west);

	\draw node (testnode1) at ($(inte1-1-1)!0.5!(inte1-3-1) + (3.5, 0)$) {};
	\matrix (inte2) [matrix of nodes, ampersand replacement=\&, right of=testnode1, node distance=7.75cm, column sep=-\pgflinewidth, outer sep=0pt, minimum height=5mm, anchor=center] {
		|[field=.22]| \begin{tabular}{c} Human \\ \textbf{Body Dynamics} \end{tabular} \&  |[field=.15]| \begin{tabular}{c} \textbf{Interface} \\ \textbf{Dynamics} \end{tabular} \\
	};
	\draw node (testnode2) at ($(inte1-4-1)!0.5!(inte1-5-1) + (3.5, 0)$) {};
	\matrix (inte3) [matrix of nodes, ampersand replacement=\&, right of=sensordiff, node distance=7.85cm, column sep=-\pgflinewidth, outer sep=0pt, minimum height=5mm, anchor=center] {
		|[field=.2, minimum height=1.3cm]| \begin{tabular}{c} Human \\ \textbf{Perception} \end{tabular} \&  |[field=.145, minimum height=1.3cm]| \begin{tabular}{c} \textbf{Display} \end{tabular} \\
	};

	\node [draw, rectangle, opacity=1, minimum width=7.4cm,minimum height=2.5cm] (hcidynamics) at ($(inte2) + (0, 0.4)$) {};
	\draw node [above of=hcidynamics, node distance=0.75cm, xshift=0cm, align=left] () {\textbf{Human-Computer System Dynamics}};
	
	\node [draw, rectangle, opacity=1, minimum width=7.4cm,minimum height=2.5cm] (hciview) at ($(inte3) + (0, 0.4)$) {};
	\draw node [above of=hciview, node distance=0.625cm, xshift=0cm, align=left] () {\textbf{Feedback}};
	
	\draw [dashed, ->] (inte1-2-1) -- (inte1-3-1);
	\draw [dashed, ->] (inte1-3-1) -- (inte1a-1-1);
	
	\draw node (split5) at ($($(testnode1.east) - (5, 0)$)!(inte1a-1-1.north)!(testnode1.east)$) {};
	\draw[-] (inte1a-1-1.north) -- (split5.center);
	\draw[->] (split5.center) -- node[xshift=0.4cm] {\begin{tabular}{c} Muscle Control \\ Signal $u_{n}$ \end{tabular}} (inte2-1-1.west);
	
	\draw node [split] (split3) at ($($(humancontroller.north east)!(testnode1.east)!(humancontroller.south east)$)!0.625!(inte2-1-1.west)$) {};
	\draw node (split4) at ($(inte1a-3-1.east)!(split3)!($(inte1a-3-1.east) + (5, 0)$)$) {};
	\draw[solid] (split3.center) -- node[right, xshift=-0.3cm, yshift=0.2cm] {} (split4.center);
	\draw[->] (split4.center) -- node[below, xshift=0.3cm, yshift=0.1cm] {\begin{tabular}{c} Efference \\ Copy \end{tabular}} ($(inte1a-3-1.north east)!(split4.center)!(inte1a-3-1.south east)$);

	\draw node (split1) at ([xshift=0.2cm]inte2.east) {};
	\draw node [split] (split2) at ($(split1)!(inte3-1-2.east)!($(split1) - (0, 10)$)$) {};
	\draw[-](inte2-1-2.east) -- node {} (split1.center);
	\draw[-](split1.center) -- node[left, yshift=0.5cm, xshift=0.1cm] (statexn1) {State $x_{n+1}$} (split2.center);
	\draw[->](split2.center) -- node {} ($(inte3-1-2.north east)!(split2.center)!(inte3-1-2.south east)$);

	\draw node[style={inner sep=0,outer sep=0}] (splitout) at ($(split2) - (1.4, 2.75)$) {\begin{tabular}{c} Optimal State Trajectory $x^{*}$ \end{tabular}};
	\draw[->, dotted](split2.center) -- node {} ([xshift=1.4cm]splitout.north);

	\draw[solid] ([xshift=-1cm]inte1a-3-1.north) -- ($(sensordiff.west)!([xshift=-1cm]inte1a-3-1.north)!($(sensordiff.west) + (-3, 0)$)$);
	\draw[->] ($(sensordiff.west)!([xshift=-1cm]inte1a-3-1.north)!($(sensordiff.west) + (-3, 0)$)$) -- (sensordiff.west);
	
	\draw[-](inte3-1-1.west) -- node[above, xshift=-0.75cm, yshift=-0.2cm] {\begin{tabular}{c} Observation \\ $y_{n+1}$ \end{tabular}}($(humanobserver.north east)!(inte3-1-1.west)!(humanobserver.south east)$);
	\draw[->]($(humanobserver.north east)!(inte3-1-1.west)!(humanobserver.south east)$) -- (sensordiff.east);
	
	\draw[->]([yshift=-0cm]humanobserver.north) -- node[left, xshift=0.2cm, yshift=-0.25cm, font=\normalsize] {\begin{tabular}{c} State \\ Estimate $\hat{x}_{n+1}$\end{tabular}} ([yshift=-0cm]inte1a-1-1.south);

	\begin{scope}[transparency group, opacity=0.2]

		\node [draw, rectangle, fill=blue, dashed, minimum width=13.8cm,minimum height=9.5cm] (humanmodel) at ($(humancontroller)!0.5!(humanobserver) + (2.255, 0.2)$) {};
		
		\node [draw, rectangle, fill=red, dashed, minimum width=3.4cm,minimum height=9.5cm] (interfacemodel) at ($(humancontroller)!0.5!(humanobserver) + (10.875, 0.2)$) {};
		
	\end{scope}

	\begin{customlegend}[legend columns=1,legend style={draw=black,column sep=1ex,at={($(inte1-2-1)-(0, 6)$)}, anchor=center},legend entries={Before Movement, During Movement, After Movement}]
		\addlegendimage{black,->,dashed}
		\addlegendimage{black,->,solid}
		\addlegendimage{black,->,dotted}
	\end{customlegend}
	
\end{tikzpicture}
	}}
	\caption{In our \textbf{optimal control framework for Human-Computer Interaction}, %
	the user is assumed to \emph{control} the state~$x$ of the interactive system, which incorporates both the body state (e.g., arm and finger position) and the interface state (e.g., mouse pointer position and velocity), and which evolves according to the Body and Interface Dynamics. 
	We assume that the user computes the controls~$u$ through \emph{optimization}, i.e., by minimizing a cost function~$J_N$ (e.g., incorporating time or effort costs) that depends on the task. \\
	\textbf{(a)} In an \textit{open-loop model}, this calculation is only based on an internal Forward Model of the Human-Computer System Dynamics. The optimal state trajectory $x^*$ is obtained by applying the resulting %
	muscle control signals $u^*$ in one forward pass. 
	The Forward Model does not have to coincide with the 
	System Dynamics. 
	\\
	\textbf{(b)} A \textit{closed-loop model} takes into account effects that appear only after execution. 
	The key difference in the Computation block is that, instead of optimal control signals $u^*$, it yields an 
	optimal Control Strategy $\pi$ that is computed before movement onset.
	At each time step $n$, this Control Strategy is used to map an arbitrary (estimated) state $\hat{x}_{n}$ to the corresponding optimal control $u_{n}$. 
	Based on the resulting state $x_{n+1}$, an observation $y_{n+1}$ is obtained via Feedback, which incorporates descriptions of both the Display and the Human Perception. 
	The Human Observer then compares this observed state to an expected state it computes using an efference copy of the current control signal $u_n$ and the Forward Model. 
	Based on the resulting difference between expected and observed signals, an internal state estimate $\hat{x}_{n+1}$ is computed and used to select the next control $u_{n+1}$, and so on. 
	}
	\label{fig:genmodel_extended}
	\Description{Fully described in caption and text.}
\end{figure}
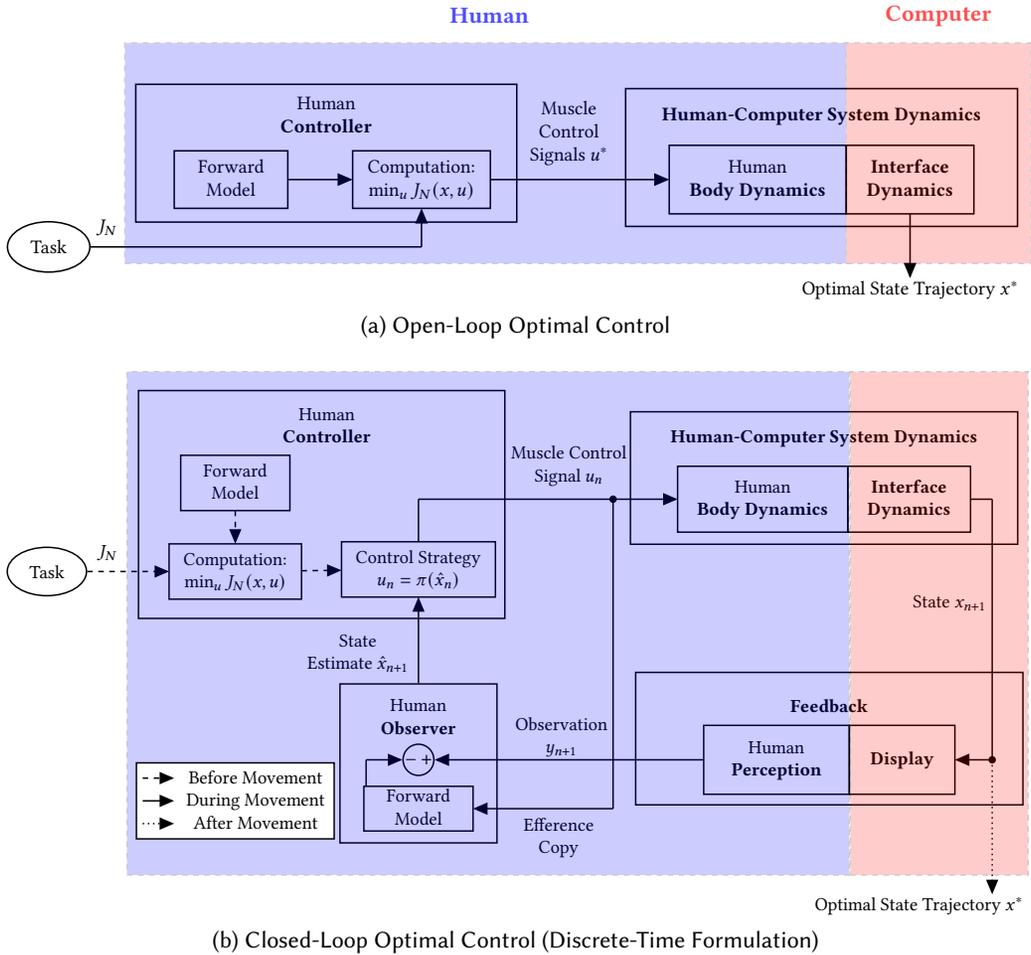

Modern motor control theory assumes that humans aim at controlling their movements optimally, given the constraints imposed by the body and environment.
Important constraints are imposed by physics, e.g., via Newton's second law, i.e., force equals mass times acceleration, %
and by the muscles, which cannot create forces instantaneously, but need to build up muscle activation (and thus force) over time.
In the case of Human-Computer Interaction, constraints are imposed not only by the human body, but also by the input device, sensor, and processing within the computer.
Further, the human perceptual system does not have direct access to the state of the world, but can only observe certain variables that depend on the state and needs to build up an internal estimate of the true world state over time.

Since these properties are characteristic for almost any HCI task, we propose a generic \textit{optimal control framework for Human-Computer Interaction}, which consists of four submodels that continuously interact with each other:

\begin{itemize}
	\item the \textbf{Human-Computer System Dynamics}, which describe the biomechanics of the considered body parts as well as the dynamics of the resulting interaction with the application interface via an input device,
	\item the \textbf{Human Controller}, i.e., the decisive part of the brain, which selects the new muscle control signals,
	\item the \textbf{Feedback}, which models how environmental information is sensed by the human,
	\item and the \textbf{Human Observer}, a cognitive model for how perceived sensory signals from the Feedback are processed and evaluated.
\end{itemize}

Since the computer operates in discrete time, we use discrete-time dynamics, i.e., we consider time steps $n\in\{0,\dots,N\}$ up to a final step $N\in \mathbb{N}$, with each time step corresponding to $h$ seconds.
However, the proposed framework is designed to be as general as possible and the following explanation also applies to the continuous-time case. %

Figure~\ref{fig:genmodel_extended} illustrates the relationship between the four components of the Human-Computer Interaction loop, specifically distinguishing between \textit{open-loop} and \textit{closed-loop} optimal control models.
In the following, we will give a brief description of the proposed framework, with focus on the differences between both variants.
Subsequently, the four submodels are explained in detail in separate subsections, introducing a more technical and mathematically rigorous notation.
Readers who are already familiar with the differences between open- and closed-loop models can proceed directly to Section~\ref{sec:human-computer-system-dynamics}.

\subsection{Open-loop vs. closed-loop models}

\textbf{Open-loop models}, as depicted in Figure~\ref{fig:genmodel_extended}(a), cannot infer any information from the \textit{System Dynamics} after applying muscle control signals $u$.
For this reason, the \textit{Human Controller} block in Figure~\ref{fig:genmodel_extended}(a) does not depend on the output of the \textit{System Dynamics} block, but depends only on an internal \textit{Forward Model}. %
In particular, the \textit{Feedback} and \textit{Human Observer} blocks are not part of the generic open-loop framework at all. %
For this reason, open-loop models allow for a strict separation between planning and execution phase of a movement, i.e., trajectories $x^{*}$ that are optimal with respect to the objective function $J_N$ can be obtained in a two-step process.
First, an (open-loop) \textit{optimal control problem} (OCP) is solved (\textit{Computation} block), i.e., a sequence of controls $u^{*}=(u_{n}^{*})_{n\in\{0,\dots,N-1\}}$ is found such that $J_{N}(x, u)$ becomes minimal among all permissible control sequences $u$, %
given an initial state (see Section~\ref{sec:human-controller} for more details). %
Second, the resulting optimal control sequence 
is applied in a \textit{forward pass}, i.e., %
the system dynamics %
are evaluated at each time step $n\in\{0, \dots, N-1\}$ to obtain the subsequent optimal state $x_{n + 1}^{*}$.

Note that if the system dynamics are deterministic and the forward model internally used to compute $u^{*}$ matches these dynamics, there is no advantage in computing the optimal controls \textit{online}, that is, only at the time they are executed during the forward pass.
However, there are a few scenarios where the internal forward model used for optimization cannot fully predict the actual behavior of the human body and interface dynamics, i.e., the outcome of the Human-Computer System Dynamics block in Figure~\ref{fig:genmodel_extended}.
This is particularly the case
\begin{itemize}
	\item if the OCP is stochastic, e.g., in the case of motor noise,
	\item if there is a mismatch between internal model and actual system dynamics, or
	\item if unexpected disturbances occur that the internal model did not account for.
\end{itemize}
If one of these assumptions holds (which usually is the case in practice), the controller will benefit from any information it receives \textit{during execution}, as this allows to condition the choice of an individual control $u_{n}$ on the true state %
in a \textbf{closed-loop} manner,
instead of using a prior state estimate. %
Note that the feedback loop immediately eliminates the strict separation between planning and execution phase, which is prevalent for open-loop models.
Instead, the optimal control sequence $u^{*}$ needs to be computed online in an iterative manner.
In particular, all information available to the controller at a time step $n\in\{0,\dots, N-1\}$ is used to compute the muscle control signal $u_{n}^{*}$ at this time step, which is applied to the \textit{Human Body Dynamics}. In combination with the \textit{Interface Dynamics}, this results in a new system state $x_{n+1}$.
The state (or partial information thereof, see Section~\ref{ssec:feedback-and-observer}) is then fed back to the controller, which again selects the next control $u_{n+1}^{*}$ based on this information, i.e., a \textit{sensorimotor} control loop between Human and Computer is established (see Figure~\ref{fig:genmodel_extended}(b)).
In particular, the optimal state trajectory $x^{*}$ does not only depend on the control sequence $u$, but also vice versa.
Since in such models, feedback is given to the controller during execution, %
these models are often denoted as %
\textbf{optimal feedback control (OFC) models}~\cite{Diedrichsen10, Crevecoeur14}.

It is important to note that with many OCP solution methods that take into account feedback during execution (including the ones presented in this paper), the actual optimization can be performed offline, i.e., before the controls are applied to the actual system dynamics. Instead of a single optimal control sequence $u^{*}$, such methods usually yield an optimal \textit{Control Strategy}, %
that is, a function $\pi:\mathcal{X}\rightarrow\mathcal{U}$ that maps an arbitrary state $x\in\mathcal{X}$ 
to a corresponding control $u\in\mathcal{U}$ that is 
optimal starting from this state.

In the later sections, we will present both an open-loop model (Section~\ref{sec:min-jerk}) as well as different variants of one of the most widely used OFC models (Sections~\ref{sec:OFC}-\ref{sec:SOFC-2}).

\subsection{Human-Computer System Dynamics}\label{sec:human-computer-system-dynamics}
The Human-Computer System Dynamics form the basis of each optimal control model of Human-Computer Interaction and consist of two parts: the Human Body Dynamics and the Interface Dynamics.

\emph{Human Body Dynamics.}
Given a vector of neural muscle control signals~$u$, 
the Human Body Dynamics describe how these signals are transformed into joint torques and accelerations. 
The corresponding joint postures are obtained via integration. 
If required, kinematic models that map joint postures to world-centered positions of arbitrary body parts can be included as well, e.g., to get more realistic movement of the wrist or the index finger based on the computed joint angles.

\emph{Interface Dynamics.}
In interaction with computers, the forces and accelerations resulting from body dynamics are applied to the physical input device, e.g., the mouse device. Inertial properties of this \textit{physical end-effector} determine the input device motion, which is sensed, filtered, and mapped to a motion of the corresponding \textit{virtual end-effector}. For example, in the case of mouse pointing, the mouse pointer position might result from the application of a pointing transfer function to the mouse device velocity measured via optical sensors~\cite{Casiez08, Casiez11}. The virtual end-effector is then used for interaction with graphical user interfaces.
These dynamics of the computer part, including physical properties of the input device as well as visualizations and animations that appear from interaction with buttons, sliders, etc., are summarized within the Interface Dynamics block in Figure~\ref{fig:genmodel_extended}.

Combined, the Human-Computer System Dynamics yield the new state vector $x$, which incorporates all relevant information about the current state of the human body, of the physical and/or the virtual end-effector, and of the applications.
In the discrete-time formulation used in this paper,
both the state and control vectors $x_{n}$ and $u_{n}$ are given at 
time steps $n\in\{0,\dots,N\}$ up to a final step $N\in \mathbb{N}$, with each time step corresponding to $h$ seconds.
The next state $x_{n+1}$ 
depends on the current state $x_n$ and control $u_n$, which in most cases can be formalized as %
\begin{equation}\label{eq:generic-nonlinear-dynamics}
	x_{n+1}=f(x_{n}, u_{n}),
\end{equation}
where $x_0$ is a given initial state.\footnote{This is closely related to a continuous-time formulation based on differential equations, where the control is assumed to be piecewise constant (i.e., it only changes once every $h$ seconds).}

This system dynamics function $f$ can be either \textit{deterministic} or \textit{stochastic}.
In the deterministic case, starting from the current state $x_{n}\in\R^{k}$ ($k\in\N$) and applying a control $u_{n}\in\R^{m}$ ($m\in\N$), the subsequent state $x_{n+1}\in\R^{k}$ is uniquely determined by the function $f:\R^{k}\times\R^{m}\rightarrow\R^{k}$.
In the stochastic case, $x_{n+1}$ randomly emerges from a set of possible successor states, according to a conditional probability distribution $p$, i.e., $x_{n+1}=f(x_{n}, u_{n}) \sim p(\cdot \mid x_{n}, u_{n})$.
Examples of stochastic systems include, e.g., \textit{Body Dynamics} with internal motor noise or \textit{Interface Dynamics} with noisy input signals.

It is important to note that the controller, which is described in Section~\ref{sec:human-controller}, is agnostic to the partitioning of the system dynamics into effects attributed to the \textit{Body Dynamics} and effects attributed to the \textit{Interface Dynamics}.
All the controller needs to know is the \textit{overall} system dynamics $f$ (or an internal approximation of it), which maps an arbitrary state-control pair $(x_{n}, u_{n})$ to the subsequent state $x_{n+1}$ reached after $h$ seconds.
Thus, an optimal control model of Human-Computer Interaction can be instantiated in two ways.
The first one is to include accurate submodels of arbitrary granularity (e.g., a separate model for each muscle activation, arm and hand dynamics, input device dynamics, and application dynamics), and combine them along the interaction loop into one aggregated system dynamics function~$f$.
However, the framework also allows to test whether some generic dynamics, such as a spring-mass-damper system or simplified muscle activation dynamics, are suitable to model the \textit{overall Human-Computer System Dynamics} for a given task setting.
The focus of this paper will be on the latter approach, since we aim to start with an easily understandable and well-established model from the field of Human Motor Control, and test whether these dynamics are applicable to the context of mouse pointing. We believe that this approach is well suited for introducing optimal control methods to HCI without going too much into (biomechanical) detail.

The system dynamics of all models considered in this paper are linear in both the state and the control, i.e.,
	\begin{equation}\label{eq:LQR Control1}
		x_{n+1}=f(x_{n}, u_{n})=Ax_{n}+Bu_{n}.
	\end{equation}
Here, the matrix $A$ describes how the dynamics evolve when no control is exerted.
The matrix $B$ describes how the control influences the system.

At first glance, this assumption seems to be very limiting, especially with regards to the complexity of the human neuro- and biomechanical system, as well as of most interaction methods and application GUIs, 
However, the tools and methods proposed in this paper for the case of linear dynamics are also beneficial for more complex models of interaction, which, for example, include muscle-driven models of the human body or non-linear pointer acceleration functions.
Using a reference trajectory, it is always possible to linearize a non-linear system around this particular trajectory in order to obtain a linear system of the form~\eqref{eq:LQR Control1}, which locally approximates the non-linear one.
While linearization-based extensions of the considered optimal control models to the non-linear case have been proposed~\cite{Li04, Todorov05b, Tassa14}, an application of these methods to typical HCI tasks would be an interesting area for future work.

The main advantage of using linear dynamics is that, when combined with quadratic costs and Gaussian noise, the resulting optimal control problem (see Section~\ref{sec:human-controller}) can be solved analytically and thus quickly and exactly. %
Finally, the linear case is easier to understand and formalize and thus well suited for the explanatory purposes of this paper.

In the case of mouse pointing, which usually requires only small movements of the arm, the hand, and the input device, linearization around a single trajectory, i.e., using constant system matrices $A$ and $B$ as in~\eqref{eq:LQR Control1}, is a reasonable initial approach to model (moderate) mouse movements.
Indeed, we will show that linear system dynamics %
can account for many phenomena that are characteristic in mouse pointing.

\subsection{Human Controller}\label{sec:human-controller}

In general, various control sequences can produce the same movement trajectory.
For example, the arm can rest on the table or stay in the air, as long as the mouse device is controlled appropriately. %
This is referred to as the \textit{joint-redundancy problem}~\cite{Bullock88}. %
For a large number of degrees of freedom,
e.g., motor signals that are applied to individual muscles, the same goal can be achieved with different controls, raising the question of which control is actually chosen by the central nervous system (CNS) and why.\footnote{Moreover, in the case of muscle-driven simulations, the set of feasible controls is relatively small compared to the total decision space, which makes it even less clear how appropriate controls are internally found~\cite{ValeroCuevas15}.}
This fundamental question, however, cannot be answered using movement dynamics alone.
Instead, the optimal control framework has been proposed to address this question~\cite{Friston11, Kording07, Todorov02}.

\subsubsection*{Optimal Control Problems (OCPs)}

Optimal control methods make use of a specific cost function, which is to be minimized.
Previous approaches include minimization of either jerk~\cite{Flash85, HoffArbib93}, peak acceleration~\cite{Nelson83}, end-point variance~\cite{HarrisWolpert98}, duration~\cite{Artstein80, Tanaka06}, or torque-change~\cite{Uno89}, among others.
Different objectives can be combined in one cost function to model trade-offs, e.g., between accuracy and effort~\cite{Todorov98_thesis, Li04}, accuracy and stability~\cite{Liu07}, or jerk and movement time~\cite{Hoff94}.
Recently, it has been argued that several abilities associated with intelligence such as knowledge, perception, or imitation naturally emerge from behaving optimally with respect to an ultimate goal~\cite{Silver21}. 
For goals that can be expressed by an adequate cost function, this particularly suggests that the optimal control framework is able to explain intelligent behavior.

In general, the (finite-horizon) \textit{discrete-time optimal control problem} is given by
\begin{subequations}\label{eq:ocp-discrete}
\begin{align}
	\begin{gathered}
		\textsl{Minimize} \quad J_{N}(x,u)= g_{N}(x_{N}) + \sum_{n=0}^{N-1} g(x_{n},u_{n}) \\
		\textsl{with respect to } u= (u_{n})_{n\in\{0,\dots,N-1\}}\subset \mathcal{U} \subset \R^{m}, %
	\end{gathered} 
\end{align}
where $x=(x_{n})_{n\in\{0,\dots,N\}} \subset \mathcal{X} \subset \R^{k}$ satisfies
\begin{align}
	\begin{gathered}
		x_{n+1}=f_{\text{int}}(x_{n},u_{n}), \enspace n\in\{0,\dots,N-1\}, \\
		x_{0}=\bar{x}_{0} %
	\end{gathered}
\end{align}
for some given initial state $\bar{x}_{0}\in\mathcal{X} \subset\R^{k}$.
\end{subequations}

Here, $f_{\text{int}}$ denotes the dynamics of the internal model used for optimization.
In Figure~\ref{fig:genmodel_extended}, this corresponds to the \textit{Forward Model} block within the Human Controller. In most optimal control models, including those considered in this work, the internal model is assumed to be exact, i.e., it matches the actual system dynamics (corresponding to the \textit{Human-Computer System Dynamics} block in Figure~\ref{fig:genmodel_extended}), which are analogously described by some function $f$. %
The \textit{objective function} $J_{N}(x, u)$ that we want to minimize consists of some \emph{terminal cost} function $g_{N}: \R^{k}\rightarrow\left[0,\infty\right[$ and the sum of \emph{running costs} $g: \R^{k}\times\R^{m}\rightarrow\left[0, \infty\right[$ accumulated over $N$ time steps.
These might be chosen dependent on the task under consideration. 
For example, in a tracking task, the distance between end-effector and target could be penalized in each step, whereas in a steering task, large costs might be applied whenever one of the bounds is reached.
The sets $\mathcal{X}$ and $\mathcal{U}$ can be used to restrict the states and controls that are permissible at each time step. 
In the following, however, we will set $\mathcal{X}=\R^{k}$ and $\mathcal{U}=\R^{m}$, i.e., we do not impose any restrictions.
The initial state $x_{0}=\bar{x}_{0}$ is assumed to be given, and the (unique) optimal solution $(x, u)$ to an OCP (assuming that it exists) is denoted by $(x^{*}, u^{*})$ in the following.

For \textit{deterministic OCPs}, both the internal model and the actual system dynamics are given by deterministic functions $f_{\text{int}}:\R^{k}\times\R^{m}\rightarrow\R^{k}$ and $f:\R^{k}\times\R^{m}\rightarrow\R^{k}$, respectively. 
For \textit{stochastic OCPs}, these dynamics are replaced by conditional probability distributions %
(see Section~\ref{sec:human-computer-system-dynamics}).
It is important to note that albeit in stochastic OCPs, the concrete successor state $x_{n+1}$ resulting from the application of a hypothetical control $u_{n}$ in the state $x_{n}$ is not available to the controller during optimization, the underlying transition probability density function $p(\cdot\mid x_{n}, u_{n})$ usually is.\footnote{In the case of unknown transition dynamics $p$ (or $f$, in the deterministic case), the controller would need to rely on sampling transitions from the environment in order to be able to estimate the expected future costs of different controls. This problem is addressed in \textit{Model-Free Reinforcement Learning}.}
Stochastic OCPs are thus capable of modeling the between-trial variability that typically occurs in human movement.

The methods to find the optimal solution of an OCP depend on its problem structure, i.e., the properties of the cost function (e.g., differentiability, convexity), the system dynamics (e.g., linearity, stochasticity), and the permissible state space $\mathcal{X}$ and control space $\mathcal{U}$. %
Often, the optimal solution can only be determined approximately, using numerical methods such as Multiple Shooting~\cite{Bock84}, Direct Collocation~\cite{Betts10}, or Reinforcement Learning~\cite{Sutton18}.
However, in some cases an explicit solution formula exists that yields the exact (and unique) optimal control sequence $u^{*}$.
In this paper, we will focus on the most widely known class of OCPs that allows for such an analytical solution scheme: those with linear system dynamics and convex, quadratic costs.%

\subsection{Feedback \& Human Observer}\label{ssec:feedback-and-observer}

In the closed-loop case of the optimal control framework for Human-Computer Interaction~(Figure~\ref{fig:genmodel_extended}(b)), the \textit{Feedback} block accounts for the fact that usually not all information included in the state $x$ are (immediately) available to the user.
First, the visual output, which is shown on the \textit{Display}, is created based on the respective state components.
This information is then sensed and processed by the \textit{Human Perception}, which describes how visual, proprioceptive, and/or auditive signals are perceived and integrated into the stream of observations $y=(y_{n})_{n\in\{0,\dots,N-1\}}$ the controller can condition on.
The same holds for information on the own body state, which is directly obtained from the \textit{Human Body Dynamics}, e.g., via proprioceptive input signals.

In general, these observations might be delayed, noisy, or incomplete.
To decide for an appropriate control $u_{n+1}$ at time step $n+1$, thus an estimate $\hat{x}_{n+1}$ of the true current state $x_{n+1}$ is required. 
This estimate is computed by the \textit{Human Observer}, which compares the observed state~$y_{n+1}$ to an expected state it computes using an efference copy of the most recent control signal $u_n$ and the Forward Model.\footnote{In this paper, we assume perfect system knowledge, i.e., the forward model consists of the same system dynamics and perception functions as used in the actual interaction loop.}
Based on the resulting difference between expected and observed signals, an internal state estimate $\hat{x}_{n+1}$ is computed, which
is then used by the Human Controller to select the next muscle control signal $u_{n+1}$, resulting in the above discussed closed interaction loop.

\subsection{Applications of the Proposed Framework}
In this paper, we analyze and compare several optimal control models of interaction.
It is important to note that not all of the considered models exploit the complexity of the generic models depicted in Figure~\ref{fig:genmodel_extended}. 
For example, the closed-loop LQR model includes a trivial perceptual model in that it assumes that the controller has complete access to the true system state $x$ immediately. 
Another example is the open-loop MinJerk model, which does not 
include task-specific system dynamics.

Each optimal control model, however, yields a continuous representation of all relevant quantities of the interactive system. In contrast to summary statistics such as Fitts' Law, this allows to simulate and predict complete movement trajectories on both the kinematic and the dynamic level.
It also allows to analyze the effects of the control $u$ and of different cost terms incorporated in the objective function $J_{N}$, on the human body and the interface (e.g., user representations, buttons, or sliders). 

Most importantly, the modularity of the proposed framework enables high flexibility and generalizability.
For example, it is possible to analyze the effects of different input devices and/or GUIs on movement trajectories and control sequences, using the same description of the human biomechanical and perceptual system. 
Additionally, a given interface can be evaluated for different tasks such as pointing, dragging, steering, etc., by modulating the internal objective function accordingly. %
The resulting continuous representations can then be evaluated with respect to different metrics, e.g., remaining distance to target~\cite{Diedrichsen10}, effort~\cite{Shadmehr16}, fatigue~\cite{Cheema20}, movement time~\cite{Tanaka06}, %
etc.

Finally, the framework can be used to reverse-engineer the internal objective function (\textit{inverse optimal control}) as well as properties of the human biomechanical system (\textit{system identification}), such that the resulting trajectories best fit some experimentally observed user trajectories.
Before we explain how to identify such model-specific parameters using a data-driven parameter fitting procedure (see Section~\ref{sec:param-fitting}), we give a brief overview of the experimental data we use to evaluate the presented models in this paper.

\section{User Trajectories and Parameter Fitting}
\subsection{The Pointing Dynamics Dataset}\label{sec:dataset}

For the evaluations in this paper, we use the {\em Pointing Dynamics Dataset}.
Task, apparatus, and experiment are described in detail in \cite{Mueller17}.
The dataset contains the mouse trajectories for a reciprocal mouse pointing task in 1D for ID 2, 4, 6, and 8 (12 participants, 8 task conditions, 7732 trajectories in total).
We use the raw, unfiltered position data in our parameter fitting process to avoid artifacts from the filtering process.
In this section, we explain how we pre-process this data for the purposes of this paper.

Pointing experiments both in the reciprocal and discrete Fitts' paradigm introduce reaction times as experimental artifacts.
In real mouse usage, users first decide on a pointing target themselves, and then start moving the mouse.
In this sense, the pointing process can be considered initiated as soon as the pointer begins to move. 
In contrast, in the experimental paradigm used in \cite{Mueller17}, the next trial started as soon as the user clicked the mouse in the previous trial.
The target given to the user appeared at that instant.
This introduces a potential confound in the starting time of each trial.
The beginning of each trial can be partly attributed to belong to the end of the previous trial, and partly to a reaction time adjusting to the new target. 
During this time, velocity and acceleration of the pointer are close to zero.
This reaction time shows considerable variation both within and between participants.

Because the focus of this paper is not on modeling reaction times, we remove them from each mouse movement. 
To this end, we drop all frames before the velocity reaches $1\%$ of its maximum/minimum value (depending on the movement direction) for the first time in each trial.\footnote{
For improved temporal alignment of the individual trajectories and to remove outliers that sometimes occur at the beginning of a movement, we additionally assume that the acceleration remains positive/negative (depending on the movement direction) for at least 40ms after the initial time. If no time step is found that satisfies both criteria, we discard the entire movement. This was the case only for a single movement (ID 2, participant 5).} %

Since the deterministic optimal control models considered in this paper can only predict average user behavior, %
we compute \textit{mean trajectories} for each user, task condition, and direction from the raw data, resulting in 192 mean trajectories. This is done as follows.

First, we remove outlier trials, where at any time step the position was more than three standard deviations from the respective mean. This was the case for 397 trajectories in total, i.e., $5.1\%$ of all trials.
We found this to be necessary as the averaging process is highly sensitive to outliers. In particular, delayed movement onsets, which, e.g., might have occurred due to a lack of attention of the participant, would inject a high bias into the computed statistics.

Second, in order to make trials of different length comparable, we assume that the pointer would not move after the mouse click. 
Given a set of trials to be averaged (with reaction times removed as described above), movements shorter than the longest one are extended by their 
last position, zero velocity, and zero acceleration 
to achieve the same length.
To avoid unnecessarily long trajectories for conditions where a few of the recorded trials were of exceptional length, we additionally remove trials with duration longer than three standard deviations from the mean duration of the respective condition, before extending the remaining trajectories to maximum length. This was the case for 87 trajectories in total (i.e., $1.1\%$ of all trials), with a maximum of two trials removed per user, task condition, and direction.
Finally, we average the resulting trajectories on a frame by frame basis. %
We also compute the respective sample covariance matrices at each time step to capture the between-trial-variability observed from user data.

\subsection{Parameter Identification}\label{sec:param-fitting}

In this section, we present a method to identify the parameters of a given instantiation of the optimal control framework introduced in Section~\ref{sec:oc-framework} that best explain experimentally observed user behavior. %
More specifically, for a given interaction model and a given dataset of user trajectories, we aim to find the model-specific parameter values such that the resulting trajectories approximate a subset of user trajectories (e.g., all trajectories of a specific user for some task condition) as closely as possible.
In the following, we will apply this procedure to each of the presented models, using the Pointing Dynamics Dataset as reference data.
Since only stochastic models can account for the between-trial variability typically observed in human movements, we need to distinguish between the deterministic and the stochastic case in the following.

In the deterministic case, we use the squared error between predicted and observed mouse pointer position, summed over time (\textit{sum squared error (SSE)}) as a measure of distance between simulation and mean user trajectories.
Given a model with parameter vector $\Lambda$, where $p_{n}^{\Lambda}$ denotes the position time series of the resulting simulation trajectory, and given a mean user position time series $p_{n}^{\text{USER}}$, the goal is to find the parameter vector $\Lambda^{*}$ such that the loss function of the parameter fitting process
\begin{equation}\label{eq:SSE-obj-fct}
	\mathcal{L}(\Lambda)=\text{SSE}(\Lambda)=\sum_{n=0}^{N} \left(p_{n}^{\Lambda} - p_{n}^{\text{USER}}\right)^{2}
\end{equation}
takes its minimum in $\Lambda^{*}$.
This is done for each mean trajectory, resulting in 192 optimal parameter vectors.

Minimizing Equation~\eqref{eq:SSE-obj-fct} with respect to $\Lambda$ can be considered a \textit{least-squares problem}~\cite{Bjoerck96}, where each function evaluation of $\mathcal{L}(\Lambda)$ requires computation of the respective model simulation trajectory to obtain $p_{n}^{\Lambda}$.
In the case of optimal control models, this particularly implies that an OCP must be solved to obtain $p_{n}^{\Lambda}$ for a given $\Lambda$, that is, the complete parameter fitting process consists of two nested optimizations.

Stochastic models, in contrast, yield a sequence of distributions of the state, such as the distributions of pointer position and velocity, over multiple trials.
These distributions can be used to sample individual trajectories.
We measure the ``similarity'' between 
two distributions
using the \textit{$2-$Wasserstein distance} (often denoted as \textit{Earth mover's distance})~\cite{Olkin82}.
Given two normal distributions $\rho_{1}$ and $\rho_{2}$ with means $\mu_{1}$ and $\mu_{2}$ and covariance matrices $\Sigma_{1}$ and $\Sigma_{2}$, respectively, the $2-$Wasserstein distance can be written as
\begin{equation}\label{eq:Wasserstein-normal}
	W_{2}\left(\rho_{1}, \rho_{2}\right) = \left(\Vert \mu_{1} - \mu_{2} \Vert_{2}^{2} + \text{tr}(\Sigma_{1}) + \text{tr}(\Sigma_{2}) - 2 \text{tr}\left((\Sigma_{1}\Sigma_{2})^{\frac{1}{2}}\right)\right)^{\frac{1}{2}},
\end{equation}
and can be interpreted as the amount of work required to transform the probability distribution $\rho_{1}$ into the probability distribution $\rho_{2}$ (and vice versa). %
In the following, we will use this formula to measure the distance between the simulation state distribution of a model with parameter vector~$\Lambda$, $\rho^{\Lambda}$, and the empirically observed state distribution, $\rho^{\text{USER}}$, at some time step $n\in\{0,\dots,N\}$, i.e., $\rho_{1}=\rho_{n}^{\Lambda}$ and $\rho_{2}=\rho_{n}^{\text{USER}}$.

One advantage of this measure is that it is only based on the relative distance between the two means and covariance matrices of the two distributions, independent of the magnitude of these quantities. 
This is in contrast to the KL divergence~\cite{Kullback51}, which increases as the variance of the reference distribution decreases.
In the special case where both distributions have the same, diagonal covariance matrix, the $2-$Wasserstein distance corresponds to the Euclidean distance between the means of both distributions.

As a measure for the similarity between complete sequences of distributions (e.g., of mouse pointer positions and velocities), %
we use the \textit{mean Wasserstein distance (MWD)} over time:
\begin{equation}\label{eq:W2-obj-fct}
	\mathcal{L}(\Lambda)=\text{MWD}(\Lambda)=\frac{1}{N+1}\sum_{n=0}^{N} W_{2}\left(\rho_{n}^{\Lambda}, \rho_{n}^{\text{USER}}\right).
\end{equation}

In both the deterministic and the stochastic case, we solve the (outer) optimization problem of minimizing $\mathcal{L}(\Lambda)$ with respect to $\Lambda$ using \textit{differential evolution}~\cite{Storn97}, which is a simple, gradient-free global optimization algorithm suitable for continuous parameter spaces.
This algorithm has proven to yield robust and reliable results even for ill-conditioned problems~\cite{Auger09}.

Of course, more efficient optimization methods, e.g., gradient-based ones, are always desirable, and algorithmic differentiation is a promising step forward in that regard. 
The main question here is the applicability of algorithmic differentiation in the case of iteratively alternating between control and estimation problems, as is required for the considered case of LQG with signal-dependent noise (see Section~\ref{sec:LQG}). 
Pursuing this endeavor, however, might very well enable real-time predictions of parameter effects on the entire interaction loop.

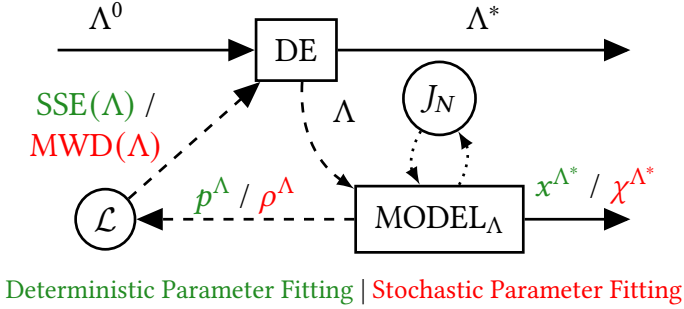
\begin{figure}
	\resizebox{0.7\linewidth}{!}{
	\begin{tikzpicture}[auto, thick, node distance=1.5cm, >=triangle 45,inner sep=2mm]
	\begin{scope}
		\draw node [input] (input) {} 
		node [block, right of=input, node distance=2.5cm] (LSQ) {DE}
		node [output,right of=LSQ, node distance=3.5cm] (output) {} 
		node [output,below of=output, node distance=1.75cm] (output2) {} 
		node [block, left of=output2, node distance=2cm] (LQR) {$\text{MODEL}_\Lambda$}
		node [sum, above of =LQR , inner sep=1mm,node distance=1.25cm] (J) {$J_N$}

		node [sum, left of=LQR, inner sep=1mm, node distance=3.5cm] (SSE) {$\mathcal{L}$}
		
		;
		\draw[->](input) --  node[near start] {$\Lambda^0$} (LSQ);
		\draw[->](LSQ) --  node[] {$\Lambda^*$} (output);
		\path  [-latex] (LSQ) edge [bend right,  dashed] node[] {$\Lambda$} (LQR);
		\draw[->, dashed](LQR) --  node[above, yshift=-0.2cm] {\begin{tabular}{c} {\color{ForestGreen} $p^\Lambda$} / {\color{red}$\rho^\Lambda$} \end{tabular}} (SSE);
		\draw[->, dashed](SSE) --  node[near start, xshift=0.4cm, yshift=-0.2cm] {\begin{tabular}{c} {\color{ForestGreen} $\text{SSE}(\Lambda)$} / \\ {\color{red} $\text{MWD}(\Lambda)$} \end{tabular}} (LSQ);
		\draw[->](LQR) --  node[xshift=0.2cm, yshift=-0.1cm] {\begin{tabular}{c} {\color{ForestGreen} $x^{\Lambda^{*}}$} /  {\color{red}$\chi^{\Lambda^{*}}$} \end{tabular}} (output2);
		\path  [-latex] (J) edge [bend right,  dotted] (LQR);
		\path  [-latex] (LQR) edge [bend right,  dotted] (J);
		
		\draw node[below of=LQR, align=center, xshift=-1cm, yshift=0.75cm] (legend) {\footnotesize {\color{ForestGreen} Deterministic Parameter Fitting} | {\color{red} Stochastic Parameter Fitting}};
		
	\end{scope}
	\end{tikzpicture}
	}
	\caption{Starting with an initial parameter vector $\Lambda=\Lambda^0$, the differential evolution (DE) algorithm obtains the loss $\mathcal{L}(\Lambda)$ (sum squared error (SSE) value in the deterministic and mean Wasserstein distance (MWD) in the stochastic case) for the currently considered parameter vector $\Lambda$.
		To do this, it calls $\text{MODEL}_\Lambda$, which computes the resulting model trajectory sequence $x^{\Lambda}$ (or a sequence of state distributions $\chi^{\Lambda}$). In case of optimal control models, this requires an inner optimization with respect to a model-specific objective function $J_{N}$.
		The resulting position time series~$p^\Lambda$ included in $x^{\Lambda}$ (or the sequence of position-velocity distributions $\rho^{\Lambda}$ included in $\chi^{\Lambda}$) is used to compute the loss $\mathcal{L}(\Lambda)$ of the current parameter vector $\Lambda$, which in turn is used by the DE algorithm.
		After obtaining the loss for a few requested parameter vectors $\Lambda$, the DE algorithm chooses the next parameter vectors $\Lambda$ until $\Lambda^{*}$ with minimum loss is found.
		Finally, $\Lambda^*$ is returned along with the respective optimal trajectory $x^{\Lambda^{*}}$ (or the optimal sequence of state distributions $\chi^{\Lambda^{*}}$).
		}~\label{fig:parameter-fitting}
	\Description{Workflow diagram of the proposed parameter fitting process, both for the deterministic and the stochastic case. Details are described in the caption and in the text.}
\end{figure}

Figure~\ref{fig:parameter-fitting} gives an overview of our parameter identification process for both the deterministic and the stochastic case.
The parameter boundaries for all models introduced below are given in Table~\ref{tab:param_bounds} in the Appendix. 
Descriptions of how discrete parameters are relaxed in order to optimize them using standard continuous optimization methods are given in the respective model sections.

\section{Pointing as a Dynamical System: The Second-Order Lag}\label{sec:dynamics}

One of the basic models of mouse pointer dynamics is the \textit{second-order lag}, which has been used as a baseline in several papers, including~\cite{Mueller17, Martin21}.
We therefore also include it as a baseline. 
The parameters $\Lambda$ of the model are the stiffness of the spring $k>0$ and the damping factor $d>0$.
In the setting described below, the mass is a redundant parameter, and we thus set it to~1. %
In continuous time, we denote the position of the mouse pointer as~$y(t)$, and its first and second derivatives with respect to time (i.e., velocity and acceleration) as~$\dot{y}(t)$ and~$\ddot{y}(t)$, respectively. 
The behavior is then described by the second-order lag equation: %
\begin{equation}\label{eq:2ol-continuous}
	\ddot{y}(t) = u(t)-ky(t)-d\dot{y}(t). 
	\tag{2OL}
\end{equation}

\begin{figure}[!ht]
	\includegraphics[width=0.6\linewidth]{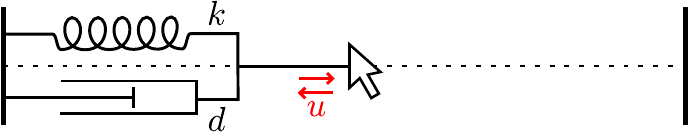}\caption{Mouse pointer model with spring and damper.}\label{fig:2OL_mouse}
	\Description{The 2OL-Eq can be interpreted as a spring-mass-damper model, with the mouse cursor attached to one edge of the screen via a spring and a damper.}
\end{figure}

An intuitive illustration of these dynamics is given in Figure~\ref{fig:2OL_mouse}:
Assuming that the mouse pointer is fixed at one edge of the screen via a spring, the parameters $k$ and $d$ correspond to the stiffness and the damping of this spring, respectively, and the control value $u$ can be interpreted as the force acting on the mouse pointer.
In particular, the pointer acceleration $\ddot{y}$ is assumed to be directly proportional to the control $u$ (apart from the damping and stiffness terms), i.e., \eqref{eq:2ol-continuous} defines a (linear) dynamical system of second order.
A control flow diagram of the model is shown in Figure~\ref{fig:2OL_controlflow} in the Appendix.

Given a target position $T\in\R$, it can be shown that for the particular choice $u\equiv kT$,
with $k, d > 0$, the position $y(t)$ approaches $T$ for large enough $t$, independent of the initial position $y(0)=y_{0}$ and initial velocity $\dot{y}(0)=\dot{y}_{0}$~\cite{Hannsgen87}.
More precisely, the state $(y,\dot{y})=(T,0)$, i.e., the desired target~$T$ is reached and the velocity is zero, is an equilibrium, meaning that once reached, that state will forever be maintained.
The resulting trajectory, which is uniquely determined given $y_{0}$, $k$, $d$, and $T$, is often referred to as \textit{second-order lag trajectory}, and can be used to model mouse pointing movements towards a given target $T$.
Since the control $u(t)$ is constant in time and converges towards the equilibrium state, we denote this variant of \eqref{eq:2ol-continuous} as \textit{2OL-Eq} in the following.

Following the general notation of~\eqref{eq:LQR Control1}, we derive a discrete-time version of~\eqref{eq:2ol-continuous}, with a step size of two milliseconds, i.e., $h=0.002$, which corresponds to the mouse sensor sampling rate.
Considering our example case of 1D pointing tasks, in which the mouse can only be moved horizontally, the state $x_n$ encodes the horizontal position and velocity of the pointer, denoted by $p_n\in\R$ and $v_n\in\R$, respectively, %
i.e.,
\begin{equation}\label{eq:x_n}
	x_n = \left(p_n,v_n\right)^\top\in\R^{2}.
\end{equation}
Using the forward Euler method\footnote{While we could also use the exact solution here, the (fairly good) approximation via forward Euler yields matrices that are more suitable for our explanatory purposes.}, we obtain the matrices $A$ and $B$ for~\eqref{eq:LQR Control1} as
\begin{equation}\label{eq:dynamics-AB}
	A = 
	\begin{pmatrix}
		1	& h \\
		-hk	& 1-hd
	\end{pmatrix},
	\quad
	B = 
	\begin{pmatrix}
		0 \\
		h
	\end{pmatrix}.
\end{equation}
This model can easily be extended to 2D or 3D pointing tasks by augmenting $x_{n}$ and $u_{n}$ with the respective components for the additional dimensions. %

\subsection{Analysis of Parameters}
Since the 2OL model is an important baseline for mouse pointing dynamics, in this section, we provide an analysis and intuitive understanding of influence of the model parameters on model behavior.

While the convergence of 2OL-Eq towards a target $T$ of fixed width $W>0$ can be easily shown under the assumptions described above, both the time until this target is reached first (i.e., the time at which the remaining distance to target is smaller than $W$) and the \textit{transient behavior} (i.e., specific characteristics of the trajectory until this time) largely depend on the parameters $k$ and $d$.
In the following, we analyze the effects of the stiffness $k$ and the \textit{damping ratio} $\zeta$, which is defined as
\begin{equation}
	\zeta = \frac{d}{2 \sqrt{k}},
\end{equation}
as this is easier to interpret than the actual damping parameter $d$.
Note that given two of the three parameters $k$, $d$, and $\zeta$, the remaining one (in this case $d$) is uniquely determined by the others and can be easily computed.

\begin{figure}[!t]
	\centering
	\includegraphics[width=\linewidth]{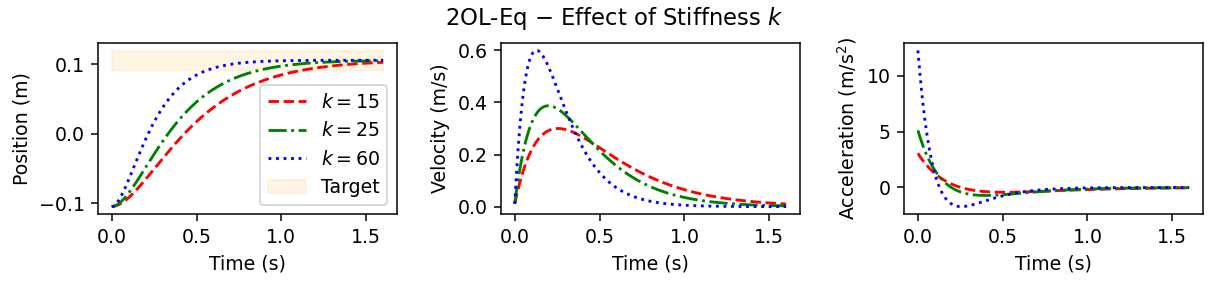}
	\\
	\includegraphics[width=\linewidth]{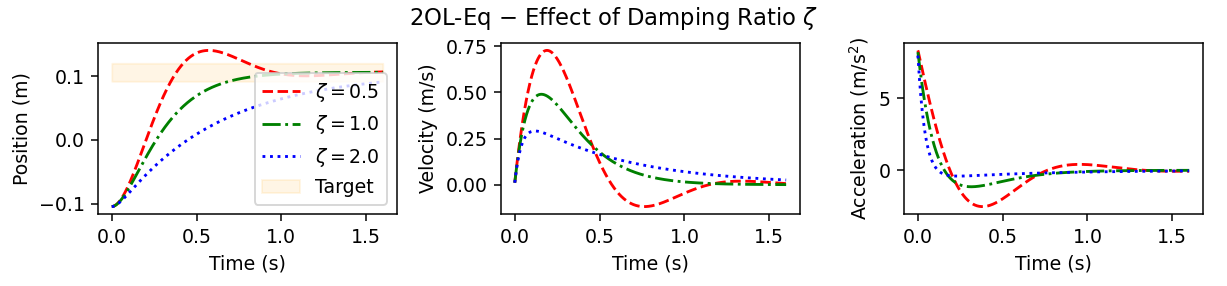}
	\caption{
		Position, velocity, and acceleration time series of typical 2OL-Eq trajectories with target shown as the orange box.
		\textbf{Top:} Effect of stiffness parameter $k$ with fixed damping ratio $\zeta=1$ (red dashed: $k=15$, green dash-dotted: $k=25$, blue dotted: $k=60$).
		\textbf{Bottom:} Effect of damping ratio $\zeta$ with fixed stiffness parameter $k=40$ (red dashed: $\zeta=0.5$, green dash-dotted: $\zeta=1$, blue dotted: $\zeta=2$). \\
		The stiffness parameter $k$ mainly affects the instantaneous initial acceleration and thus the speed at which the target is approached. The parameter $\zeta$ determines the relative amount of damping, with $\zeta<1$ (underdamped) resulting in oscillations around the target and $\zeta>1$ (overdamped) leading to trajectories that reach the target later.
	}
	\label{fig:2OL-Eq_effects}
	
	\Description{Fully described in caption and text.}
\end{figure}

The position, velocity, and acceleration time series of typical 2OL-Eq trajectories are shown in Figure~\ref{fig:2OL-Eq_effects}.
The initial and target values stem from an ID 4 task condition from the Pointing Dynamics Dataset.
The most characteristic feature of 2OL-Eq trajectories is the large positive acceleration at the beginning of the movement. This is due to the model being second-order, i.e., the control $u$ is proportional to the acceleration $\ddot{y}$ (apart from the damping and stiffness terms).
The velocity profile is typically left-skewed, since the deceleration phase is considerably longer than the acceleration phase. As a consequence, the peak velocity is reached relatively early during the movement.

Given a constant damping ratio $\zeta$, the stiffness parameter $k$ mainly affects the initial acceleration, and consequently the peak velocity and the time at which the target is reached first. As can be seen in the top row of Figure~\ref{fig:2OL-Eq_effects}, a large stiffness (blue dotted line) leads to a high initial acceleration and peak velocity, and thus the target is reached earlier than with lower stiffness values (red dashed line).
For damping ratios $\zeta<1$ (red dashed line in the bottom row of Figure~\ref{fig:2OL-Eq_effects}), i.e., the damping $d$ is small compared to the stiffness $k$, oscillations occur in the trajectories, leading to multiple peaks in velocity and acceleration time series and to \textit{overshooting} in the position time series (the so-called \textit{underdamped} case).
For $\zeta=0$, the pointer does not even converge towards the target, but oscillates indefinitely (not shown).
If $\zeta>1$ (the so-called \textit{overdamped} case, blue dotted line in the bottom row of Figure~\ref{fig:2OL-Eq_effects}), the pointer converges towards the target slowly, without oscillations. %
If $\zeta=1$, the trajectory is \textit{critically damped}, which means that the pointer reaches 
(and stays at) %
the equilibrium (i.e., the target $T$) in minimum time.%

\subsection{Results of Parameter Fitting}
For each combination of participant, task, and direction, we identify the parameters $\Lambda = (k, \zeta)$ that best explain the corresponding mean trajectory from the Pointing Dynamics Dataset, using the deterministic parameter fitting process described in Section~\ref{sec:oc-framework}.
The loss function is the SSE on position, see~\eqref{eq:SSE-obj-fct}.
Note that the results obtained from our parameter fitting do not exactly match those presented in~\cite{Mueller17}, since we apply a different pre-processing to the experimental user trajectories and optimize the parameters with respect to the positional error only.

\begin{figure}
	\centering
	\includegraphics[width=\linewidth]{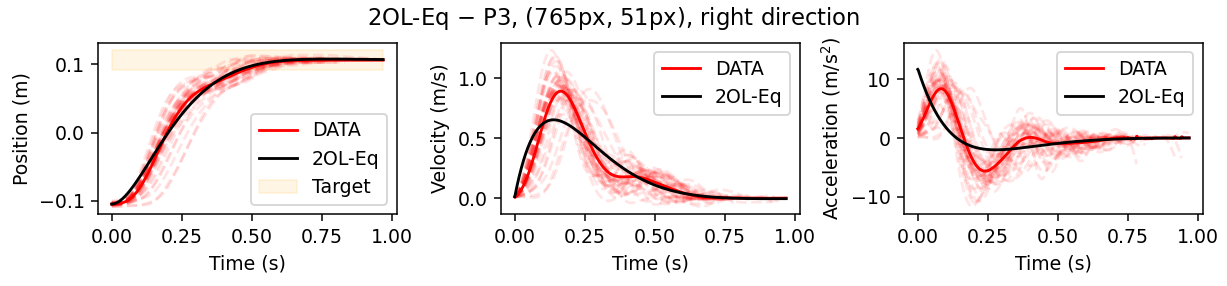}
	\caption{While the position time series visually matches the observed user data quite well, 2OL-Eq yields a much less symmetric velocity and acceleration profile during the surge (here: up to 0.36s). This is due to the assumption of constant equilibrium control, which results in a physically implausible instantaneous peak acceleration.
	}
	\label{fig:2OL-Eq_data_ID4}%
	\Description[Position, velocity, and acceleration time series of 2OL-Eq vs. user data (ID 4 task)]{Position, velocity, and acceleration time series of both the Pointing Dynamics Dataset (Participant 3, ID 4 (765px distance, 51px width), right direction) and the corresponding 2OL-Eq simulation trajectory.}
\end{figure}

The fitted trajectory for a representative ID 4 task condition is shown in Figure~\ref{fig:2OL-Eq_data_ID4}.
As discussed in~\cite{Mueller17}, the main differences between model and human behavior are the less symmetric velocity profile and the large initial accelerations produced by the 2OL-Eq model.
In particular, the user trajectories exhibit velocity profiles that are close-to-symmetric and bell-shaped, at least for the initial ballistic movement towards the target (the ``surge'' \cite{Mueller17}), which is consistent with previous findings \cite{Morasso81}.
The differences can be explained with the physical interpretation of the 2OL-Eq as a spring-mass-damper system:
Since $u$ is constant in this model, as the system is released, the spring instantaneously accelerates the system with a force that is proportional to the extension of the spring.
Because human muscles cannot build up force instantaneously~\cite{Schmidt05}, this behavior is not physically plausible.

\begin{figure}
	\centering
	\subfloat{\includegraphics[width=0.4\linewidth]{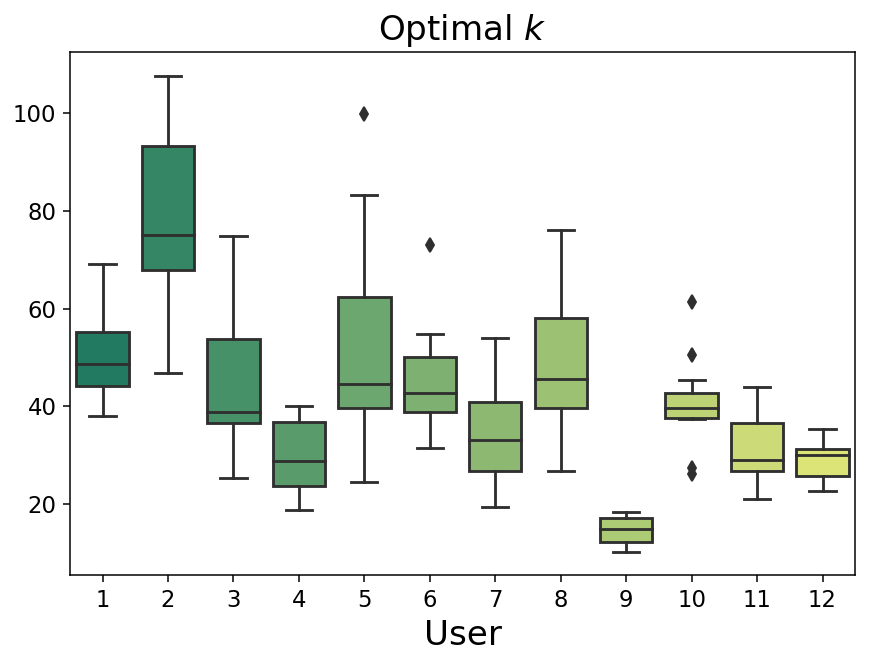}}
	\subfloat{\includegraphics[width=0.4\linewidth]{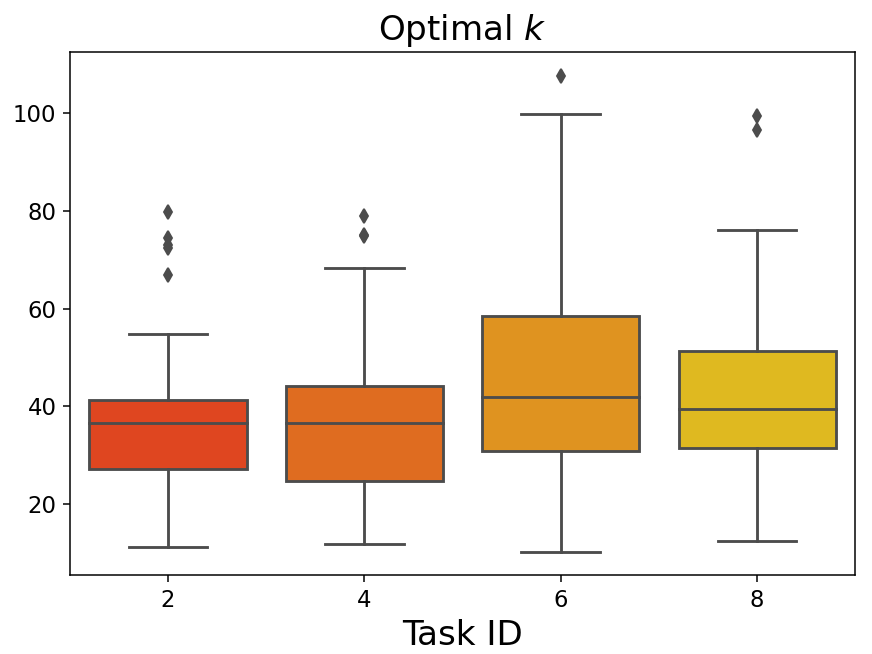}} 
	\subfloat{\includegraphics[width=0.1125\linewidth]{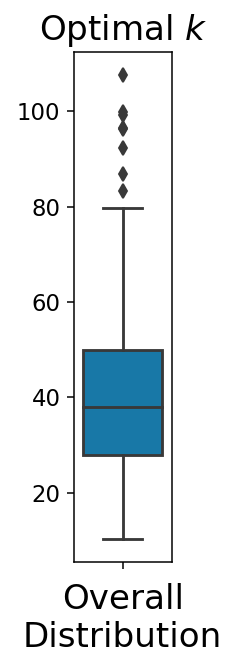}} 
	\\
	\subfloat{\includegraphics[width=0.4\linewidth]{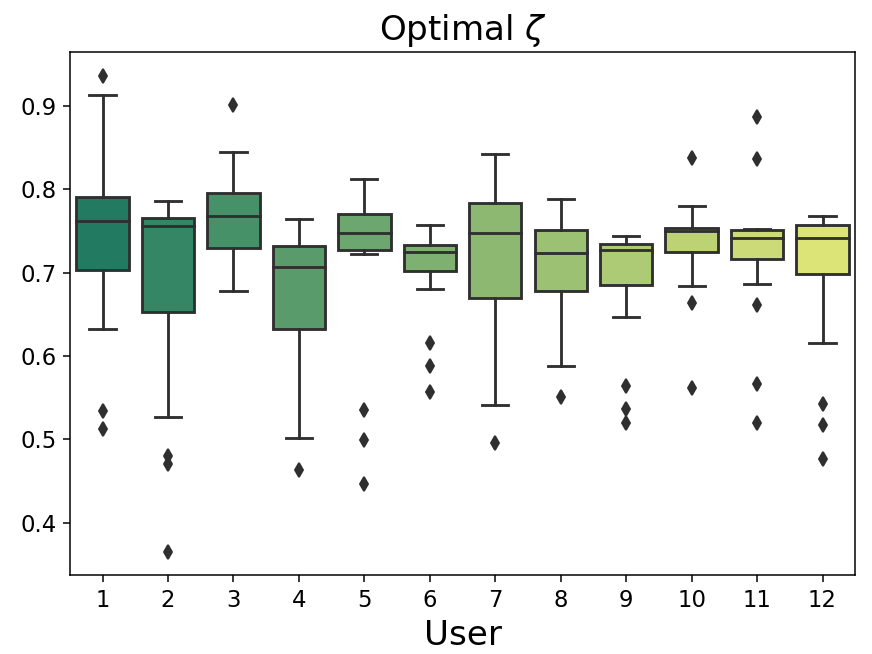}}
	\subfloat{\includegraphics[width=0.4\linewidth]{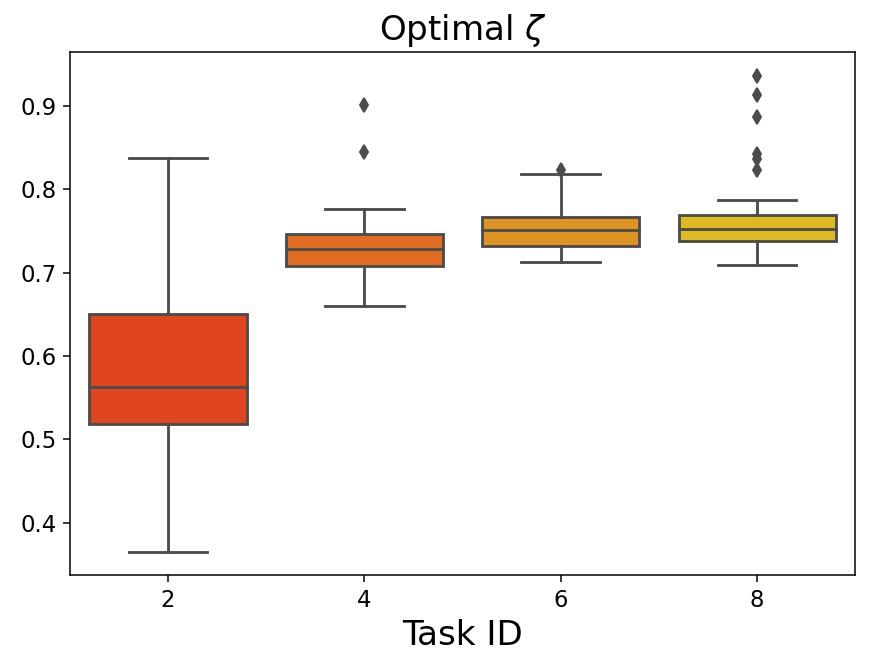}} 
	\subfloat{\includegraphics[width=0.1125\linewidth]{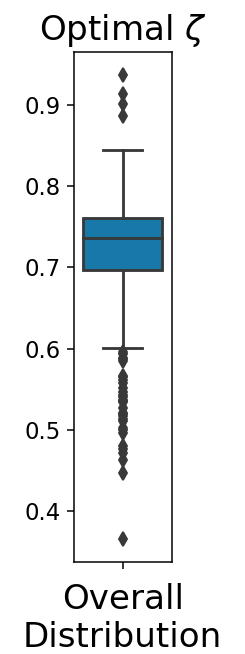}} 
	\caption{Parameters of 2OL-Eq, optimized for the mean trajectories of all participants, tasks, and directions, grouped by participants (left) and by ID (middle), and as overall distribution (right). While the optimal stiffness parameter $k$ considerably differs between participants, the damping ratio $\zeta$ is mainly affected by the task ID. Interestingly, all resulting simulation trajectories are underdamped (mean $\zeta=0.71$), with $\zeta$ increasing as ID increases.
	}~\label{fig:2OL-Eq_opt}
	\Description{Fully described in caption and text.}
\end{figure}

In Figure~\ref{fig:2OL-Eq_opt}, the optimal values of $k$ and $\zeta$ are given for all participants, tasks, and movement directions, both grouped by participants (left) and by ID (right).
Interestingly, different behavior between individual users is mainly captured by different stiffness values $k$. 
Participant 2, for instance, seems to be characterized best by a large stiffness (between $46.8$ and $107.6$, with mean $78.3$), while the trajectories of participant 9 are best explained by a considerably lower stiffness (between $10.3$ and $18.3$, with mean $14.4$) in the 2OL-Eq model. %
In contrast, the damping ratio $\zeta$ seems to center around $0.7$-$0.76$, independent of the participant.
Instead, it is mainly influenced by the index of difficulty of the task. Lower IDs tend to result in a lower damping ratio, with trajectories of ID 2 tasks being considerably more underdamped than others.
This is consistent with previous findings~\cite{Guiard93, Bootsma04, Mueller17} and might be explained by the reciprocal nature of the considered pointing task, where participants alternately moved between two given targets (which we denote as initial and target position for a given movement direction) without dwell time.

In summary, the stiffness $k$ mostly seems to account for movement strategies that are characteristic of specific users, whereas the damping ratio $\zeta$ mainly differs between indices of task difficulty. %

\subsection{Discussion}

The main shortcomings of the 2OL-Eq as a model of mouse pointer movements are the unrealistically high initial acceleration and the resulting skewed velocity profile.
This is mainly due to the assumption of equilibrium control, while the literature suggests that the motor control signals are actively changed during the movement~\cite{Bizzi92, Georgopoulos82, Todorov98_thesis}.
From a conceptual standpoint, the 2OL-Eq only describes the passive dynamics of the mouse pointer as a differential equation.
It does not separately model the user's ``brain'' or intention in controlling these dynamics.
In particular, it does not describe what the user is trying to achieve.
This can be explained by optimal control models. 

\section{Pointing as Optimal Open-Loop Control: The Minimum Jerk Model}\label{sec:min-jerk}

An elementary model of aimed movements that assumes that users behave optimally according to an internal cost function is the \textit{minimum jerk model} by Flash and Hogan~\cite{Flash85}. 
This model, which we will refer to as \emph{MinJerk} in the following, assumes that the objective of users is to generate smooth movements by minimizing the jerk of the end-effector, i.e., the time derivative of the end-effector's acceleration, while reaching the target exactly at a prescribed movement time with zero final velocity and acceleration.
Within HCI, this model has been successfully used by Quinn and Zhai~\cite{Quinn18} to model the shape of gestures on a word-gesture keyboard. %

The model assumes that the movement is controlled open-loop, and thus cannot explain how users would correct for disturbances or inaccurate execution. 
Similar to 2OL-Eq, the choice of parameters and boundary values already determines the complete trajectory. %
However, there are a few important differences to 2OL-Eq.
First, MinJerk does not only require information about the initial, but also about the final state (as a third-order model, initial and final position, velocity, and acceleration need to be specified). %
Second, the overall movement time, which is denoted by $N_{MJ}$ in the following, needs to be known in advance.
This is in contrast to the discrete-time formulation of 2OL-Eq. %

In discrete time, minimizing jerk corresponds to minimizing the differences between subsequent accelerations.
While in principle, the MinJerk optimal control problem could be transformed into a closed-loop discrete-time system similar to~\eqref{eq:LQR Control1} (but with time-dependent matrices $A_{n}$ and $B_{n}$, see ~\cite{HoffArbib93}), here we make use of the analytical solution of the original continuous-time problem, and afterwards discretize the resulting 5th-degree polynomial with respect to time.
For arbitrary initial state $x_{0}=\left(\bar{p}_{0}, \bar{v}_{0}, \bar{a}_{0}\right)$ and final state $x_{N_{MJ}}=\left(\bar{p}_{N_{MJ}}, \bar{v}_{N_{MJ}}, \bar{a}_{N_{MJ}}\right)$, where the third component is the respective end-effector acceleration, the (discrete-time) MinJerk trajectory is given by
\begin{subequations}\label{eq:minjerk-polynomial-discrete}
	\begin{gather}
		x_{n} = \sum_{i=0}^{5}c_{i}\left(\frac{n}{N_{MJ}}\right)^{i}, \\
	 	\begin{pmatrix}
	 		c_{0} \\ c_{1} \\ c_{2} \\ c_{3} \\ c_{4} \\ c_{5}
	 	\end{pmatrix} =
	 	\begin{pmatrix}
	 	1 & 0 & 0 & 0 & 0 & 0 \\
	 	0 & t_f & 0 & 0 & 0 & 0 \\
	 	0 & 0 & 0.5t_f^2 & 0 & 0 & 0 \\
	 	-10 & -6t_f & -1.5t_f^2 & 10 & -4t_f & 0.5t_f^2 \\
	 	15 & 8t_f & 1.5t_f^2 & -15 & 7t_f & -t_f^2 \\
	 	-6 & -3t_f & -0.5t_f^2 & 6 & -3t_f & 0.5t_f^2
	    \end{pmatrix}
	 	\begin{pmatrix}
	 		\bar{p}_{0} \\
	 		\bar{v}_{0} \\
	 		\bar{a}_{0} \\
	 		\bar{p}_{N_{MJ}} \\ 
	 		\bar{v}_{N_{MJ}} \\ 
	 		\bar{a}_{N_{MJ}}
	 	\end{pmatrix},
	\end{gather}
\end{subequations}
where $t_{f}=N_{MJ}h$ denotes the final time, and $h$ is the fixed time interval between two consecutive time steps.

\subsection{Extension to Complete Trajectories}

\begin{figure}
	\centering
	\includegraphics[width=\linewidth]{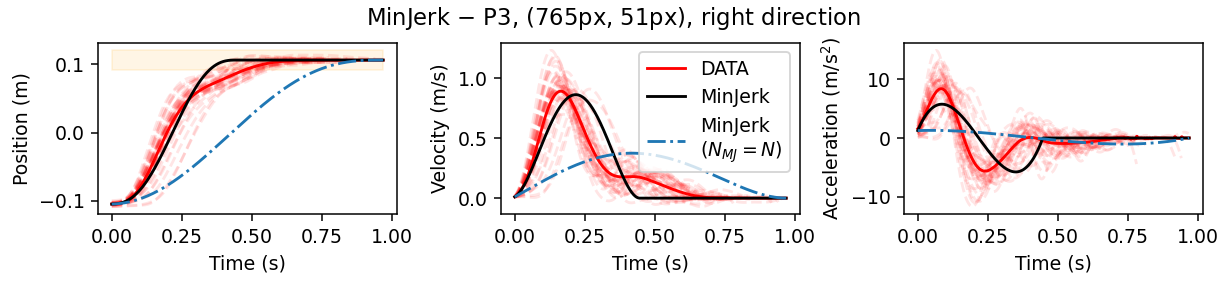}
	\caption{Visually, MinJerk trajectories show a relatively good fit to observed user data, as long as the duration parameter $N_{MJ}$ is chosen to cover only the surge (black solid line; here, $N_{MJ}=223$ corresponds to 0.446s).
	However, the MinJerk model predicts zero velocity at the end of the surge, which results in a considerably worse overall fit of user trajectories that exhibit clear submovements. 
	If $N_{MJ}$ is set as total movement duration $N$, the resulting simulation trajectory (dash-dotted blue line) does not fit the data at all.
	}
	\label{fig:MinJerk_data_ID4}%
	\Description[Position, velocity, and acceleration time series of MinJerk vs. user data (ID 4 task)]{Position, velocity, and acceleration time series of both the Pointing Dynamics Dataset (Participant 3, ID 4 (765px distance, 51px width), right direction) and the corresponding MinJerk simulation trajectories for $N_{MJ}<N$ and $N_{MJ}=N$.}
\end{figure}

The MinJerk model has been derived from data of an experiment that did not involve any corrective submovements~\cite{Flash85}.
However, in mouse pointing tasks with large ID, submovements occur regularly. %
If MinJerk is used for modeling of the entire movement, i.e., until the final time step~$N$, the fit is thus very poor (see blue dash-dotted lines in Figure~\ref{fig:MinJerk_data_ID4}).
Instead of a quick movement towards the target with corrective submovements, as in our data, the model predicts a slow, smooth movement, reaching the target only at the final time. %

A much better fit is obtained by using MinJerk only for the first, rapid movement towards the target (the ``surge''), and assuming that the pointer does not move afterwards.
To this end, we define the (extended) MinJerk trajectory as follows.
For $n \leq N_{MJ}$, $x_{n}$ corresponds to the minimum jerk polynomial~\eqref{eq:minjerk-polynomial-discrete} with %
$\bar{p}_{0}$, $\bar{v}_{0}$, and $\bar{a}_{0}$ taken from user data, $\bar{p}_{N_{MJ}}=T$, and $\bar{v}_{N_{MJ}}=\bar{a}_{N_{MJ}}=0$.
For $n > N_{MJ}$, the trajectory is constantly extended by the final state of the MinJerk polynomial, i.e., $x_{n}=x_{N_{MJ}}=(T, 0, 0)^{\top}$.

\begin{figure}
	\centering
	\includegraphics[width=\linewidth]{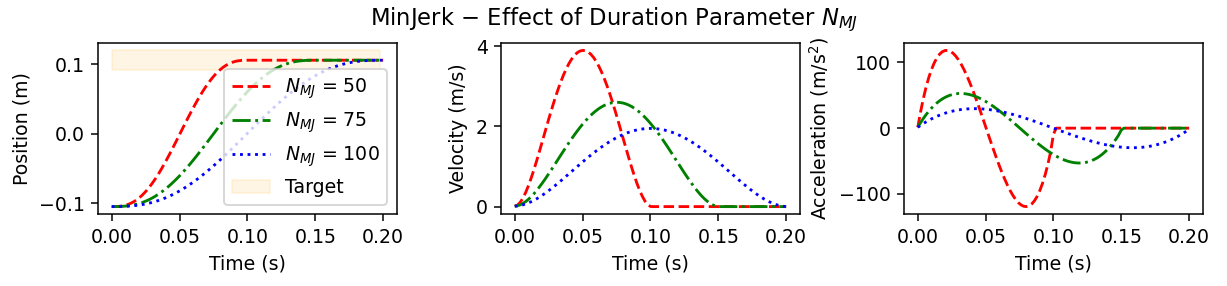}
	\caption{
		Position, velocity, and acceleration time series of different MinJerk trajectories with target shown as the orange box.
		The parameter $N_{MJ}$ determines the end time of the symmetric and smooth jerk-minimizing movement, after which the trajectory is constantly extended by its final position value and zero velocity and acceleration (red dashed: $N_{MJ}=50$, green dash-dotted: $N_{MJ}=75$, blue dotted: $N_{MJ}=N=100$). \\
	}
	\label{fig:MinJerk_effects}
	\Description{Fully described in caption and text.}
\end{figure}

The effect of the parameter $N_{MJ}$ in our MinJerk model is shown in Figure~\ref{fig:MinJerk_effects}.
Varying this parameter allows to model variable peak velocities and accelerations.
However, with $N_{MJ}$ being a pure scaling parameter of the trajectory, the velocity profile is always bell-shaped and the acceleration profile N-shaped.
Moreover, the target center is reached at time step $n=N_{MJ}$ by definition, i.e., the model naturally cannot account for corrections that typically occur after the surge.
As illustrated in Figure~\ref{fig:MinJerk_data_ID4} for the same ID 4 task as above (black solid lines), this can result in a considerably worse overall fit of the trajectory, at least for movements that consist of several submovements. %

\subsection{Results of Parameter Fitting}

Similar to the parameter fitting process for 2OL-Eq, we identify the optimal duration parameter $N_{MJ}$ with respect to positional SSE for each participant, task, and movement direction, using the respective mean trajectory as our reference.
To improve convergence of the used optimization algorithm, we relax this parameter by allowing for continuous values of $N_{MJ}$ in \eqref{eq:minjerk-polynomial-discrete} and compute the discrete-time MinJerk states $x_{n}$ for $n\leq\left\lceil N_{MJ} \right\rceil$.

\begin{figure}
	\centering
	\subfloat{\includegraphics[width=0.33\linewidth]{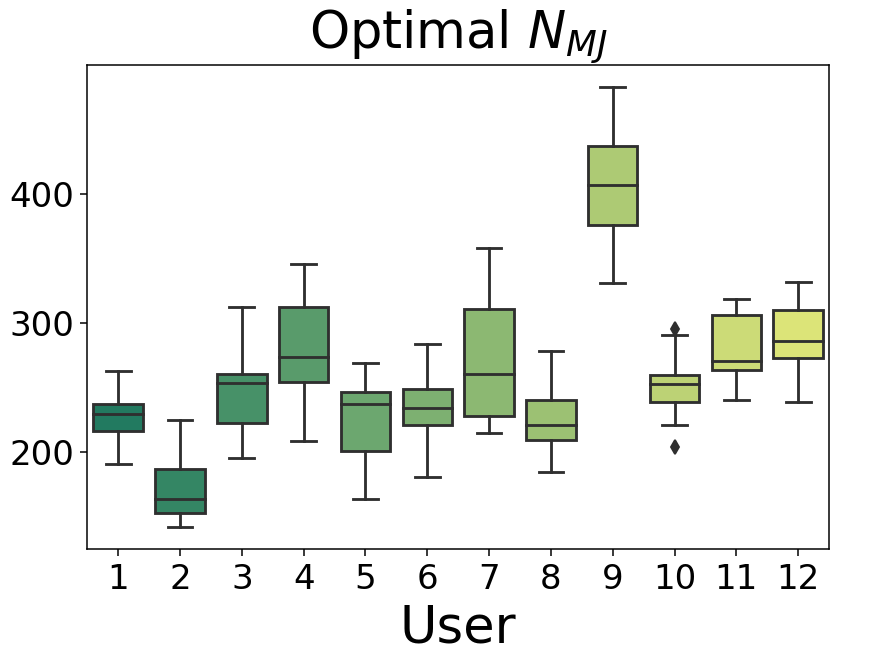}}
	\subfloat{\includegraphics[width=0.33\linewidth]{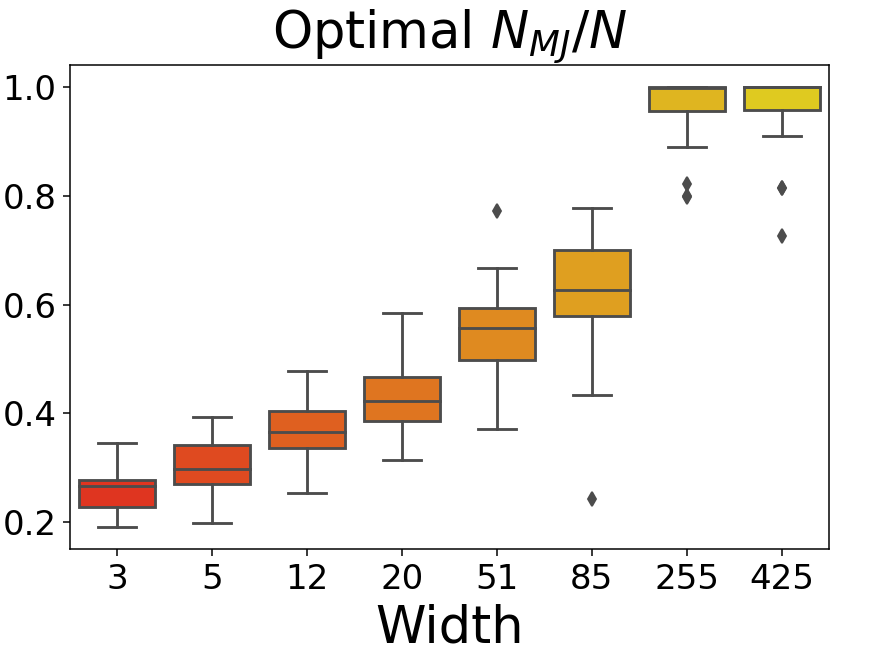}}
	\subfloat{\includegraphics[width=0.33\linewidth]{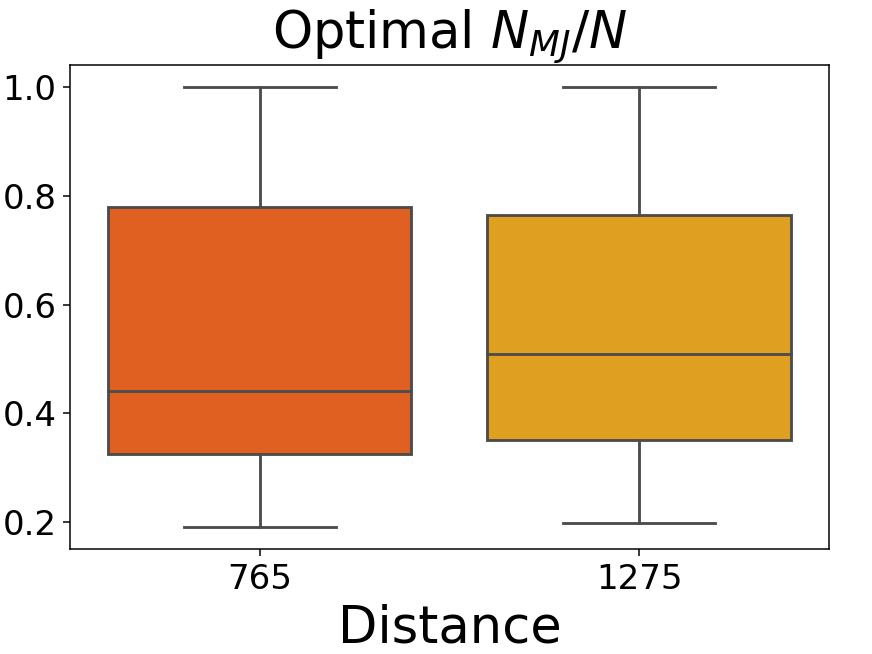}}
	\caption{Duration parameter $N_{MJ}$ of MinJerk, optimized for the mean trajectories of all participants, tasks, and directions. 
		\textbf{Left:} Absolute parameter values, grouped by participants. 
		\textbf{Middle:} Parameter values relative to the total movement duration $N$, grouped by target width. 
		\textbf{Right:} Parameter values relative to the total movement duration $N$, grouped by distance to target.
	}~\label{fig:MinJerk_opt}
	\Description{Fully described in caption and text.}
\end{figure}

The optimal values of $N_{MJ}$ grouped by participants are shown in the left plot of Figure~\ref{fig:MinJerk_opt}. %
Different users can be characterized by different values for $N_{MJ}$. 
Interestingly, the user-specific effects match those observed for the stiffness parameter $k$ in the 2OL-Eq model fairly well.
Indeed, there is an inverse-linear relationship between $k$ and $N_{MJ}$, as illustrated in Figure~\ref{fig:MinJerk_2OL-Eq_params} in the Appendix.
Comparing the effects of $k$ (Figure~\ref{fig:2OL-Eq_effects}, top row) and $N_{MJ}$ (Figure~\ref{fig:MinJerk_effects}) on the respective model trajectories, however, this is not very surprising.
Both a higher stiffness $k$ and a lower surge duration $N_{MJ}$ lead to a faster movement that reaches the target earlier (note that participant 9 moved considerably slower (average duration: 1.7s) than the rest of the participants (average duration: 0.95s)).
Differences between the effects of these two parameters are mainly related to the initial acceleration, which scales with $k$ in 2OL-Eq, but is always fixed in MinJerk, and to the skewness of the velocity profile, which is only affected by $k$. %

In contrast, the average surge duration does not differ much between different task IDs (not shown). %
However, the \textit{relative time spent in the surge}, i.e., $\nicefrac{N_{MJ}}{N}$, clearly increases as the target width increases, while it is unaffected by the distance between initial and target center (see middle and right plot of Figure~\ref{fig:MinJerk_opt}).
This is consistent with previous findings, which suggest that the skewness of the velocity profile and thus the duration of corrective submovements is mainly determined by the target size, while the distance to target has a much smaller effect on the relative duration of the initial ballistic movement ~\cite{Thompson07, Bohan03, Mueller17}. %

\subsection{Discussion}
The minimum jerk model can explain the shape of the initial ballistic movement (the surge) towards the target very well.
However, both initial and terminal conditions need to be known in advance. 
The same holds for the overall movement time, unless it is identified through a parameter fitting process, using experimentally observed user data. %
It should also be noted that it is difficult to explain conceptually why users should aim to minimize the jerk of the movement (see, e.g., Harris and Wolpert~\cite{HarrisWolpert98}).
 
The main limitation of the model is that it is a pure kinematic model.
That is, the trajectory of the mouse pointer is uniquely defined given the initial and terminal conditions.
The underlying reasons for the movement, such as the acting forces, are not explained.
In particular, the model does not involve any explanation of the underlying biomechanics of the user, not even as a point-mass model such as 2OL-Eq.
Due to its deterministic nature, it cannot account for the between-trial variability typically observed in user movements (see red dashed lines in Figure~\ref{fig:MinJerk_data_ID4}).
Further, 
as an open-loop model,
the movement trajectory is completely specified at the beginning of the movement, and in its standard form, the model cannot react to perturbations or inaccuracies in the movement.
In order to explain %
how users react to visual or proprioceptive feedback, models based on optimal feedback control are required.

\section{Pointing as Optimal Feedback Control: The LQR}\label{sec:OFC}

In general, the optimal control problem given by~\eqref{eq:ocp-discrete} is very difficult to solve, since no solution method is known that guarantees convergence towards the global optimum without imposing (fairly strong) assumptions on the costs and system dynamics (e.g., convexity, continuous differentiability, etc.). %
One important subclass of problems where an analytic solution method exists is the case of \textbf{linear dynamics} and \textbf{quadratic costs}.
The solution in this case is given by the Linear-Quadratic Regulator (LQR). %

These optimal control problems usually are of the following form:
\begin{subequations}\label{eq:ocp-lq-discrete}
	\begin{align}\label{eq:ocp-lq-discrete-objective}
		\begin{gathered}
			\textsl{Minimize} \quad J_{N}(x,u)= \sum_{n=0}^{N}x_{n}^{\top}Q_{n}x_{n} + \sum_{n=0}^{N-1} u_{n}^{\top}R_{n}u_{n} \\
			\textsl{with respect to } u= (u_{n})_{n\in\{0,\dots,N-1\}}\subset \R^{m}, %
		\end{gathered} 
	\end{align}
	where $x= (x_{n})_{n\in\{0,\dots,N\}} \subset \R^{k}$ satisfies 
	\begin{align}\label{eq:discrete-control_2}
		\begin{gathered}
			x_{n+1}=A x_{n} + B u_{n}, \quad n\in\{0,\dots,N-1\}, \\
			x_{0}=\bar{x}_{0}
		\end{gathered}
	\end{align}
	for some given initial state $\bar{x}_{0}\in\R^{k}$.
\end{subequations}

As before, $x_n$ is the state of the human-computer system, $u_n$ is the control (e.g., muscle excitation), the matrix $A$ describes the dynamics of the human-computer system if no control is applied, i.e., $u\equiv0$, and $B$ describes how the control influences the system (e.g., the force generated by muscles). 
The matrices $Q_n$ and $R_n$ can be interpreted as coefficients or weights for the state and control costs, respectively, where, e.g., the former formalizes that users aim to reach the target and the latter formalizes that they aim to do so with minimal effort.
Note that in our case, the controls $u_{n}$ are one-dimensional ($m=1$), i.e., the matrix $R_{n}$ only consists of a single entry. 

Regarding the optimal control framework for Human-Computer Interaction, the minimization in Equation~\eqref{eq:ocp-lq-discrete-objective} corresponds to the \textit{Human Controller} block in Figure~\ref{fig:genmodel_extended}(b), and Equation~\eqref{eq:discrete-control_2} corresponds to the \textit{Human-Computer System Dynamics} block.
The \textit{Feedback} and \textit{Human Observer} blocks in Figure~\ref{fig:genmodel_extended}(b) are assumed trivial, i.e., observation and internal state estimate both equal the actual state of the system ($y_{n+1}=\hat{x}_{n+1}=x_{n+1}$).
This is clearly different from the open-loop case depicted in Figure~\ref{fig:genmodel_extended}(a), where the controls are independent from the state estimates.

In the following, we use the muscle model and the cost function that have been used by Todorov~\cite{Todorov05} in the case of the (stochastic, and therefore significantly more complex) Linear-Quadratic Gaussian regulator, which we discuss in Section~\ref{sec:SOFC}.
To understand the stochastic case, it is useful to first consider the deterministic case, which we introduce in this section.
The simplified second-order muscle model that we use has been proposed by van der Helm~\cite{vanderHelm00}, and obtains control signals as input $u(t)$ and yields the forces applied to the end-effector $F(t)$ 
as output:
\begin{equation}\label{eq:2I-M2OL}
	F(t) = u(t) - \tau_{1}\tau_{2}\ddot{F}(t) - (\tau_{1} + \tau_{2})\dot{F}(t),
\end{equation}
with time constants $\tau_{1}$, $\tau_{2}>0$.
Throughout this section, we use  $\tau_{1}=\tau_{2}=0.04$. 

A discrete-time approximation of these dynamics is obtained by the \textit{Forward Euler method} with time interval $h>0$,
\begin{align}\label{eq:2I-M2OL-discrete}
	\begin{gathered}
		f_{n+1} = f_{n} + \frac{h}{\tau_{2}} \left(g_{n} - f_{n}\right), \\
		g_{n+1} = g_{n} + \frac{h}{\tau_{1}} \left(u_{n} - g_{n}\right),
	\end{gathered}
\end{align}
where $f_{n}$ and $g_{n}$ denote the muscle activation (corresponding to force) and excitation at time step $n$, respectively.
Recall that $h=0.002$ corresponds to the two millisecond sampling rate of the mouse sensor.
Following Todorov~\cite{Todorov05}, we assume a unit mass of 1 kg of the hand-mouse system.

Combining these muscle dynamics with a second-order integrator, %
this leads to the following %
system dynamics matrices: %
 \begin{equation}\label{eq:lqg-dynamics-AB}
 	A = 
 	\begin{pmatrix}
		1	& h	& 0 & 0 & 0 \\
		0	& 1	& h & 0 & 0 \\
		0	& 0	& 1 - \frac{h}{\tau_{2}}  & \frac{h}{\tau_{2}} & 0 \\
		0   & 0	& 0	& 1 - \frac{h}{\tau_{1}} & 0 \\
		0	& 0	& 0	& 0	& 1
 	\end{pmatrix},
 	\quad
 	B = 
 	\begin{pmatrix}
		0 \\
		0 \\
		0 \\
		\frac{h}{\tau_{1}} \\
		0
 	\end{pmatrix}.
 \end{equation}
Here, the state $x_{n} = (p_n, v_n, f_n, g_n, T)^{\top}\in\R^{5}$ (i.e., $k=5$ in~\eqref{eq:discrete-control_2}) consists of the pointer position $p_n$ and velocity $v_n$, as well as muscle force %
$f_n$ and muscle excitation $g_{n}$. 
Moreover, the fixed target $T$ is included in the state for technical reasons. %
Note that, in contrast to the 2OL-Eq model, the controls $u_{n}$ do not equal a fixed, target-dependent value (that is, we do not prescribe pure equilibrium control), but are chosen to minimize the cost $J_{N}$.

In principle, the cost matrices $Q_{n}$ and $R_{n}$ can be chosen freely.
Based on~\cite{Todorov05}, we derive them from the following assumptions: %
\begin{itemize}
	\item Users aim to minimize the distance between pointer and target.
	\item Users aim to stay inside the target after reaching it. 
	\item Users aim to minimize the effort required to fulfill the task.
\end{itemize}

Ideally, no distance costs should occur within the target, which is assumed to be a box of width $W$.
However, a fundamental limitation of the LQR setting is that cost terms need to be quadratic in the 
states and controls.
Therefore, costs that are zero everywhere within the target are unfortunately infeasible within the LQR setting.
To approximate such costs, %
we construct the distance costs such that we have lower costs inside the target and higher costs outside.
In particular, we penalize the squared remaining Euclidean distance between the end-effector position $p_{n}$ and the desired target position $T$, which is given as
\begin{equation}
	D_{n}^2=\vert p_{n}-T \vert^2.
	\label{eq:distanceerrorcosts}
\end{equation}

To create an incentive to keep the end-effector inside the target once it is reached, the squared velocity $v_{n}^{2}$ and the squared force $f_{n}^{2}$ (which can be interpreted as acceleration, since unit mass is assumed) are penalized as well, weighted with $\omega_{v}, \omega_{f}\geq0$. 
All these cost terms are quadratic with respect to the state $x_{n}$, i.e., a positive semi-definite matrix $Q_{n}$ can be found such that
\begin{equation}\label{eq:state-cost-terms}
	x_{n}^{\top}Q_{n}x_{n}=D_{n}^{2} + \omega_{v}v_{n}^{2} + \omega_{f}f_{n}^{2}.
\end{equation}

At the same time as minimizing the distance to the target, we assume that users aim at minimizing their effort. 
This assumption is well-established in motor control theory, motivated by both neuroscientific findings and mathematical requirements~\cite{Todorov02, Li04, Guigon07}.\footnote{In optimal control problems, penalizing the controls $u$ acts as a \textit{regularizer}, i.e., it constrains the subspace of optimal solutions, which often results in a unique optimal solution.} %
Moreover, we assume that the effort cost matrices are constant in time, i.e., $R_{n}=R$ holds for all $n\in\{0,\dots,N-1\}$, and normalized with respect to the duration of the movement for better comparability between conditions. %
In particular, we choose
\begin{equation}
	R = \frac{\omega_{r}}{N-1}, %
\end{equation}
where the weight parameter $\omega_{r}>0$ determines how much effort the user is willing to invest to reduce the distance to the target more quickly. %

\begin{figure}
	\centering
	\includegraphics[width=\linewidth]{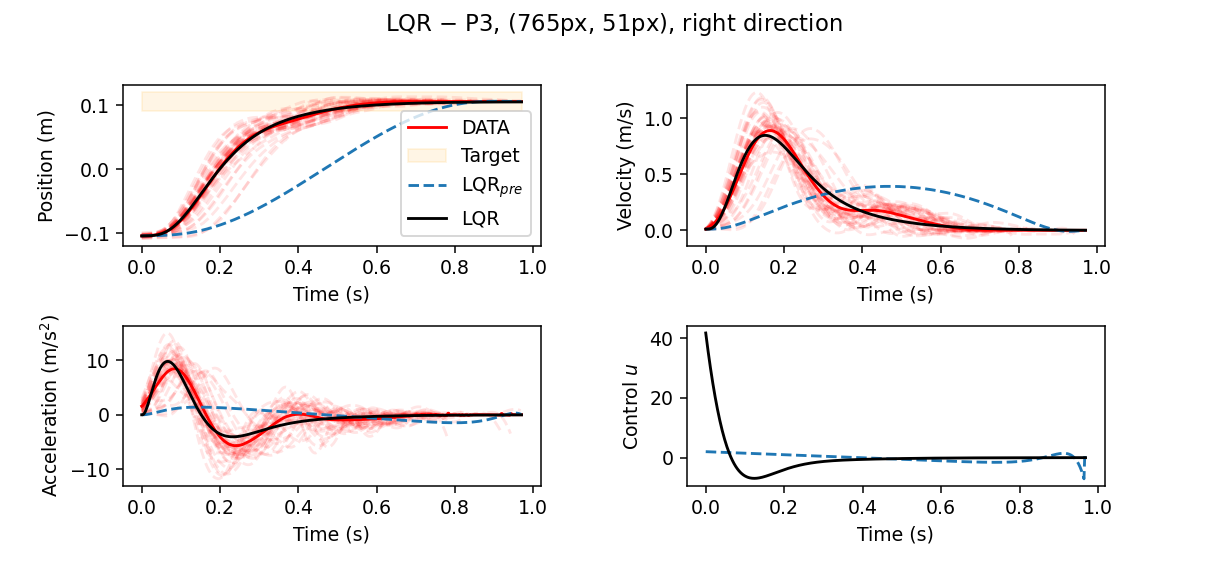}
	\caption{If state costs are only applied at the final time step $N$ (blue dashed line), the LQR cannot capture the correction phase, which is highly pronounced for tasks with sufficiently small target. With continuous state costs (black solid line), LQR trajectories visually exhibit a good fit, although the peak acceleration is slightly higher. %
		In contrast to MinJerk, the end of the surge phase does not need to be prescribed, but emerges implicitly from the model.
	}
	\label{fig:LQR_data_ID4}%
	\Description[Position, velocity, and acceleration time series of the LQR vs. user data (ID 4 task)]{Position, velocity, and acceleration time series of both the Pointing Dynamics Dataset (Participant 3, ID 4 (765px distance, 51px width), right direction) and the corresponding LQR simulation trajectories, each with continuous state costs and with terminal state costs only. The control time series of both LQR variants are shown as well.}
\end{figure}

While in~\cite{Todorov05}, the state costs~\eqref{eq:state-cost-terms} were only applied at the final time $n=N$, which must be known in advance, in the deterministic LQR case without neuromotor noise, this does not work. 
As can be seen in Figure~\ref{fig:LQR_data_ID4}, for tasks with fairly small targets %
that require a considerable correction phase, the resulting LQR trajectory (dashed blue lines) does not resemble the observed user behavior at all. 
The main reason for this is that distance costs that only occur at a single time step do not create an incentive to reach this state earlier than necessary. 
The constant incurrence of control costs adds to this, resulting in an optimal policy that chooses relatively small controls during most of the movement, while shortly before the final time step $N$, larger controls are applied to reach the target ``just in time'' with low velocity and acceleration. %

This problem can be addressed in two ways.
Either neuromotor noise can be included, as it is done in~\cite{Todorov05} and in the next section.
Or, in the deterministic case, behavior more similar to humans can be achieved by assuming that the state costs from Equation~\eqref{eq:state-cost-terms} are applied \textit{during the entire movement}. %
This clearly creates an incentive to move the pointer towards the target from the start, in order to reduce the sum of the distance costs.

In summary, the objective function of our LQR model is given by
\begin{equation}\label{eq:lqr-objective-explicit}
	J_{N}^{\text{(LQR)}}(x,u)= \sum_{n=0}^{N} \left( D_{n}^{2} +  \omega_{v} v_{n}^{2} + \omega_{f} f_{n}^{2} \right) + \frac{\omega_{r}}{N-1} \sum_{n=0}^{N-1}u_{n}^{2}.
\end{equation}

We assume that the user computes the \emph{optimal} control, denoted by $u^*_n$, in a feedback manner, based on the current state (i.e., the model is \textit{closed-loop}).
It has been proven that for these kinds of problems the optimal control $u^*_n$ depends linearly on the state~\cite{Dorato71}, i.e.,
\begin{equation}\label{eq:lqr-control-law}
	u_{n}^{*}=\pi(x_{n})=-L_{n}x_{n}
\end{equation}
holds for some uniquely determined \textit{feedback gain matrices} $L_{n}$.
Recall that in our case, the control $u_n$ is one-dimensional.
Since the state $x_n$ is a vector in $\R^5$, $L_n$ is thus a $1\times 5$ matrix.

Given the matrices~$A$ and $B$ as well as the cost function~$J_N^{\text{(LQR)}}$, these feedback gain matrices can be computed once before movement onset (i.e., at the planning stage) by iteratively solving the corresponding \emph{Discrete Riccati Equation} (see Appendix; details are given in~\cite[Theorem~7]{Todorov98_thesis}). %
This results in a very fast on-line computation of the optimal controls, yet taking into account the most recent state observations.

\subsection{Analysis of Parameters}\label{ssec:lqr-analysis-of-parameters}

\begin{figure}
	\centering
	\includegraphics[width=.88\linewidth]{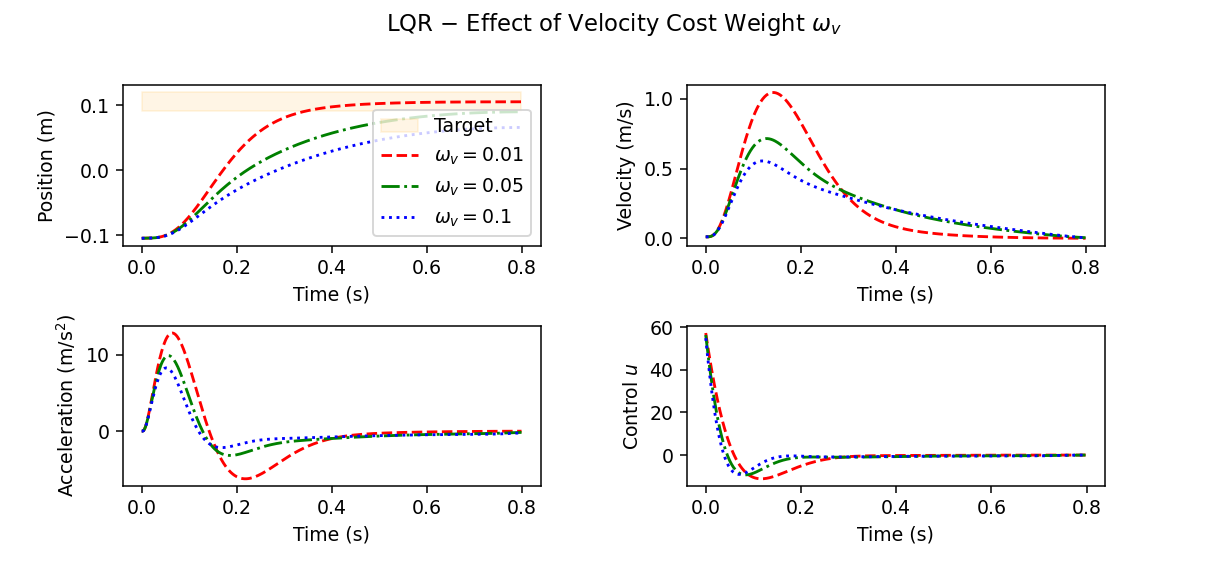}
	\\
	\includegraphics[width=.88\linewidth]{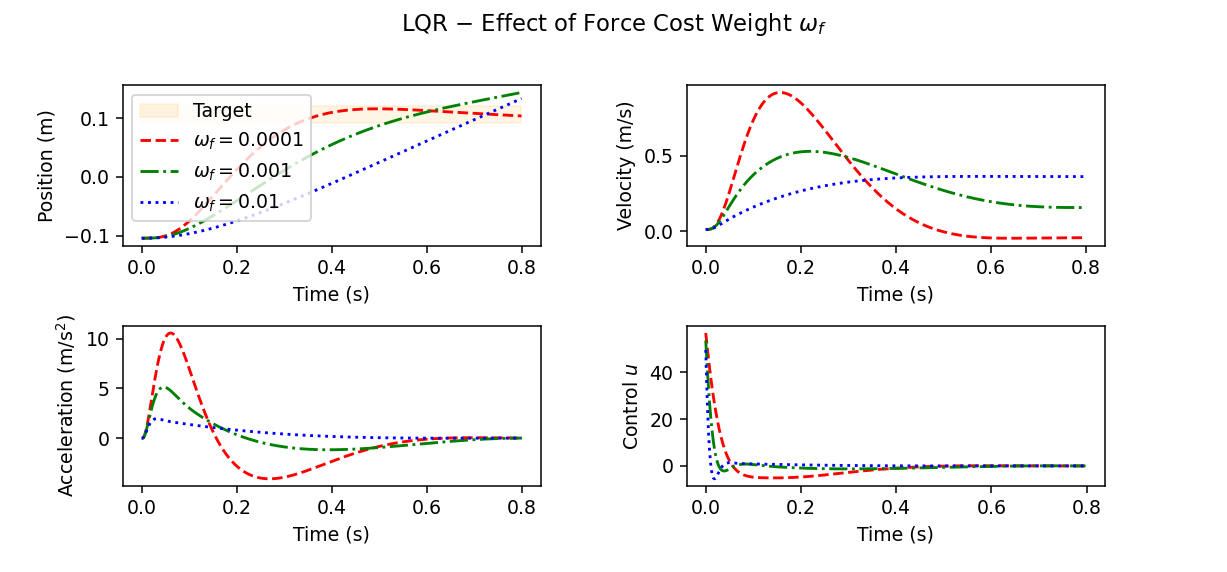}
	\\
	\includegraphics[width=.88\linewidth]{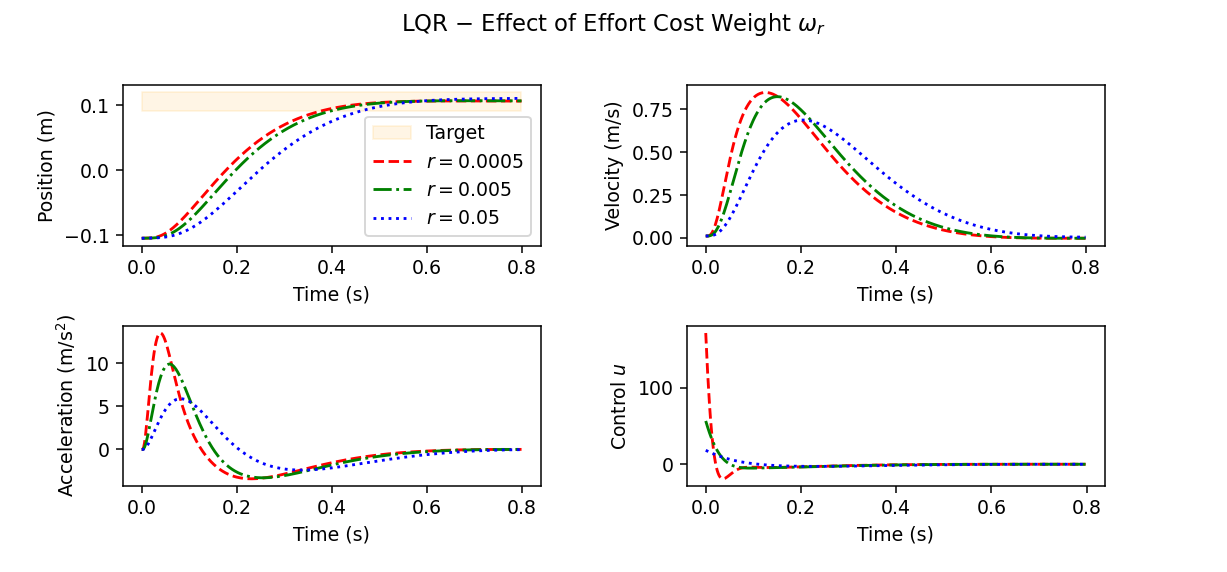}
	\caption{
		Position, velocity, acceleration, and control time series %
		of typical LQR trajectories with target shown as the orange box.
		\textbf{Top:} Effect of velocity cost weight $\omega_{v}$ with fixed cost weights $\omega_{f}=0$, $\omega_{r}=~$5e-3 (red dashed: $\omega_{v}=0.01$, green dash-dotted: $\omega_{v}=0.05$, blue dotted: $\omega_{v}=0.1$). %
		\textbf{Middle:} Effect of force cost weight $\omega_{f}$ with fixed cost weights $\omega_{v}=0$, $\omega_{r}=~$5e-3 (red dashed: $\omega_{f}=~$1e-4, green dash-dotted: $\omega_{f}=~$1e-3, blue dotted: $\omega_{f}=~$1e-2). %
		\textbf{Bottom:} Effect of effort cost weight $\omega_{r}$ with fixed cost weights $\omega_{v}=0.01$, $\omega_{f}=~$1e-4 (red dashed: $\omega_{r}=~$5e-4, green dash-dotted: $\omega_{r}=~$5e-3, $\omega_{r}=0.05$). %
	}
	\label{fig:LQR_effects}
	\Description{Fully described in caption and text.}
\end{figure}

In Figure~\ref{fig:LQR_effects}, the individual effects of the cost function weights $\omega_{v}$, $\omega_{f}$, and $\omega_{r}$ are shown. 
A higher velocity cost weight $\omega_{v}$ (top plots, blue dashed lines) results in a lower peak velocity, as expected, since velocity is penalized quadratically. 
Keeping the remaining parameters constant, this leads to a less symmetric velocity profile, as higher velocities towards the end of the movement are necessary to compensate for the lower peak. %
Moreover, the target is reached later. 
Note that this only occurs as long as the velocity cost weight is not dominant.
Otherwise, for large enough $\omega_{v}$, there is no incentive to reach the target at all. %

Similar effects (peak values, symmetry) can be observed in the acceleration profile for the force cost weight $\omega_{f}$ (since the forces are applied to a unit mass and thus can be interpreted as acceleration) and in the control profile for the effort cost weight $\omega_{r}$.
Moreover, large force cost weights lead to a constant, positive velocity at the end of the movement (middle plots, velocity profile), i.e., the pointer moves across the target instead of staying inside (middle plots, position profile).
In contrast, the magnitude of the effort cost weight $\omega_{r}$ mainly affects the duration of the surge (bottom plots, position profile), while the target is still reached with relatively low velocity and acceleration for moderate values of~$\omega_{r}$. %

\subsection{Results of Parameter Fitting}

Analogously to the parameter fitting process for the 2OL-Eq and MinJerk models, we identify optimal values for the cost weights $\omega_{v}$, $\omega_{f}$, and $\omega_{r}$ collected in the paramter vector $\Lambda$ for each mean trajectory.
Since these parameters only define the objective function $J_{N}^{\text{(LQR)}}$ of the OCP, and the system matrices $A$ and $B$ are uniquely determined given the above fixed values of $m$, $\tau_{1}$, and $\tau_{2}$, the parameter fitting for the LQR can be regarded as an inverse optimal control problem~\cite{Jin21}.\footnote{Including the system dynamics parameters $m$, $\tau_{1}$, and $\tau_{2}$ in the parameter fitting process did not improve the fit to observed user behavior.} %
As usual, we use the %
bi-level approach 
described in Section~\ref{sec:param-fitting}, i.e., at each iteration of the parameter fitting method, the OCP subject to the parameter vector $\Lambda$ is solved %
as described above.

\subsubsection{Qualitative Results}

The resulting optimal LQR trajectory (for the same ID 4 task considered for the previous models) is shown as the black solid line in Figure~\ref{fig:LQR_data_ID4}. %
The position time series is approximated fairly well at first glance, even better than by the MinJerk trajectory (see Figure~\ref{fig:MODELS_data_ID4}; quantitative comparisons between all models are given in Section~\ref{sec:quantitative-comparison}).
Peak velocity and acceleration are also very close to the values of the respective user trajectory, albeit the maximum acceleration is higher and the timing of the minimum and maximum acceleration does not match exactly. %
However, it is important to note that the duration of the surge phase was not explicitly built into the LQR model\footnote{In theory, it would be possible to include prior knowledge about submovements and thus induce time-dependent behavior by choosing the time-dependent cost matrices $Q_{n}$ and $R_{n}$ appropriately.}, but emerges naturally from the interplay of the optimal parameters $\omega_{v}$, $\omega_{f}$, and $\omega_{r}$. %
In contrast, in the MinJerk model, the duration of the surge phase needs to be either known in advance or determined using the parameter fitting process.

\subsubsection{Quantitative Results}

\begin{figure}
	\centering
	\subfloat{\includegraphics[width=0.45\linewidth]{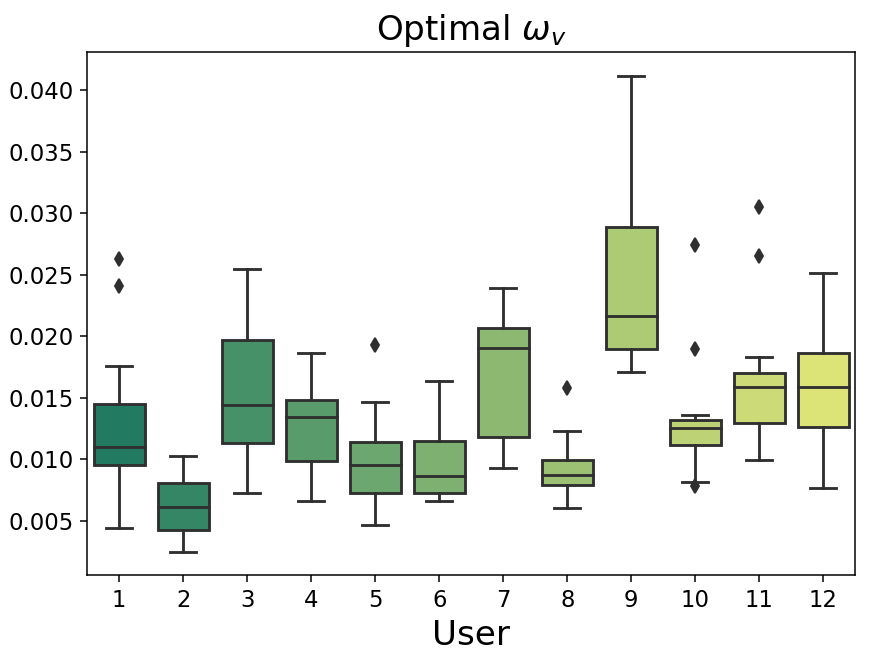}}
	\subfloat{\includegraphics[width=0.45\linewidth]{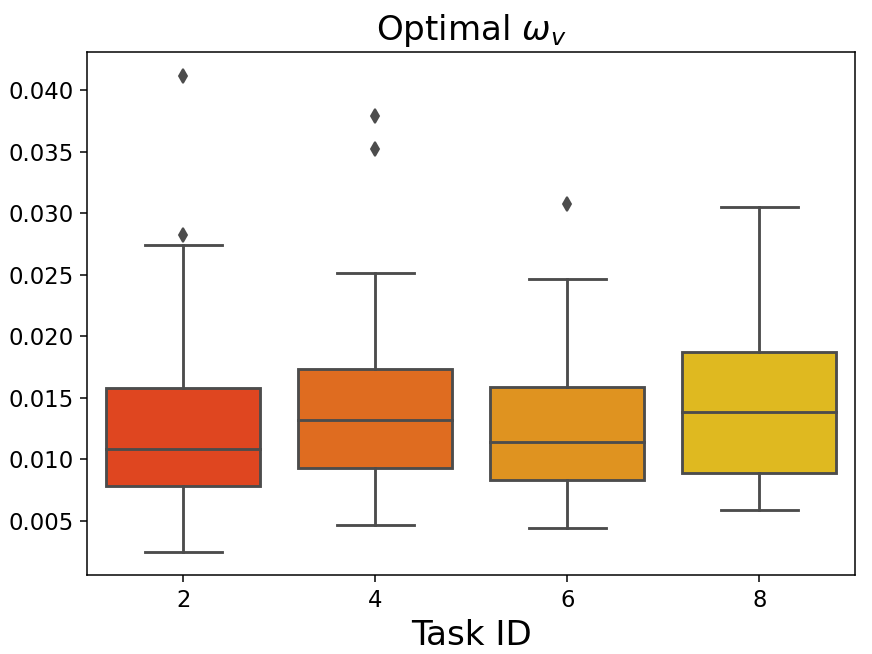}}
	\\
	\subfloat{\includegraphics[width=0.45\linewidth]{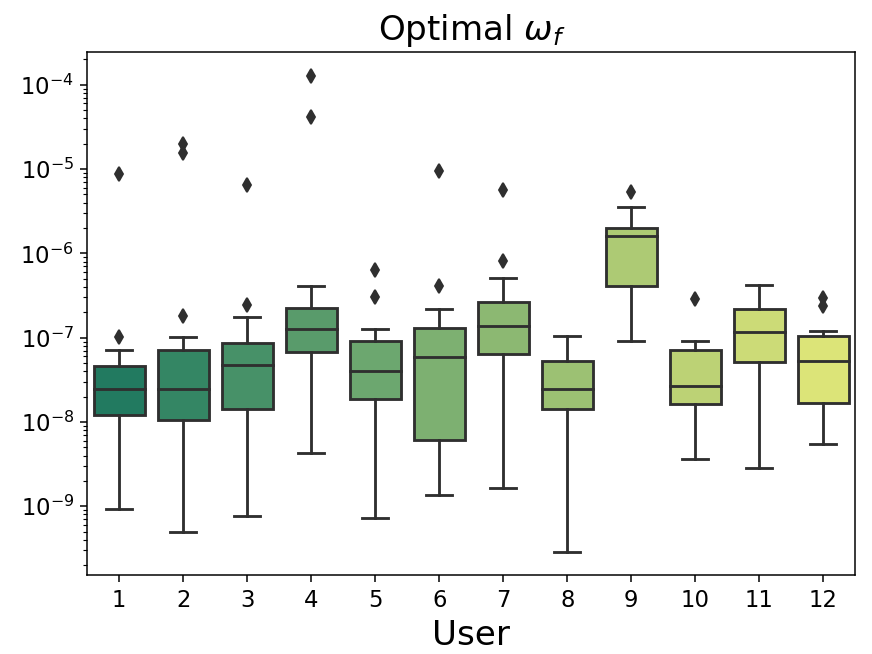}}
	\subfloat{\includegraphics[width=0.45\linewidth]{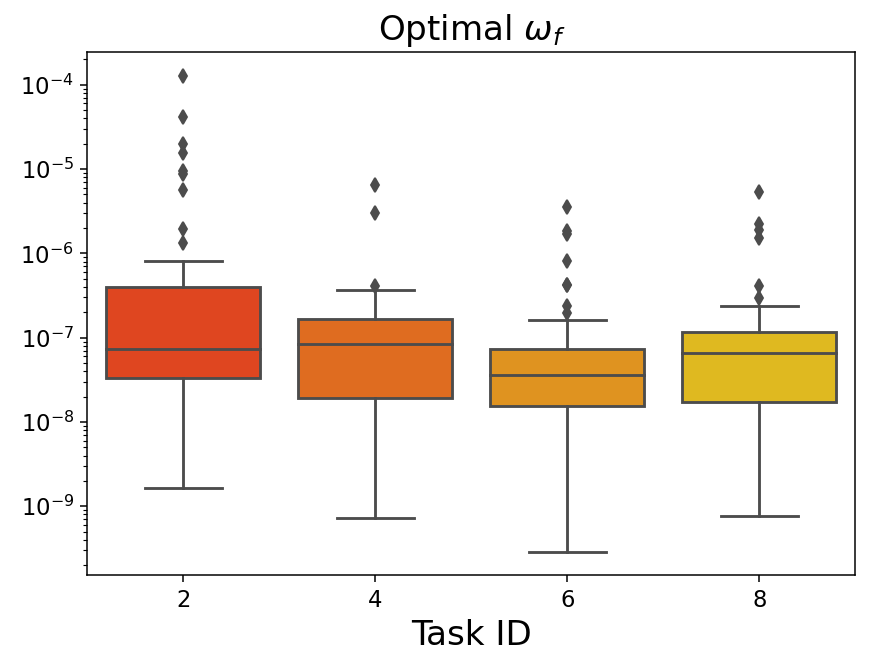}}
	\\
	\subfloat{\includegraphics[width=0.45\linewidth]{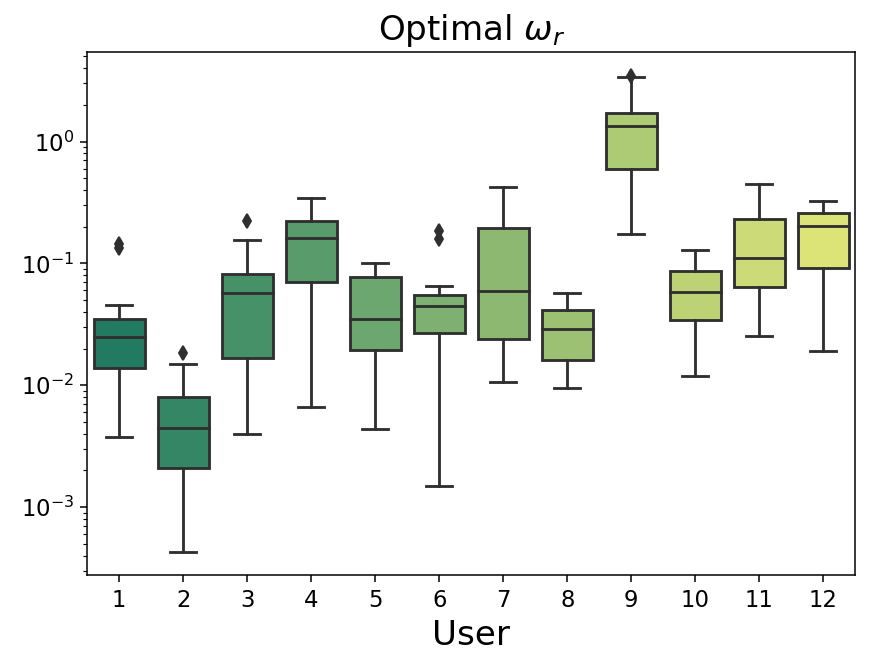}}
	\subfloat{\includegraphics[width=0.45\linewidth]{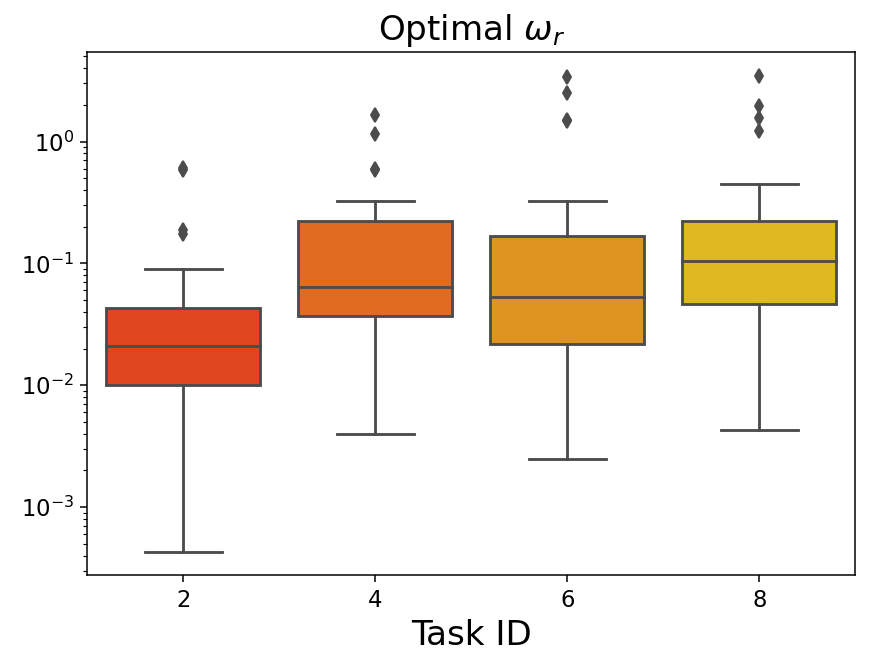}}
	\caption{Parameters of the LQR model, optimized for the mean trajectories of all participants, tasks, and directions, grouped by participants (left) and by ID (right).
		For better visibility, %
		the optimal values of $\omega_{f}$ and $\omega_{r}$ are plotted on a logarithmic scale.
	}~\label{fig:LQR_opt}
	\Description{Fully described in caption and text.}
\end{figure}

All optimal parameter values, grouped by both user and task ID, are shown in Figure~\ref{fig:LQR_opt} (note the logarithmic scale for $\omega_{f}$ and $\omega_{r}$).

Each of the three parameters exhibit a large between-user variability.
Interestingly, the between-user effects show similar trends and characteristics for all cost weights.
For example, trajectories of participant 9 are characterized by very large weights (i.e., relatively small distance costs, since the weights affect every term in the cost function but the distance costs), while the behavior of participant 2 is explained best by considerably lower weights (i.e., relatively high distance costs).
Moreover, these characteristic differences in the behavior of individual users are similar to those identified for the surge duration $N_{MJ}$ in MinJerk (see Figure~\ref{fig:MinJerk_opt}~(left)).
In contrast, the task difficulty does not seem to have a clear effect on the optimal values of the cost weights.
Hence, our findings suggest that the identified optimal parameters rather encode user-specific characteristics than the task under consideration. %

\subsection{Discussion}
The LQR combines beneficial features of both 2OL-Eq (movement duration emerges from the model) and MinJerk (smooth movements with ``close-to-bell-shaped'' velocity profiles, as observed in many user studies).
With a continuous penalization of remaining distance to target, velocity, and force \textit{during the entire movement}, the LQR model is able to explain \textit{average} user behavior in terms of position, velocity, and acceleration profiles.
However, whether the modified cost terms are plausible from a biomechanical and neuroscientific perspective is debatable~\cite{Berret11, Richardson02}.
Moreover, as the MinJerk model, due to its deterministic nature, the LQR model cannot account for the considerable between-trial variability, which is typically observed in user movements.

In the next section, we will thus introduce signal-dependent control noise, which in combination with continuous effort costs and only terminal distance, velocity, and force costs, allows to replicate user behavior similarly well as the proposed LQR variant, with the additional benefit of obtaining \textit{a distribution of} optimal trajectories.

\section{Pointing as Optimal Feedback Control Subject to Signal-Dependent Motor Noise: The LQG}\label{sec:SOFC}

While the LQR model%
 visually captures typical \textit{average} user behavior relatively well (see Figure~\ref{fig:LQR_data_ID4}; for a quantitative comparison between different models, see Section~\ref{sec:model-comparison}), it has one major drawback: the proposed optimal trajectory is necessarily deterministic and thus cannot account for the large variability %
observed among multiple trials of the same participant for the same task~\cite{Dhawale17, Todorov02}. 

An extension %
that allows to consider different sources of variability, which might occur during the sensorimotor control loop, is the \textit{Linear-Quadratic Gaussian Regulator (LQG)}~\cite{Hoff92, Loeb90, Todorov05}. 
It belongs to the class of \textit{Stochastic Optimal Feedback Control (SOFC)} models, as it takes into account the variance resulting from various noise terms.

In the following, we will present and analyze the LQG model introduced by Todorov for reaching movements~\cite{Todorov05}.
In comparison to the LQR model from Section~\ref{sec:OFC}, this model includes signal-dependent Gaussian control noise and an observation model with additive Gaussian noise. Moreover, all state costs (i.e., distance, velocity, and force costs) are only applied at the final time step $N$. %

\subsection{Linear-Quadratic Gaussian Regulator (LQG)}\label{sec:LQG}

Starting from the linear-quadratic control problem~\eqref{eq:ocp-lq-discrete}, we extend the deterministic LQR model presented in the previous section with stochastic noise terms.
In principle, there are two main types of noise terms that can be included in the system dynamics: additive noise and multiplicative noise.\footnote{In this paper, we only consider \textit{white noise}, i.e., the noise terms that occur at different time steps are assumed to be independent. However, the LQG model could be extended to incorporate temporarily correlated noise by augmenting the state space accordingly (for details, see~\cite[Section 3.4.3]{Todorov98_thesis}).}
While the former introduces the same level, or variability, to the system at each time step, the variance of multiplicative noise depends on the system variables themselves. 
In the considered case of \textit{signal-dependent (multiplicative) control noise}, a larger magnitude of the applied control $u_{n}$ thus results in a higher uncertainty about the subsequent state $x_{n+1}$.
From a neuroscientific perspective, this dependency can be justified by the empirically observed effect of neural motor commands on the variance of the resulting motor-neuronal firing~\cite{Sutton67, Schmidt79, Clamann69, Jones02}. %
In particular, signal-dependent noise can account for well-established phenomena such as cosine tuning, muscle synergies, smooth movements, and the trade-off between speed (or duration) and the end-point accuracy of a movement~\cite{Todorov02b, HarrisWolpert98, Todorov04}.

Introducing the %
\textit{control noise level} $\sigma_{u}>0$ and the sequence $(\eta_{n})_{n\in\{0,\dots,N-1\}}$ of (univariate) Gaussian random variables $\eta_{n}\sim \mathcal{N}(0; 1)$, the resulting discrete-time system dynamics can be written as %
\begin{align}\label{eq:stochastic-control}
	\begin{gathered}
		x_{n+1}=A x_{n} + (1 + \sigma_{u} \eta_{n})B u_{n}, \quad n\in\{0,\dots,N-1\}, \\
		x_{0}\sim \mathcal{N}(\bar{x}_{0}, \Sigma_{0}),
	\end{gathered}
\end{align}
where the initial state $x_{0}$ is drawn from a multivariate Gaussian distribution with given mean $\bar{x}_{0}$ and covariance matrix $\Sigma_{0}$.
The specific values for $\bar{x}_{0}$ and $\Sigma_{0}$ can be extracted from data, see Section~\ref{ssec:lqg-analysis-of-parameters}.
Following~\cite{Todorov05}, we do not include additive control noise, albeit this would be easily possible in the proposed framework. %

As in the previous section, the state $x_{n}\in\R^5$ contains pointer position, velocity, force, and muscle excitation, as well as the target position. 
We assume that the controller, which needs to decide for a control $u_{n}\in\R$ at each time step $n\in\{0,\dots,N-1\}$, does not have complete access to the current state of the system $x_{n}$. 
The main reason is that usually, not all information stored in $x_{n}$ %
is observable. %
This means that the control $u_{n}$ at time step $n\in\{0,\dots,N-1\}$ may depend only on some observation $y_{n}\in\R^{l}$ ($l\in\N$), but not on the true state $x_{n}$. %
Moreover, the observation can be noisy, i.e., errors during observation may occur. 
More specifically, to maintain linear model dynamics, we assume that the (noisy) observation $y_{n}$ is linear in $x_{n}$, i.e.,
\begin{equation}\label{eq:lqg-observation-dynamics}
	y_{n} = H_{n}x_{n} + G\xi_{n}.
\end{equation}
For simplicity, we assume that position, velocity, and force values can be observed immediately in global coordinates, while muscle excitation and target position cannot be directly observed, i.e.,
\begin{equation}
	H_{n} =
	\begin{pmatrix}
		1	& 0	& 0 & 0	& 0 \\
		0	& 1	& 0 & 0	& 0 \\
		0	& 0	& 1 & 0	& 0
	\end{pmatrix},
\end{equation}
which results in
\begin{equation}\label{eq:lqg-observation-space}
	H_{n}x_{n} = \left(p_{n}, v_{n}, f_{n}\right)^{\top}.
\end{equation}
The error that occurs during observation is modeled by the additive noise term $G\xi_{n}$.
The $3\times 3$ matrix $G$ determines the individual noise levels.
As in~\cite{Todorov05}, it is given by
\begin{equation}\label{eq:G-matrix-LQR}
	G = \sigma_{s} ~ diag(0.02, 0.2, 1),
\end{equation}
where $\sigma_{s} > 0$ is a scaling parameter and the constants were chosen in order to reflect the different magnitudes between position, velocity, and force (i.e., acceleration). 
The vector~$\xi_{n}$ is a three-dimensional Gaussian random variable, i.e., $\xi_{n}\sim \mathcal{N}(0;I_{3})$, where $I_3$ denotes the $3\times 3$ identity matrix.
In particular, the observation $y_n$ is a three-dimensional vector.

The objective function is related to the one used in the LQR model, \eqref{eq:lqr-objective-explicit}, i.e., a weighted combination of distance, velocity, and force costs, plus effort costs.
The two differences are: (i) distance, velocity, and force costs incur only at the final time step $N$, while only the effort costs incur continuously throughout the movement, and (ii) since state and control are not deterministic due to noise, we use the expected value of these terms, denoted by $\mathbb{E}[\cdot]$. 
The objective function is thus given by
\begin{equation}\label{eq:lqg-objective-explicit}
	J_{N}^{\text{(LQG)}}(x,u) = \mathbb{E} \left[ D_{N}^{2} + \omega_{v} v_{N}^{2} + \omega_{f} f_{n}^{2} + \frac{\omega_{r}}{N-1} \left(\sum_{n=0}^{N-1}u_{n}^{2}\right) \right].
\end{equation}
In total, this results in the following stochastic optimal control problem:
\begin{subequations}\label{eq:ocp-lqg-discrete}
	\begin{align}\label{eq:ocp-lqg-discrete-objective}
		\begin{gathered}
			\textsl{Minimize} \quad J_{N}^{\text{(LQG)}}(x,u) \\ %
			\textsl{with respect to } u= (u_{n})_{n\in\{0,\dots,N-1\}}\subset \R^{m}, %
		\end{gathered} 
	\end{align}
	where %
	$y=(y_{n})_{n\in\{0,\dots,N-1\}} \subset \R^{l}$ satisfies 
	\begin{align}\label{eq:discrete-observation_1}
		\begin{gathered}
			y_{n}=H_{n} x_{n} + G \xi_{n}, \quad n\in\{0,\dots,N-1\}, %
		\end{gathered}
	\end{align}
	and $x= (x_{n})_{n\in\{0,\dots,N\}} \subset \R^{k}$ satisfies 
	\begin{align}\label{eq:discrete-control_4}
	\begin{gathered}
		x_{n+1}=A x_{n} + (1 + \sigma_{u} \eta_{n})B u_{n}, \quad n\in\{0,\dots,N-1\}, \\ %
		x_{0}\sim \mathcal{N}(\bar{x}_{0},\Sigma_{0}).
	\end{gathered}
	\end{align}
\end{subequations}
We specifically use the dimensions $m$, $l$, and $k$ throughout to underline the generality and easy expandability of the LQG model. 
In our case, we recall that $m=1$, $l=3$, and $k=5$.

Since the true state $x_{n}$ %
at time step $n$ is not available to the controller, it needs to compute internal state estimates $\hat{x}_{n}$ %
based on the information available. 
Under the LQG assumptions (linear dynamics, quadratic costs, Gaussian noise), the state estimates $\hat{x}_{n}$ can be computed using a linear estimator:
\begin{equation}\label{eq:lqg-estimator}
	\hat{x}_{n+1}=A \hat{x}_{n} + B u_{n} + K_{n} (y_{n} - H_{n}\hat{x}_{n}), \quad n\in\{0,\dots,N-1\}.
\end{equation}
Here, $y_{n}$ and $H_{n}\hat{x}_{n}$ denote the obtained and the expected sensory input at time step $n$, respectively (recall the \textit{Human Observer} block in Figure~\ref{fig:genmodel_extended}(b)), where initially, at $n=0$, we set $\hat{x}_{0}=\bar{x}_{0}$. 
The matrix $K_{n}$, which is to be determined, specifies to which degree the observed differences between these quantities should be taken into account for the computation of the subsequent state estimate $\hat{x}_{n+1}$. %
As a trivial example, consider $K_{n}=0$, which corresponds to the open-loop case, i.e., no sensory information is used by the controller.
When it comes to choosing a ``good'' matrix $K_{n}$, the above LQG assumptions allow to analytically derive the matrix that is optimal in the sense that it minimizes the 
objective function $J_{N}^{\text{(LQG)}}$.
The resulting optimal estimator %
is known as the \textit{Kalman Filter}.\footnote{In the case of state-dependent observation noise, i.e., if $G$ is replaced by $G_{n}(x_{n})$, the Kalman Filter is still optimal among all linear estimators~\cite{Todorov98_thesis}.}

Moreover, it can be shown that, similar to the LQR case, the optimal closed-loop solution $u^{*}=(u_{n}^{*})_{n\in\{0,\dots,N-1\}}$ is linear in the state estimate $\hat{x}_{n}$, %
i.e., there exist unique \textit{feedback gain matrices} $L_{n}$, $n\in\{0,\dots,N-1\}$ such that
\begin{equation}\label{eq:lqg-control-law}
	u_{n}^{*}=\pi(\hat{x}_{n})=-L_{n}
	\hat{x}_{n}
\end{equation}
holds. %
As in the previous section, $L_n$ is a $1\times 5$ matrix.
For more details, we refer to~\cite{Todorov05}. 

It is important to note that in the considered case of signal-dependent control noise, the feedback gain matrices $L_{n}$ and the Kalman gain matrices $K_{n}$ non-trivially depend on each other.
More precisely, each feedback gain matrix $L_{n}$ depends on the subsequent Kalman filter matrices $(K_{i})_{i\in\{n+1,\dots,N-1\}}$, and each Kalman filter matrix $K_{n}$ depends on the previous feedback gain matrices $(L_{i})_{i\in\{0,\dots,n-1\}}$.
Thus, the feedback gain matrices $L_{n}$ can only be computed backward in time, starting at the final time step $n=N$, whereas the Kalman filter matrices $K_{n}$ can only be computed forward in time, starting at the first time step $n=0$.
In particular, one must be given in order to optimally determine the other.
Fortunately, iterating alternately between the two optimizations results in a coordinate descent algorithm, which can be shown to converge towards a (local) minimum~\cite{Todorov05}.\footnote{Numerically, it has been shown that this algorithm even converges towards the global minimum~\cite{Todorov05}.}
This iterative computation of the matrices $K_n$ and $L_n$ can be done offline, i.e., before movement onset.

As suggested by Todorov~\cite{Todorov05}, we therefore initialize the Kalman gain matrices to zero, i.e., we first compute the optimal open-loop control strategy matrices $L_{n}$, and iteratively compute $K_{n}$ and $L_{n}$ until the resulting objective function value %
converges towards its optimum. %
In particular, we terminate the iterative optimization procedure in the $i$-th step if the relative improvement of $J_{N}^{\text{(LQG)}}$ falls below a predefined threshold $\epsilon_J>0$, i.e.,
\begin{equation}\label{eq:lqg-iterations}
	\left\vert \frac{\left(J_{N}^{\text{(LQG)}}\right)_{i + 1} - \left(J_{N}^{\text{(LQG)}}\right)_{i}}{\left(J_{N}^{\text{(LQG)}}\right)_{i}} \right\vert \leq \epsilon_J,
\end{equation}
or after a maximum number of iterations (set to 20; we found that this value sufficed to obtain parameters that did not change considerably afterwards).
In our implementation, we use $\epsilon_J=~$1e-3. %

We do not make use of the \textit{adaptive Kalman filter},
as it was proposed for an LQG model designed for via-point movements~\cite{Todorov98_thesis}.
The adaptive Kalman filter computes Kalman gain matrices $K_{n}$ that explicitly depend on the observations $(y_{i})_{i\in\{0,\dots,n-1\}}$ received so far, i.e., this needs to be done online. 
Moreover, since these observations are stochastic, the matrices $K_{n}$ (and thus the expected behavior) differ between several runs.
In addition, it was empirically observed that the adaptive Kalman filter might be unstable~\cite{Todorov05}. %
In contrast, the non-adaptive Kalman filter from~\cite{Todorov05} %
computes matrices $K_{n}$ that only depend on information available \textit{before movement onset}. 
The resulting a priori expectations over the state sequences %
can be directly used in the loss function of the parameter fitting process to compare the expected outcome of different parameter vectors. %

In principle, using an adaptive filter only during runtime (with $L_{n}$ optimized with respect to the non-adaptive filter) could further reduce the expected total costs. However, this effect has shown to be minor~\cite{Todorov05}, which is why we refrain from using an adaptive filter at all.

\subsection{Analysis of Parameters}\label{ssec:lqg-analysis-of-parameters}

The above mentioned 
a priori expectations
over state trajectories, 
which are used to compute the non-adaptive Kalman filter, 
allow to compare the stochastic results of the LQG controller between different sets of parameters.
In total, there are 5 parameters we aim to optimize:
\begin{itemize}
	\item the (terminal) velocity cost weight $\omega_{v}$,
	\item the (terminal) force cost weight $\omega_{f}$,
	\item the effort cost weight $\omega_{r}$,
	\item the signal-dependent control noise level $\sigma_{u}$, and
	\item the observation noise level $\sigma_{s}$.
\end{itemize}

The first three parameters correspond to those from Section~\ref{ssec:lqr-analysis-of-parameters}. 
In particular, setting all other parameters as well as the initial variance to zero and using exact initial state estimates $\hat{x}_{0}=\bar{x}_{0}=x_{0}$ and $\hat{\Sigma}_{0}=\Sigma_{0}=0$, the stochastic optimal control problem~\eqref{eq:ocp-lqg-discrete} equals the deterministic optimal control problem~\eqref{eq:ocp-lq-discrete}, i.e., the LQR %
is a special
case of the LQG. %
However, in contrast to the LQR model from Section~\ref{sec:OFC}, distance, velocity, and force costs are only applied at the final time step $N$.
The effect of the individual cost weight parameters is thus much smaller. %

In principle, the same fixed noise level parameters $\sigma_{u}$ and $\sigma_{s}$ could be used for all participants and task conditions.
However, choosing them from literature is difficult since their effects strongly depend on the assumptions and system dynamics of the considered model.

\begin{figure}
	\centering
	\includegraphics[width=0.87\linewidth]{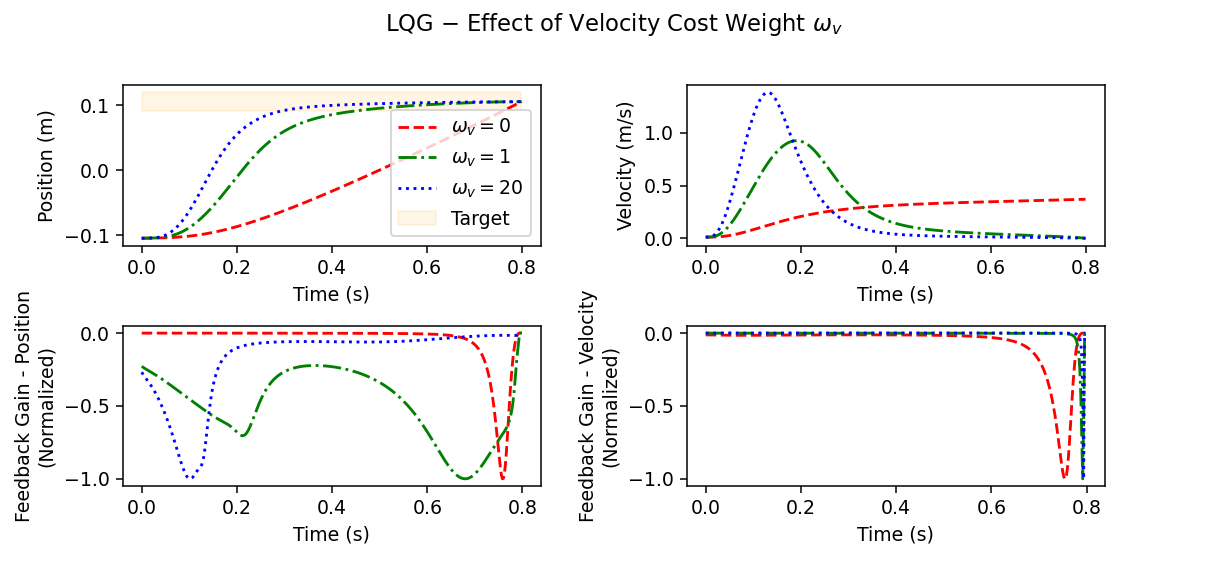}
	\\
	\includegraphics[width=0.87\linewidth]{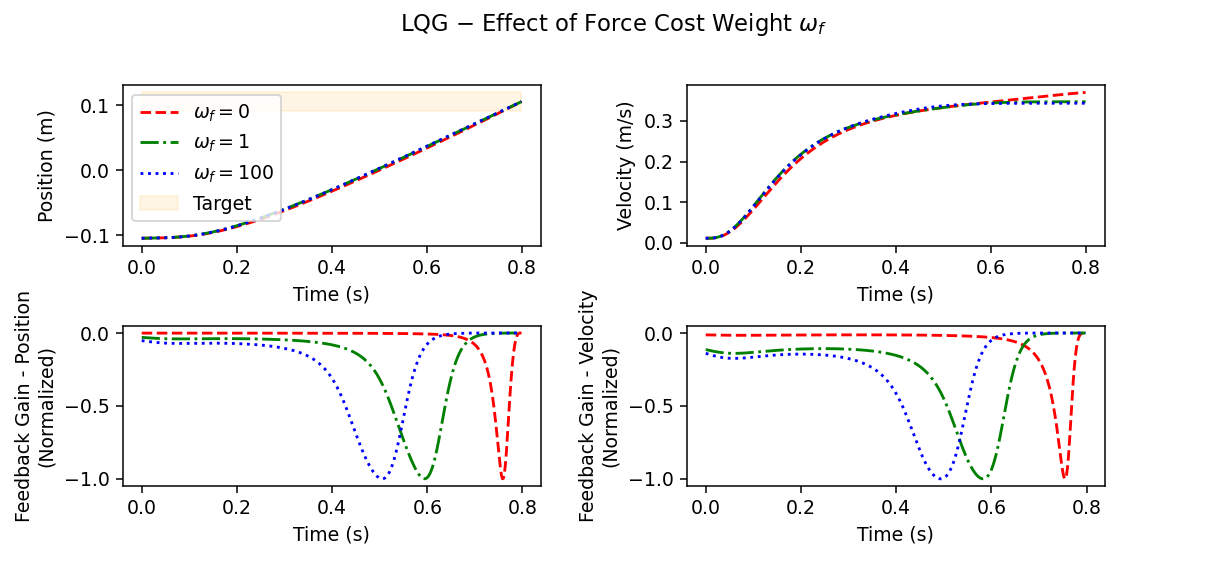}
	\\
	\includegraphics[width=0.87\linewidth]{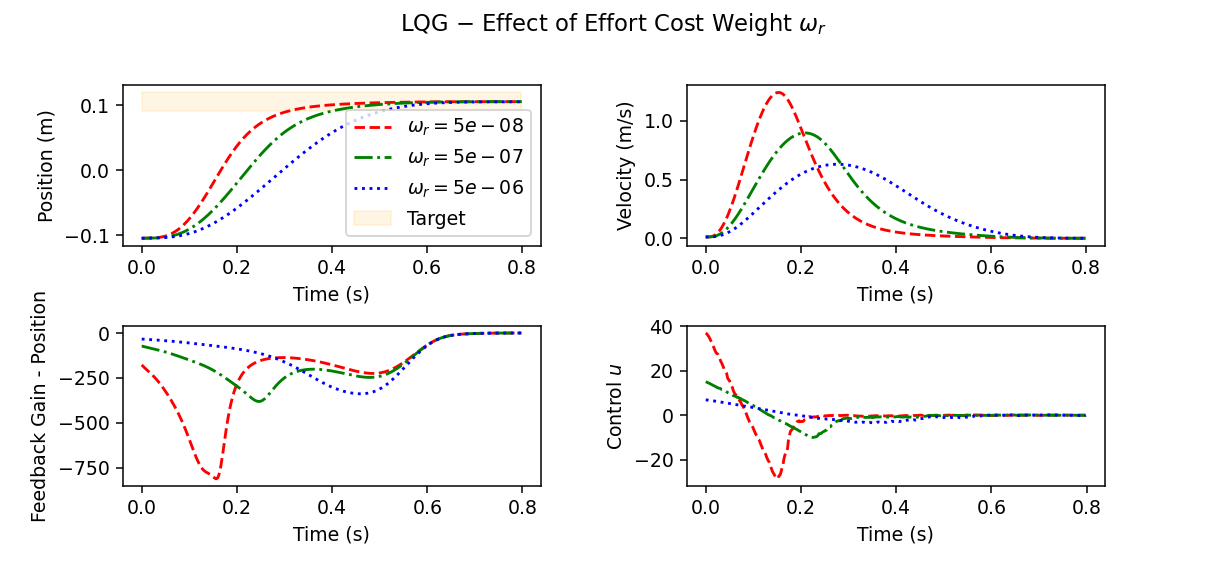}
	\caption{
		Typical LQG trajectories with target shown as the orange box, as well as selected entries of the corresponding feedback gain matrices, and control time series, with noise levels $\sigma_u=0.2$, $\sigma_s=0.5$.
		\textbf{Top:} Effect of velocity cost weight $\omega_{v}$ with fixed cost weights $\omega_{f}=0$, %
		$\omega_{r}=~$1e-7 
		(red dashed: $\omega_{v}=0$, green dash-dotted: $\omega_{v}=1$, blue dotted: $\omega_{v}=20$).
		\textbf{Middle:} Effect of force cost weight $\omega_{f}$ with fixed cost weights $\omega_{v}=0$, %
		$\omega_{r}=~$1e-7 
		(red dashed: $\omega_{f}=0$, green dash-dotted: $\omega_{f}=1$, blue dotted: $\omega_{f}=100$).
		\textbf{Bottom:} Effect of effort cost weight $\omega_{r}$ with fixed cost weights $\omega_{v}=\omega_{f}=2$ %
		(red dashed: $\omega_{v}=~$5e-8, green dash-dotted: $\omega_{v}=~$5e-7, blue dotted: $\omega_{v}=~$5e-6). %
	}
	\label{fig:LQG_effects_0}
	\Description{Fully described in caption and text.}
\end{figure}

As shown in Figure~\ref{fig:LQG_effects_0} (top plots), an increase in the velocity cost weight $\omega_{v}$ results in a higher peak velocity, which is attained earlier. 
This can be explained by the corresponding decrease of the end-point %
velocity, which effectively reduces the total costs. %
This is also visible from the entries of the optimal feedback gain matrices $L_{n}$, which are depicted in Figure~\ref{fig:LQG_effects_0} as well. 
The magnitude of the position component\footnote{We recall that to compute the resulting control $u_{n}$, each entry of the matrix $L_n$ is multiplied by the corresponding entry of the state $x_{n}$, see~\eqref{eq:lqg-control-law}. 
In our case, $L_n$ is a $1\times 5$ matrix, which allows for easy interpretation of the matrix entries.
For example, the first entry of $L_n$ is multiplied by the first entry of $x_n$, which is the position, and hence referred to as the \emph{position component} of $L_n$.}
of $L_{n}$ determines the importance of correcting the error %
between end-effector and target at time step $n$, whereas the magnitude of the velocity component of $L_{n}$ determines the importance of adjusting the end-effector velocity.
For the LQG model, an increase in the velocity cost weight $\omega_{v}$ results in a later peak of the velocity component, i.e., the controller would correct for a large velocity that occurs very shortly before the end of the movement, in order to reduce the terminal velocity costs.
In the position component of the feedback gain matrices,  %
a second, earlier peak occurs for moderate values ($\omega_{v}=1$, green dash-dotted line).
This suggests that it is particularly important to the controller to eliminate deviations from the target both at the end of the surge phase (peak at 0.212s for the considered case)
and at the end of the movement (peak at 0.68s).
For large velocity cost weights ($\omega_{v}=20$, blue dotted line), the earlier peak becomes more dominant and remains the only peak, i.e., end-point errors that occur shortly before the end of the movement are not corrected anymore, as this would result in a larger end-point velocity and thus a higher total expected cost.
In addition, the (relative) length of the correction phase increases as the terminal velocity cost weight increases. 
It is important to note that these effects only occur under both observation noise and signal-dependent control noise, i.e., for $\sigma_u$, $\sigma_s>0$. %

Terminal force costs only, i.e., without terminal velocity costs, cannot account for the typically observed corrective movements following the surge (see Figure~\ref{fig:LQG_effects_0}, middle plots).
In particular, there is no incentive to reduce the velocity towards the end of the movement. Instead, the acceleration is reduced, which is visible from the constant velocity at the end of the movement for large $\omega_{f}$ (green dash-dotted and blue dotted lines).
Similarly to the velocity cost weight $\omega_{v}$, the force cost weight $\omega_{f}$ also affects the time period at which deviations from the desired terminal position and velocity are corrected. 
As $\omega_{f}$ increases, this period shifts forward, i.e., late-occurring deviations are not corrected anymore.

For positive velocity and force cost weights $\omega_{v}$ and $\omega_{f}$, an increase of the effort cost weight $\omega_{r}$ leads to a trajectory that reaches the target later, with lower peak velocity (see Figure~\ref{fig:LQG_effects_0}, bottom plots).
This is intuitive, since higher effort costs reduce the magnitude of the optimal control signals, resulting in lower accelerations and velocities.
The terminal velocity is close to zero for any moderate $\omega_{r}$. %

Note that the shape of the optimal control sequences differs considerably from %
the deterministic LQR model (see Figure~\ref{fig:LQR_effects}).
In the LQG model, the control usually attains its maximum at the beginning, then linearly decreases towards its minimum, and increases again towards zero.
This is mainly due to the velocity and the acceleration being penalized only at the final time step $N$, which allows to reach a higher peak velocity and acceleration (achieved through larger control signals at the beginning of the movement; %
also note the large (non-normalized) positional feedback gain values in bottom left plot).
Under the assumption of signal-dependent control and constant observation noise, the control is very close to zero during the correction phase (which, for ID $\geq4$, makes up a considerable part of the movement).
In Figure~\ref{fig:LQG_effects_0} (bottom right plot), this is shown for moderate control and observation noise levels $\sigma_{u}=0.2$, $\sigma_{s}=0.5$. %

If the control noise $\sigma_{u}$ is set large enough, similar effects can be observed without any observation noise, i.e., $\sigma_{s}=0$ (not shown). 
In contrast, large observation noise levels $\sigma_s$ cannot compensate a missing control noise.
This is intuitively plausible -- %
since the controller is aware that there is no control noise, the deterministic system states resulting from the closed-loop system become perfectly predictable through the internal (correct) forward model, i.e., there is no need to rely on noisy observations at all.

\begin{figure}
	\centering
	\includegraphics[width=\linewidth]{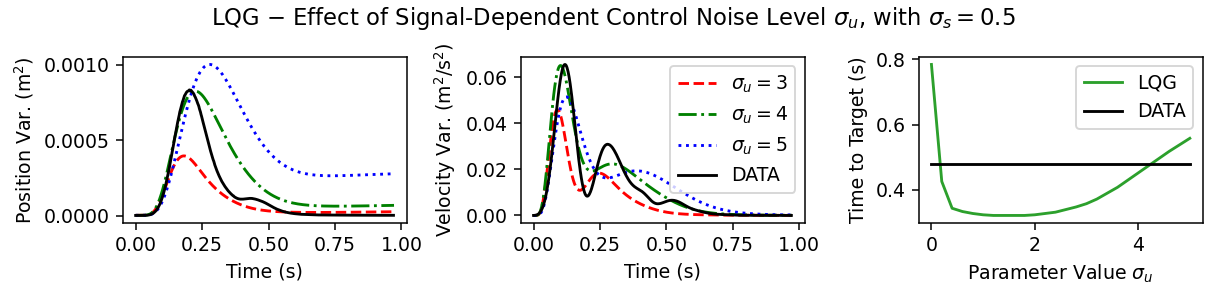}
	\\[0.1cm]
	\includegraphics[width=\linewidth]{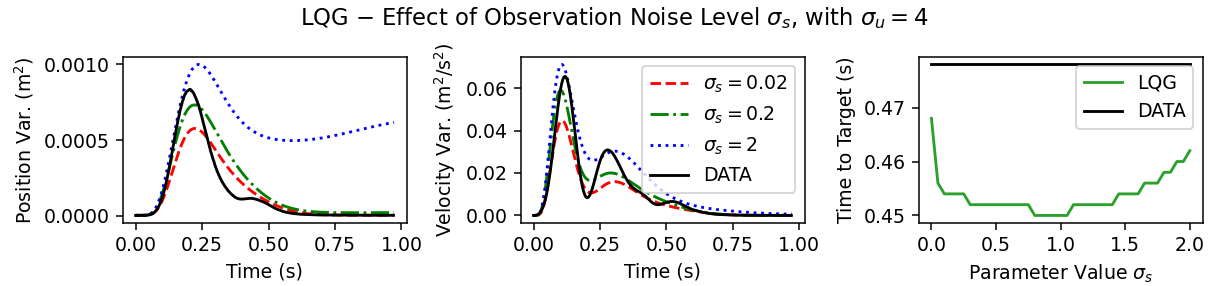}
	\caption{
		\textbf{Top:} Effect of the signal-dependent noise level $\sigma_u$ on the variance time series of the LQG position and velocity profiles (red dashed: $\sigma_u=3$, green dash-dotted: $\sigma_u=4$, blue dotted: $\sigma_u=5$) %
		and on the time until the target is reached; all simulations with observation noise level $\sigma_s=0.5$.	
		\textbf{Bottom:} Effect of the observation noise level $\sigma_{s}$ on the variance time series of the LQG position and velocity profiles (red dashed: $\sigma_s=0.02$, green dash-dotted: $\sigma_s=0.2$, blue dotted: $\sigma_s=2$) 
		and on the time until the target is reached; all simulations with signal-dependent noise level $\sigma_u=4$.
	}
	\label{fig:LQG_effects_1}
	\Description{Fully described in caption and text.}
\end{figure}

The effects of the control noise level $\sigma_{u}$ (with $\sigma_{s}=0.5$)
and the observation noise level $\sigma_{s}$ (with $\sigma_{u}=4$)
are shown in Figure~\ref{fig:LQG_effects_1}.
In the left and in the middle plots, the
variance of the resulting LQG position and velocity profiles is plotted against time.
The black solid lines show a representative variance profile from the Pointing Dynamics Dataset. %
Visually, the combination of signal-dependent control noise and constant observation noise can explain the characteristic variance profiles relatively well. %
Under signal-dependent noise, the application of smaller controls  in the second half of the movement (during the correction phase, which follows the surge) results in a decreasing movement variability towards the end of the movement.
This is in accordance with the two-phase positional variance profiles that are typically observed in aimed movements~\cite{Gutman92, Lai05, Gori20}.
From an information-theoretic perspective, the decaying rate of these profiles %
can be explained by a user-specific channel capacity~\cite{Gori20}.
In the considered LQG model, the idea of user-specific control and observation noise levels affecting the (expected) end-effector variability provides a different interpretation.
However, very large and physically implausible noise levels ($\sigma_u\approx1-4$) %
are required to account for the substantial variance observed in the position and velocity profiles of mouse pointing movements. %
The velocity variance profiles, 
which are typically bimodal~\cite{Gutman92, Darling88},
can also be replicated by %
the noise model of the LQG, 
albeit the second peak is usually less pronounced in the simulation. %

The effect of the signal-dependent noise level $\sigma_{u}$ on the (average) time until the target is reached is depicted in the top right plot of Figure~\ref{fig:LQG_effects_1}, 
with total movement duration $N$ corresponding to 0.97s.
Note that the time a target is reached does not have to coincide with the total movement duration. 
This is because the movement does not end upon reaching the target, but instead is the experimentally observed time between two mouse clicks.
To distinguish these two, we will abbreviate the average time until the target is reached by \textit{time to target} in the following.

Without signal-dependent noise, i.e., $\sigma_{u}=0$, there is little incentive to reach the target much earlier than at time step $N$, which is the only time step at which the positional error, velocity, and acceleration are penalized. 
As $\sigma_{u}$ increases, the movement duration rapidly decreases.
This can be again explained by the need to apply lower controls towards the end, in order to reduce the uncertainty regarding the final end-effector position.
Signal-dependent noise thus induces a strategy that could be described as ``doing most of the work at the beginning to be on the safe side''.
However, if $\sigma_{u}$ becomes too large, its effect on the movement time reverses: the more signal-dependent noise, the more time is required to reach the target.
This is intuitive, since higher control noise levels result in a larger positional variance of the end-effector, which in turn forces the controller to more rely on the received observations when updating the internal state estimate.
In combination with a positive observation noise level $\sigma_{s}$, an increase in $\sigma_{u}$ thus results in a larger uncertainty regarding the own position, which makes it difficult to reach the target at the same time. %
Such an increase of the time to target with $\sigma_{u}$ cannot be observed if observation noise is omitted (not shown).

\begin{figure}
	\centering
	\includegraphics[width=\linewidth]{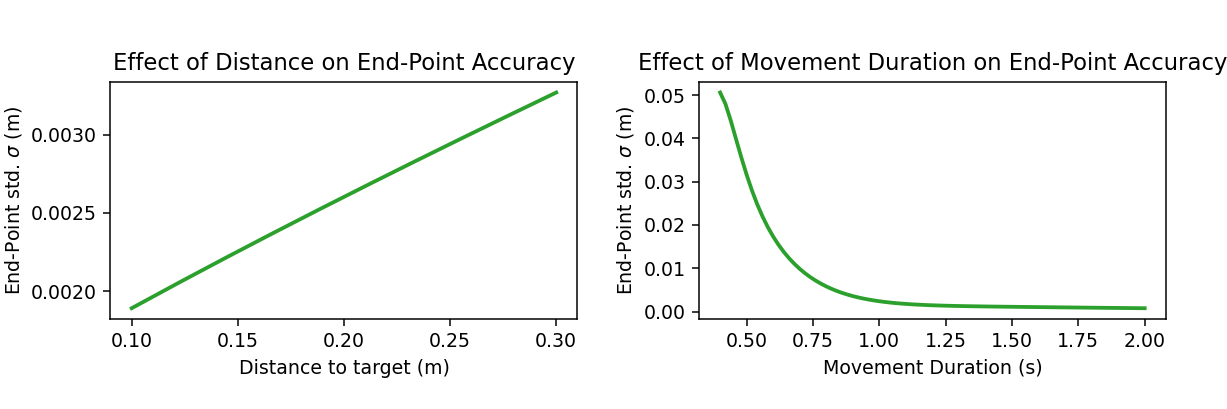}
	\caption{Under signal-dependent noise and given a fixed movement duration (here: 0.97s), the end-point standard deviation of the LQG linearly increases with distance. %
		Given a fixed distance (here: 0.21m), the end-point standard deviation inverse-quadratically decreases with movement duration. %
	}~\label{fig:E-LQG_speedaccuracy}
	\Description[LQG end-point standard deviation is affected by both distance and movement time]{The standard deviation of the LQG end-effector position plotted against the initial distance to target $D$ and plotted against the movement duration $N$.}
\end{figure}

The assumption of signal-dependent control noise is also in accordance with the well-known speed-accuracy trade-off, which suggests that faster movements %
result in a larger end-point variance~\cite{Woodworth99, Todorov02, HarrisWolpert98}. %
As can be seen in Figure~\ref{fig:E-LQG_speedaccuracy}~(left), for a fixed movement duration $N$, the positional end-point accuracy (i.e., the standard deviation of the terminal end-effector position) linearly increases with the initial distance to target.
This is plausible, as larger movements require higher controls $u_{n}$, %
and is consistent with empirical findings~\cite{Gordon94, Todorov98_thesis, Wright83, Todorov05}.
Similarly, an inverse relationship between movement duration and end-point standard deviation can be observed (see Figure~\ref{fig:E-LQG_speedaccuracy}~(right)).

For the observation noise level $\sigma_{s}$, similar effects on the variance profiles and the time to target can be observed, given a fixed control noise level $\sigma_{u}$ (bottom row of Figure~\ref{fig:LQG_effects_1}).
The increasing variance in position and velocity as $\sigma_{s}$ increases can also be explained by the increasing uncertainty regarding the own end-effector position as $\sigma_{s}$ increases, as this uncertainty in turn affects the variance %
of the applied muscle control.
For moderate observation noise levels $\sigma_{s}$, this effect decreases towards %
the end of the movement, since smaller controls are applied then, i.e., the future system states become less dependent on the internal state estimates.
However, if $\sigma_{s}$ becomes too large, the internal state estimates are not precise enough to reliably move the end-effector towards the target. In this case, the positional variance remains constant or even increases towards the end of the movement (blue dotted line in the bottom left of Figure~\ref{fig:LQG_effects_1}).
Note, however, that \textit{on average}, the target is still reached early during the simulation (after 0.46s for $\sigma_{s}=2$, which is even slightly faster than the representative user trajectory for the considered task shown as black line in the bottom right plot of Figure~\ref{fig:LQG_effects_1}).

In summary, the mutual dependency between observations and applied controls, i.e., between the Kalman gains $K_{n}$ and the feedback gains $L_{n}$, implies a positive effect of both signal-dependent control noise and constant observation noise on the movement variability, whereas the effect on the time to target depends on the absolute value of both noise levels.
However, the effects of $\sigma_{s}$ are considerably smaller than the effects of $\sigma_{u}$ (note the logarithmic scale of $\sigma_{s}$ for both variance profiles, and the smaller linear scale in the time to target).
In particular, observation noise only has an effect \textit{in combination with control noise}, because otherwise the system is deterministic, i.e., the controller does not need to rely on sensory input at all.

\subsection{Results of Parameter Fitting}

Using our parameter fitting process, we identify the optimal values of all five parameters for each combination of participant, task condition, and direction within the Pointing Dynamics Dataset. 
Since the LQG model yields a sequence of state \textit{distributions}, 
we use the stochastic parameter fitting variant described in Section~\ref{sec:param-fitting}, with $2-$Wasserstein distance applied to the position-velocity components of the respective state distributions as the loss function.

The parameters of the initial distribution, $\bar{x}_{0}$ and $\Sigma_0$, from~\eqref{eq:discrete-control_4} are specified as follows.
We choose
$\bar{x}_{0}=(p_{0}^{\text{USER}}, v_{0}^{\text{USER}}, 0, 0, T)^\top$, where $p_{0}^{\text{USER}}$ and $v_{0}^{\text{USER}}$ denote the average initial position and velocity, respectively, of all trials for the considered combination of participant, task condition, and direction.
The covariance $\Sigma_{0}$ of the initial state is defined as follows.
The components for position and velocity correspond to the sample covariance matrix empirically observed from these user trajectories, as described in Section~\ref{sec:dataset}. All other components are set to zero. 

\subsubsection{Qualitative Results}

\begin{figure}
	\centering
	\includegraphics[width=\linewidth]{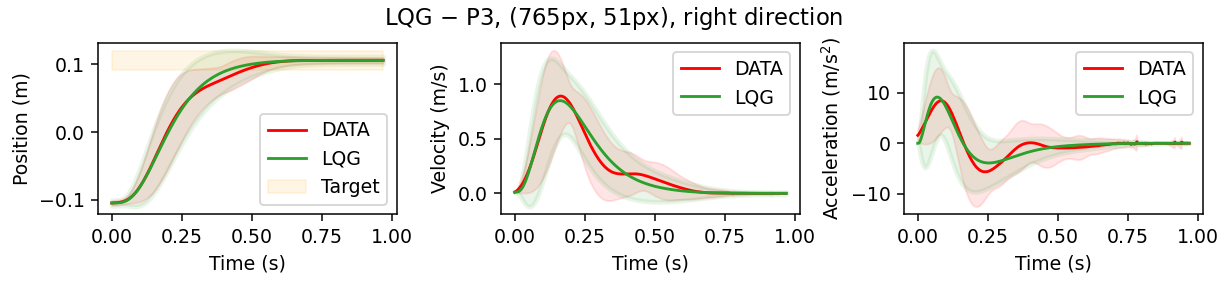}
	\caption{Visually, LQG trajectories show a good fit to typical user trajectories. The mean trajectories (solid lines) as well as the $95\%$ point-wise confidence bands match very well at first glance. Some simulation movements exhibit a slightly too low velocity at the beginning of the movement.
	The LQG acceleration profiles are obtained by applying a Savitzky-Golay filter of degree 3 and frame size 15 to the respective velocity profile and differentiating the corresponding polynomials.
	}
	\label{fig:LQG_data_ID4}%
	\Description[Position, velocity, and acceleration time series of the LQG vs. user data (ID 4 task)]{Position, velocity, and acceleration time series of both the Pointing Dynamics Dataset (Participant 3, ID 4 (765px distance, 51px width), right direction) and the corresponding LQG simulation trajectory, in terms of mean values and $95\%$ point-wise confidence intervals.}
\end{figure}

The optimal LQG solution for the same representative ID 4 task as used for the previous models is shown in Figure~\ref{fig:LQG_data_ID4}, where both the mean trajectories and the $95\%$ point-wise confidence bands are plotted. The mean position and velocity profiles (green solid lines) visually match those from user data (red solid lines) very well.
The same holds for the acceleration time series, apart from the second submovement (starting around 0.3s), which is slightly less pronounced in the simulation.
The most notable differences are that in the LQG model, there is a slightly larger positional variance at the transition from the ballistic to the corrective phase of the movement (see also Figure~\ref{fig:MODELS_data_ID4}), and a larger variance in velocity and acceleration at the beginning of the movement. %

\subsubsection{Quantitative Results}

\begin{figure}
	\centering
	\subfloat{\includegraphics[width=0.33\linewidth]{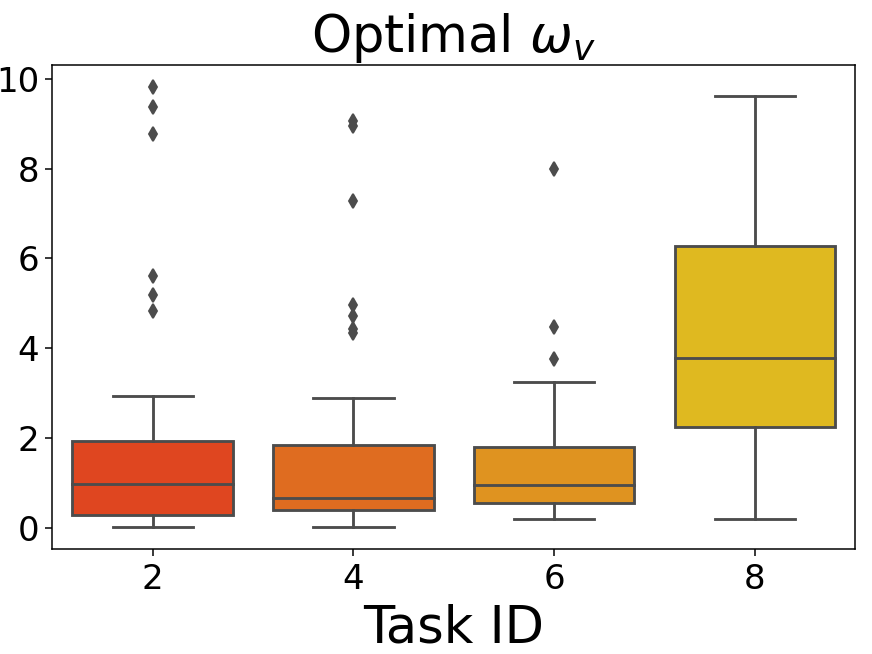}}
	\subfloat{\includegraphics[width=0.33\linewidth]{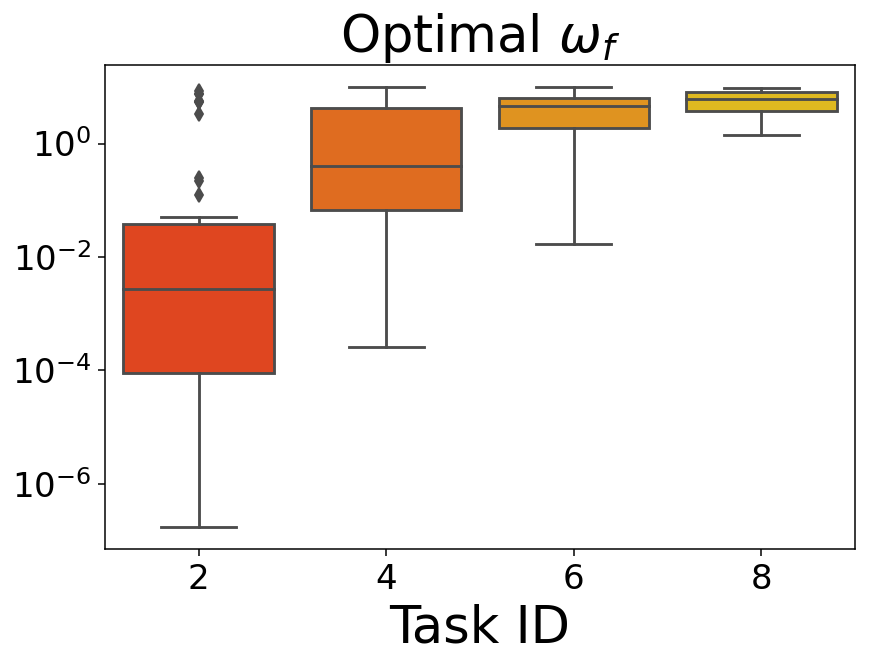}}
	\subfloat{\includegraphics[width=0.33\linewidth]{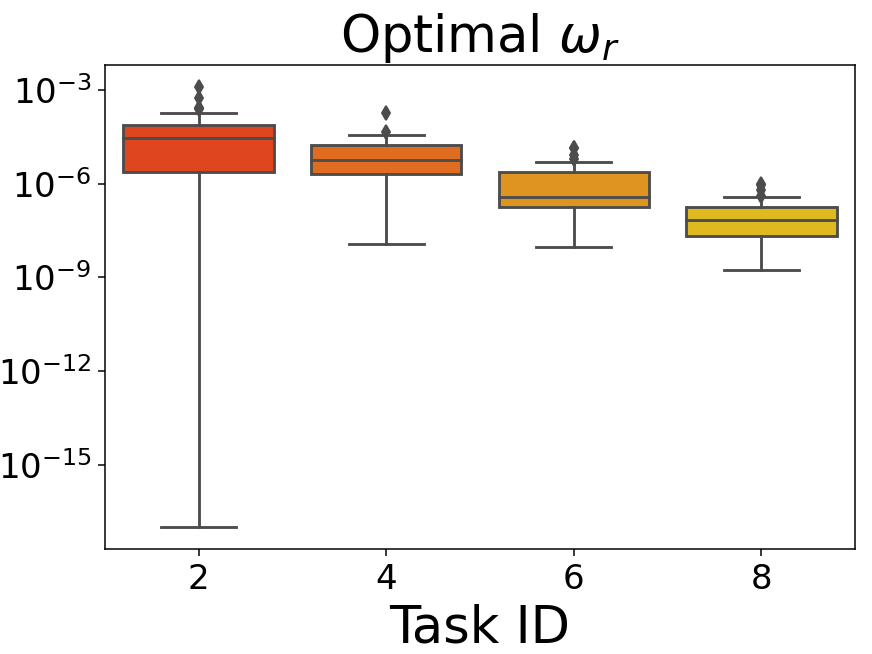}}
	\\
	\subfloat{\includegraphics[width=0.24\linewidth]{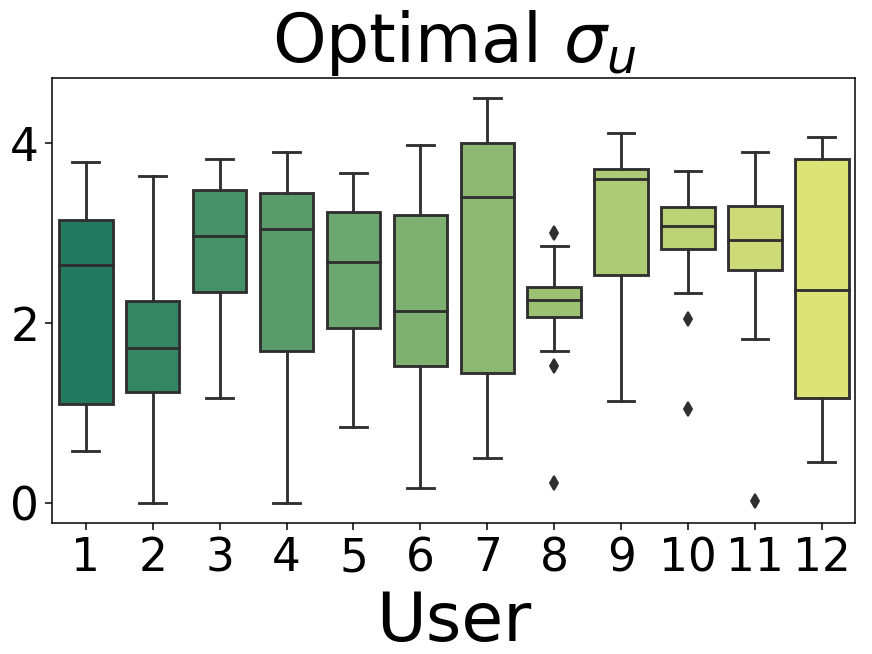}}
	\hfill
	\subfloat{\includegraphics[width=0.24\linewidth]{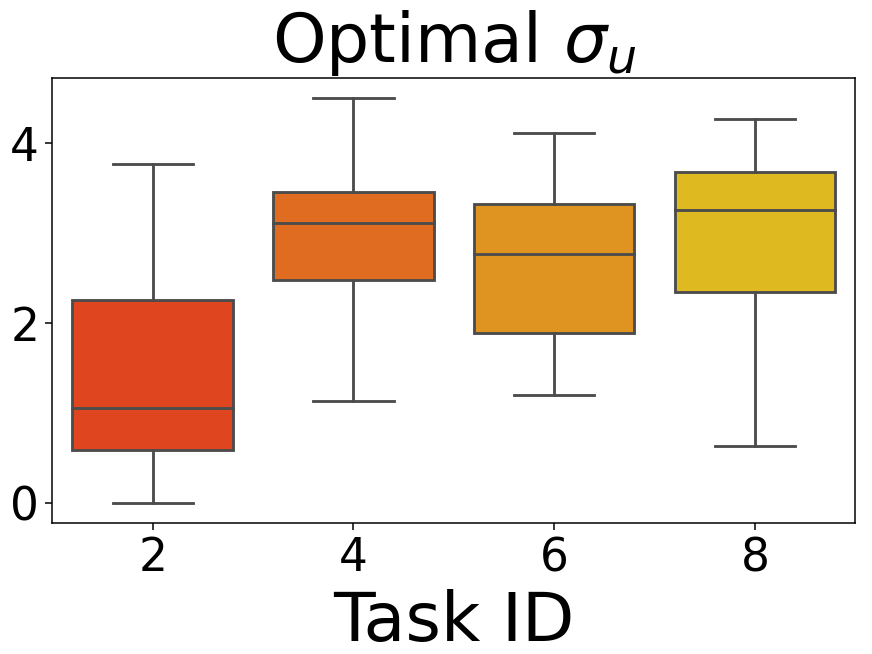}}
	\hfill
	\subfloat{\includegraphics[width=0.24\linewidth]{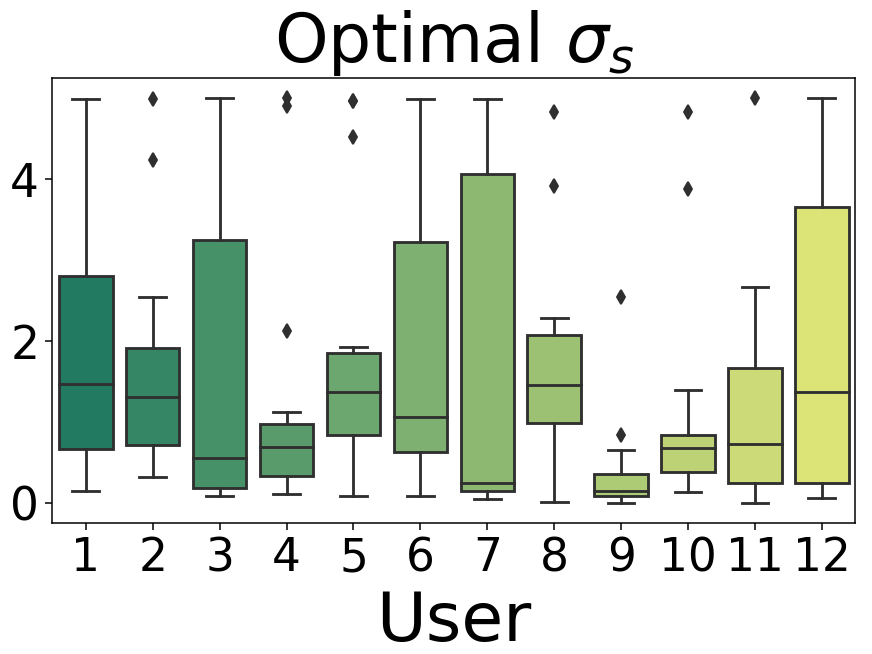}}
	\hfill
	\subfloat{\includegraphics[width=0.24\linewidth]{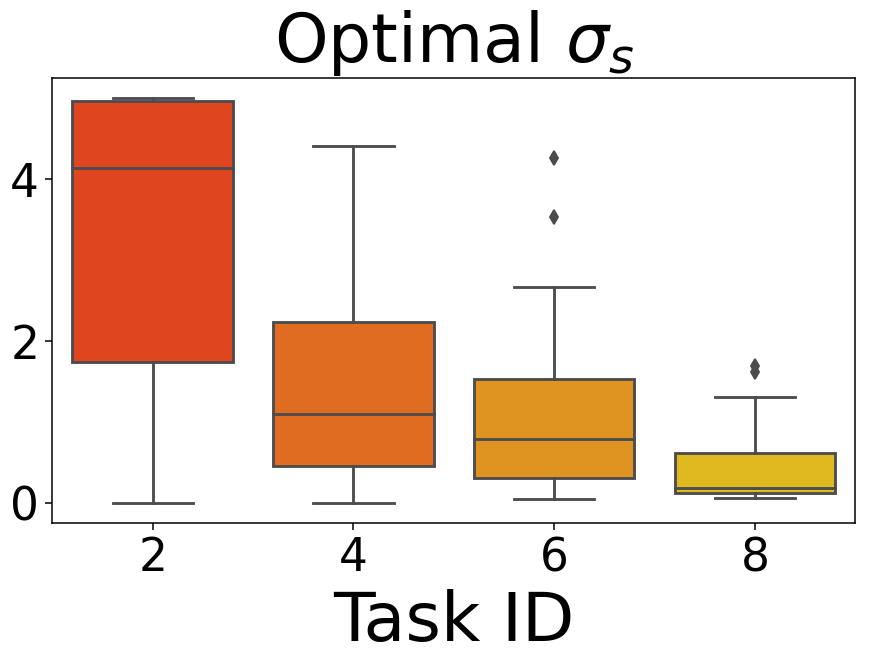}}
	\caption{Parameters of the LQG model, optimized for the trajectory sets of all participants, tasks, and directions, grouped by participants (left) and by ID (right).
	Note that the optimal values of $\omega_{f}$ and $\omega_{r}$ are plotted on a logarithmic scale. %
	}~\label{fig:LQG_opt}
	\Description{Fully described in caption and text.}
\end{figure} 

In Figure~\ref{fig:LQG_opt}, optimal parameter values of the LQG are shown, grouped by both user and task ID. For better visibility, the values of $\omega_{f}$ and $\omega_{r}$ are plotted on a logarithmic scale.

In contrast to the LQR model, the optimal cost weights are more affected by the task ID than by the individual participants.
Since in the LQG model, distance, velocity, and force costs are only applied in the final time step $N$, the corresponding weights can be interpreted as importance of the endpoint accuracy constraint relative to keeping the required effort low.

The force cost weight $\omega_{f}$ monotonously increases with task ID, whereas the velocity cost weight $\omega_{v}$ takes similar values for ID 2, 4, and 6 tasks, and is considerably larger for ID 8 tasks.
Since the velocity cost weight $\omega_{v}$ mainly affects the relative time spent in the surge phase (see Figure~\ref{fig:LQG_effects_0}, top plots), the latter finding suggests that movements for very difficult tasks (ID 8) exhibit a considerably longer correction phase.
Similarly, the force cost weight $\omega_{f}$ determines the time period at which positional errors and large velocities are corrected (see Figure~\ref{fig:LQG_effects_0}, middle plots).
The monotonous increase of $\omega_{f}$ with task ID thus implies that the more difficult the task, the less attention is paid to deviations near the end of the movement.

The effort cost weight $\omega_{r}$ exponentially decreases as the task ID increases. %
This is not surprising, since the controller does not have explicit knowledge on the target width $W$, but only on the distance to target $D$ (via the initial end-effector and the target position, which are both included in~$\bar{x}_{0}$). Instead, the desired increase of end-point accuracy as $W$ decreases needs to be implemented via the cost weights.
The observed decrease of the effort cost weight $\omega_{r}$ as the task becomes more difficult, which is equivalent to a relative increase of the terminal costs, can thus be interpreted as a higher importance of keeping the end-effector inside the target (higher accuracy) with small velocity and force at the final time step $N$ (higher stability).

The task ID also has an effect on the two noise level parameters $\sigma_{u}$ and $\sigma_{s}$. The signal-dependent control noise level $\sigma_{u}$ takes considerably lower values for ID 2 tasks, while the observation noise level $\sigma_{s}$ decreases as the task becomes more difficult. %

While in contrast to the LQR model, the effect of the participants on the optimal cost weights is less pronounced %
(see Figure~\ref{fig:LQG_opt_2} in the Appendix), different users are clearly characterized by different noise levels. %
For example, the trajectories of participant 8 can be explained by a small control noise level $\sigma_{u}$ (and a rather large observation noise level $\sigma_{s}$), whereas participant 9 is best explained by a larger control noise level $\sigma_{u}$ (and a small observation noise level $\sigma_{s}$).

\subsection{Discussion}

The LQG model assumes that users behave optimally with respect to a combination of terminal distance, velocity, and force costs, as well as continuous effort costs, within the constraints imposed by the human-computer system dynamics, and subject to signal-dependent motor noise and constant observation noise. %
Using the presented stochastic parameter fitting, this allows for an excellent replication of user trajectories.

As shown in Figure~\ref{fig:LQG_data_ID4}, trajectories resulting from the presented parameter fitting process capture both the average behavior and between-trial variability that is typically observed in mouse pointing movements.

However, the observation model from~\cite{Todorov05} has some shortcomings, as it assumes that all relevant quantities are perceived instantaneously in global coordinates and perturbed by additive Gaussian noise only. Moreover, the target position is assumed to be perfectly known during the entire movement.
Thus, online comparisons between the target signals obtained from an appropriate observation model (i.e., $H_{n}$ such that $H_{n}x_{n}$ includes $T$, see Equation~\eqref{eq:lqg-observation-space}) and those predicted by the internal model would not yield any additional benefit. %
This, however, is in contradiction to many empirical observations, which suggest that visual stimuli are internally used whenever available, even in the considered case of serial movements between the same two targets~\cite{Sarlegna15, Todorov98_thesis}. %

In the following section we thus extend the LQG model by considering both fixation-centered and world-centered sensory input signals, %
inspired by~\cite{Todorov98_thesis}. %

\section{Pointing as Optimal Feedback Control Subject to Signal-Dependent Noise and Saccades: The E-LQG}\label{sec:SOFC-2}

The observation model of the LQG model in Section~\ref{sec:SOFC}, which was taken from~\cite{Todorov05}, has some major drawbacks.
In particular, it assumes that the position, velocity, and force of the end-effector can be directly observed in world-centered coordinates, only perturbed by additive Gaussian noise whose magnitude is also known to the controller.

In the following, we will thus present an extension of this LQG model, denoted by \textit{E-LQG}, which includes a more complex and physiologically plausible human observation model. %
The main concepts are taken from~\cite{Todorov98_thesis}, with an adaption to the considered case of mouse pointing.
In contrast to~\cite{Todorov98_thesis}, we assume that the end-effector position cannot be observed via proprioception (i.e., in world-centered coordinates), %
but only from visual input (i.e., relative to the current eye fixation position), 
since it corresponds to the mouse cursor position in our case.

Compared to the LQG model from Section~\ref{sec:SOFC}, the E-LQG model
\begin{itemize}
	\item models eye movement based on accurate saccades between the initial and the target position,
	\item distinguishes between visual input (in fixation-centered coordinates) and observations of the eye fixation position (in world-centered coordinates), %
	and
	\item works with an imperfect initial target estimate, which is updated during the movement based on sensory input.
\end{itemize}
In the following, these differences are described in more detail.

\subsubsection*{Eye Saccades}

For the considered goal-directed movements, sensory input %
can be assumed to be based on \textit{eye saccades}, i.e., fast movements of the fovea between two fixation points~\cite{Kowler11, Krauzlis17, Zelinsky08}. %
The number and choice of fixation points usually depends on the complexity of the observed scene, the underlying goal (i.e., which information should be extracted from visually input), and the salience of individual objects, among others~\cite{Schuetz11, Tatler11, Gegenfurtner16}.
However, previous experiments on via-point tasks suggest that a single and precise movement of the fovea towards the aimed target is sufficient for the considered case of reciprocal pointing towards clearly delimited target areas~\cite{Todorov98_thesis}.
Following~\cite{Todorov98_thesis}, we thus can assume that at the beginning of the movement, the eye fixation corresponds to the initial position (regarding the repetitive movements from the Pointing Dynamics Dataset, this is equivalent to the target position of the previous movement). 
At a certain time during the arm movement, the gaze is assumed to move towards the target, which is then fixated until the end of the trial.

However, the eye saccades are decoupled from the rest of the movement in the sense that the controller can neither modify the time of the saccade nor the fixation points.
Instead, we assume that both fixations are accurate (which can be argued by combining ``possible corrective saccades in one `saccade' moving the eyes from one target to another'' \cite{Todorov98_thesis}), and optimize the saccade time within the outer parameter fitting process.\footnote{In the future, it will be interesting to include the eye position in state space and make it controllable via some (simplified) eye dynamics, similar to the simplified muscle dynamics that are used to control the end-effector.}

\subsubsection*{Visual Input}

We assume that visual input signals yield information regarding the position of the end-effector (i.e., the mouse pointer), the target, and the initial position, each \textit{relative to the eye fixation position} (i.e., in fixation-centered coordinates).\footnote{Note that depending on whether the saccade has already taken place, the relative initial position (before the saccade) or the relative target position (after the saccade) equals zero and thus can be discarded from the observation space.}

Based on the principles of foveal and peripheral vision~\cite{Strasburger11, Li10}, these observations are assumed to be disturbed by noise that linearly increases in the distance between the respective object (pointer, initial/target box) and the eye fixation point. %
This is a major difference to the LQG model from Section~\ref{sec:LQG}, which included additive observation noise only.
In addition, both the end-effector velocity and acceleration (which corresponds to force due to the assumption of unit mass) are perceived from visual input channels with additive noise, i.e., the magnitude of the observation noise is assumed independent of the distance to the eye fixation. %
The rationale for using additive noise here is that the minimum detectable difference between velocities is known to hardly differ between the peripheral and the foveal field. %
This particularly implies that useful observations of the end-effector velocity can be obtained independent of whether the end-effector moves close to the fixation point~\cite{Mckee84}. %

\subsubsection*{Proprioceptive and Eye Fixation Input}

Proprioceptive signals are signals that refer to the own body position and movement.
While the visual input channel yields fixation-centered observations of the end-effector and the target position,  %
proprioception could be used to obtain world-centered estimates of the own body position and orientation, e.g., of the arm, hand, or head. %
The used Human-Computer System Dynamics, however, do not explicitly distinguish between quantities corresponding to the human body and quantities corresponding to the input device or interface. 
Instead, it gives %
the overall dynamics that are directly applied to the virtual end-effector (see the discussion in Section~\ref{sec:human-computer-system-dynamics}).
In the considered case of mouse pointing, the end-effector corresponds to the mouse cursor shown on the display, which cannot be perceived proprioceptively.
In contrast to~\cite{Todorov98_thesis}, we thus do not include world-centered observations of the end-effector position in the proposed model.

Besides feedback on the end-effector position and orientation, the human body usually also provides information about the eye position.
In recent years, a large debate has evolved about whether the cortical eye position is rather obtained via proprioception or using internal efference copies of ``outflow'' signals~\cite{Wang07, Medendorp11}.
For details on the neurophysiological mechanisms underlying coordinated eye-hand movements, we refer the interested reader to the excellent overview given in~\cite{Shadmehr05}. 
Regarding the observation model used for the E-LQG model, we assume that the eye fixation point (perturbed by additive noise) can be perceived in world-centered coordinates.
Note that the eye fixation dynamics are not part of the Human-Computer System Dynamics, but are implicitly modeled via the following workaround. 
The two attainable values are included in the state, and the time of the instantaneous switch is determined via some parameter (more details are given in Section~\ref{sec:E-LQG}).

In summary, the eye fixation is assumed to take only two different values during an aimed movement: the initial position (before the saccade) and the target position (after the saccade). 
In particular, the (perturbed) target position can be observed in world-centered coordinates, as soon as the saccade has taken place.

\subsubsection*{Internal Target Estimate}
We assume that at the beginning of the movement, the controller is not aware of the exact target position.
This is intuitively plausible, since even in the considered case of reciprocal tasks, where the users know that the two same targets will appear alternately, visual input signals are known to be used to improve the (rough) prior target estimates during the movement~\cite{Sarlegna15, Todorov98_thesis}.

In both the LQG and the E-LQG models, the internal state estimates $\hat{x}_{n}$ include an estimate of the desired target.
However, the controller in the LQG model is given an \textit{exact} initial estimate $\hat{x}_{0}$, i.e., $\hat{x}_{0}$ includes the correct (mean) initial position, velocity, force, muscle excitation, and target position.
Since the target is known to be constant during the movement, 
this immediately implies a correct target estimation during the complete movement.
In contrast, in the E-LQG model, we assume that the target component of the initial estimate %
differs from the actual target position $T$.
For simplicity, we assume that this initial target estimate %
corresponds to the \textit{initial position} $T_{0}$ (see Section~\ref{sec:E-LQG}), that is, the center of the target box of the previous movement in the Pointing Dynamics Dataset; note that both initial and target box were permanently displayed in the experiment~\cite{Mueller17}. %
This is in agreement with the reciprocal nature of the movements from the Pointing Dynamics Dataset, where the initial position should in turn equal the target position of the preceding movement.
During the mouse movement, the internal target estimate, i.e., the target component of $\hat{x}_{n}$, %
is then updated based on the perceived sensory input signals $y_{n}$.

\subsection{LQG with Extended Observation Model (E-LQG)}\label{sec:E-LQG}

Based on the above assumptions, %
we modify the LQG model presented in Section~\ref{sec:SOFC} as follows. 

In order to model eye fixation of the initial position $T_{0}$, we first need to include $T_{0}$ in the state space. 
We thus define $x_{n} = (p_n, v_n, f_n, g_n, T_0, T)^{\top}\in\R^{6}$, i.e., the state vector now consists of the end-effector position, velocity, force, muscle excitation, the (fixed) initial position, and the (fixed) target position.
Moreover, we introduce the \textit{saccade time step} $n_{s}$, which defines the time at which the eye fixation switches from initial to target position. In order to be able to optimize this saccade time step within the stochastic parameter fitting process from Section~\ref{sec:param-fitting}, we relax this parameter by allowing continuous values, i.e.,  $n_{s}\in\left[0,N\right]$.

The observations $y_{n}$ are more complex than in the LQG model, see~\eqref{eq:lqg-observation-dynamics}. 
They are given by the following observation model:\footnote{Note that the generality and flexibility of the proposed framework in principle allows to incorporate multiple and imprecise saccades, as well as more sophisticated approaches to model visual input~\cite{Enderle10}. However, this is beyond the scope of this work.}
\begin{equation}\label{eq:e-lqg-observation-dynamics}
	y_{n} = H_{n}x_{n} + G_{n}(x_{n})\xi_{n},
\end{equation}
where the observation matrix $H_{n}$ depends on whether the saccade has taken place or not. 
Before the saccade has taken place, the unperturbed observations $H_{n}x_{n}$ include the end-effector velocity, the end-effector acceleration (corresponding to muscle activation and force), the initial position (i.e., the center position of the initial boundary box, which corresponds to the eye fixation position), as well as the end-effector and target position, both relative to the the initial position.
After the saccade has taken place, the unperturbed observations $H_{n}x_{n}$ include velocity, acceleration, the target position (which now corresponds to the eye fixation position), as well as the end-effector and initial position, both relative to the the target position.
At saccade time step $n=\lfloor n_{s}\rfloor$, we use a convex combination of the two observations.\footnote{This relaxation ensures that the gradient of the state sequence $x$ with respect to $n_{s}$ becomes non-zero, which is necessary to apply standard continuous optimization methods within the parameter fitting process.
}
To this end, we formally define
the two matrices,
\begin{equation}
	H^{(T_{0})} =
	\begin{pmatrix}
		0	& 1	& 0 & 0 & 0 & 0 \\
		0	& 0	& 1 & 0 & 0 & 0 \\
		0	& 0	& 0 & 0 & 1 & 0 \\
		1	& 0	& 0 & 0 & -1 & 0 \\
		0	& 0	& 0 & 0 & -1 & 1
	\end{pmatrix}, \quad\text{and}\quad
	H^{(T)} =
	\begin{pmatrix}
		0	& 1	& 0 & 0 & 0 & 0 \\
		0	& 0	& 1 & 0 & 0 & 0 \\
		0	& 0	& 0 & 0 & 0 & 1 \\
		1	& 0	& 0 & 0 & 0 & -1 \\
		0	& 0	& 0 & 0 & 1 & -1
	\end{pmatrix}, 
\end{equation}
and introduce $\{n_{s}\}$, which denotes the fraction part of $n_{s}$, i.e., $\{n_{s}\} = n_{s} - \lfloor n_{s}\rfloor$.
Then we can define the observation matrix $H_{n}$ as:
\begin{equation}
	H_{n} = 
	\begin{cases}
		H^{(T_{0})}, &\text{ if } n < \lfloor n_{s}\rfloor, \\
		H^{(T)}, &\text{ if } n > \lfloor n_{s}\rfloor, \\
		\{n_{s}\} H^{(T_{0})} + (1 - \{n_{s}\}) H^{(T)}, &\text{ if } n = \lfloor n_{s}\rfloor.
	\end{cases}
\end{equation}

The signal-dependent observation noise is introduced by the second term on the right-hand side of~\eqref{eq:e-lqg-observation-dynamics}.
Here, the vector~$\xi_{n}$ is a five-dimensional Gaussian random variable, i.e., $\xi_{n}\sim \mathcal{N}(0;I_{5})$, where $I_5$ denotes the $5\times 5$ identity matrix.
The observation noise matrix $G_{n}$ is defined by
\begin{equation}\label{eq:e-lqg-observation-noise}
	G_{n}(x_{n}) = \begin{cases} 
	diag(\sigma_{v}, \sigma_{f}, \sigma_{e}, \gamma\vert p_{n} - T_{0}\vert, \gamma\vert T - T_{0}\vert), &\text{if } n < \lfloor n_{s}\rfloor \\
	\begin{aligned}[c]
	diag(\sigma_{v}, \sigma_{f}, \sigma_{e}, \gamma\left(\{n_{s}\}\left(\vert p_{n} - T_{0}\vert\right) + (1-\{n_{s}\})\left(\vert p_{n} - T\vert\right)\right), \\ \gamma\left(\{n_{s}\}\left(\vert T - T_{0}\vert\right) + (1-\{n_{s}\})\left(\vert T_{0} - T\vert\right)\right)),
	\end{aligned} &\text{if } n = \lfloor n_{s}\rfloor \\
	diag(\sigma_{v}, \sigma_{f}, \sigma_{e}, \gamma\vert p_{n} - T\vert, \gamma\vert T_{0} - T\vert), &\text{if } n > \lfloor n_{s}\rfloor\\
	\end{cases}. %
\end{equation}
The end-effector velocity and force are perturbed by visual noise levels $\sigma_{v}$ and $\sigma_{f}$, respectively.
Similarly, the eye fixation position is perturbed by the gaze noise level $\sigma_{e}$.
The magnitude of the visual position observations depends on the respective distance to the eye fixation point, scaled by the parameter~$\gamma$.
This is consistent with Weber's Law~\cite{Cooper12}, which claims that the minimum required stimulus changes that lead to a considerable change in the visual perception (that is, changes that are larger than the perceptual noise) are linear in the absolute value of the respective signal, suggesting that the perceptual noise linearly depends on this absolute value as well~\cite{Li18}.

\subsection{Analysis of Parameters}

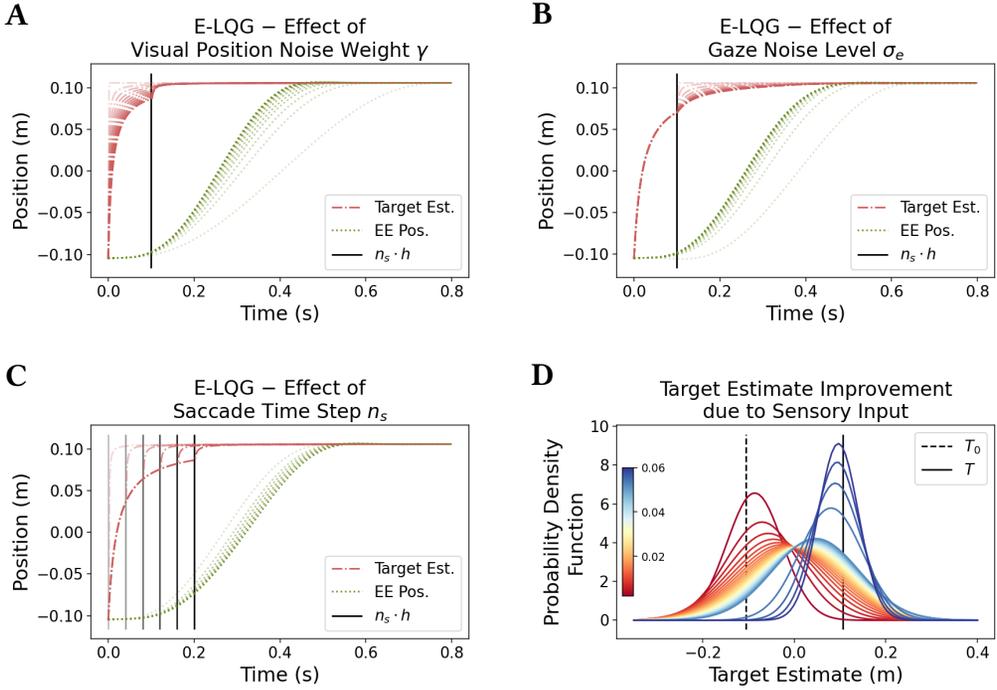
\begin{figure}[!ht]
	\centering
	
	\begin{tikzpicture}
		\node (img1) at (0,0) {\includegraphics[width=.45\linewidth]{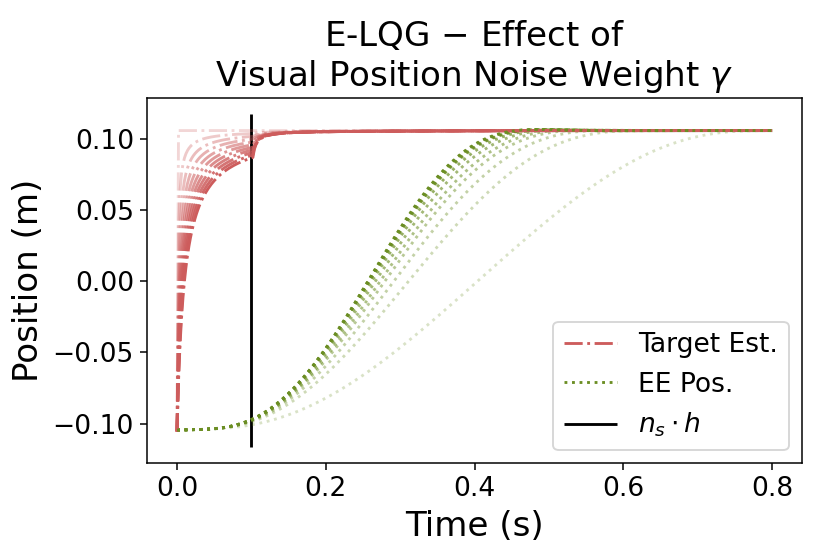}};
		\node[xshift=-3cm, yshift=2cm] at (img1) {{\Large\textbf{A}}};
		\node (img2) at (7,0)
		{\includegraphics[width=.45\linewidth]{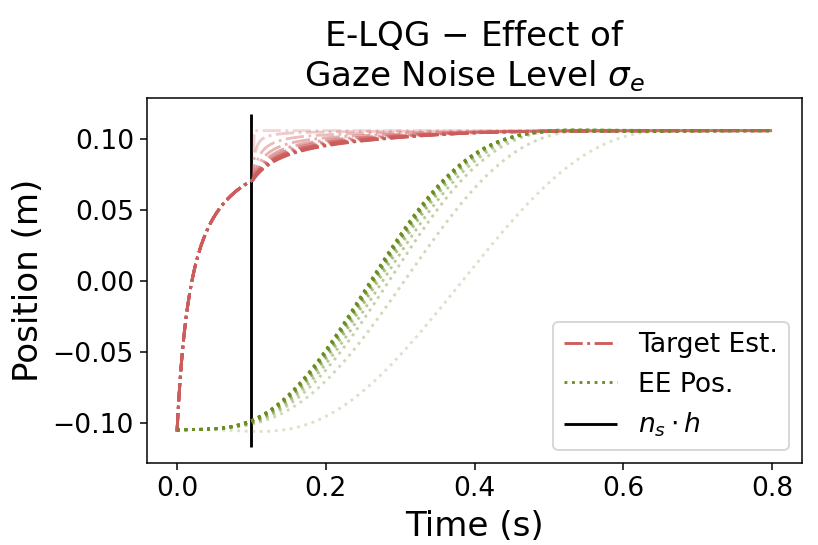}};
		\node[xshift=-3cm, yshift=2cm] at (img2) {{\Large\textbf{B}}};
		\node (img3) at (0,-4.75)
		{\includegraphics[width=.45\linewidth]{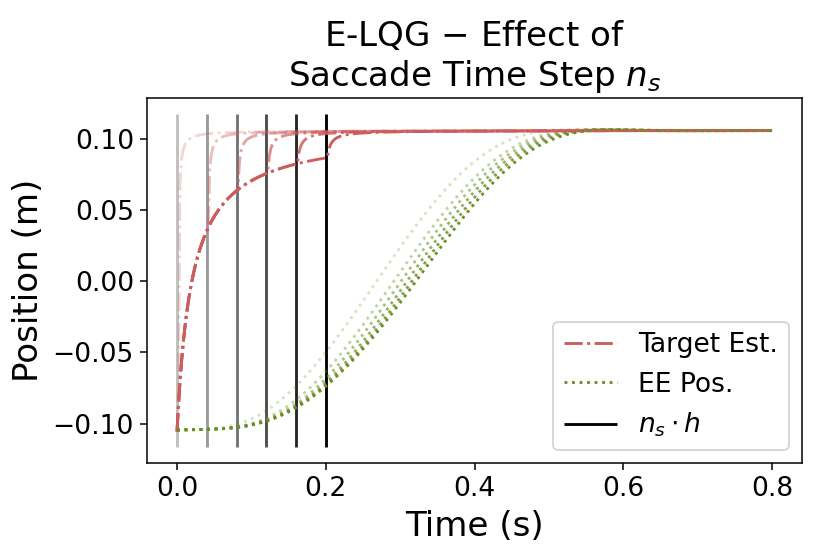}};
		\node[xshift=-3cm, yshift=2cm] at (img3) {{\Large\textbf{C}}};
		\node (img4) at (7,-4.75)
		{\includegraphics[width=.45\linewidth]{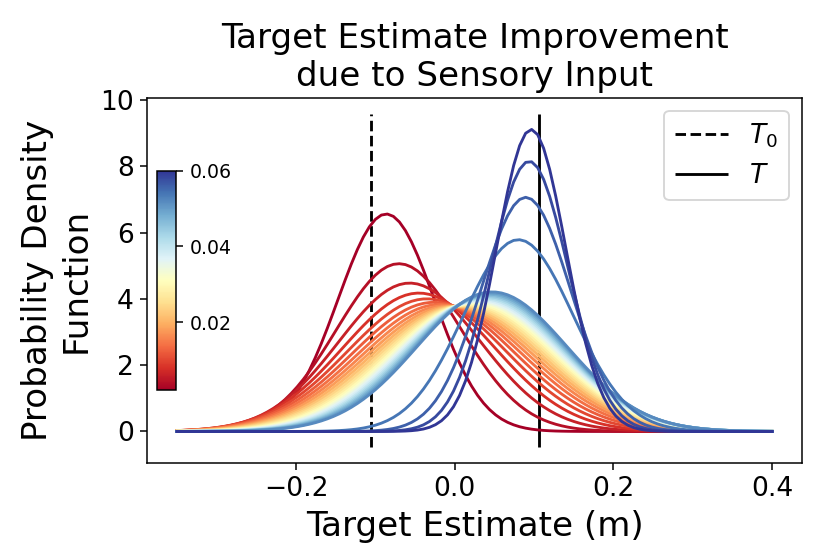}};
		\node[xshift=-3cm, yshift=2cm] at (img4) {{\Large\textbf{D}}};
	\end{tikzpicture}
	\caption{
		Internal estimates of the target position in the E-LQG model, as well the resulting (expected) position profile of the end-effector. In \textbf{A-C}, darker lines correspond to larger parameter values. 
		We set $\omega_{v}=2$, $\omega_{f}=0.02$, $\omega_{r}=~$1e-7, $\sigma_{v}=5$, $\sigma_{f}=1$,
		$\sigma_u=\sigma_c=0$, $h=0.002$,
		and performed at least 3 iterations to compute the Kalman and feedback gain matrices $K_n$ and $L_n$, see~\eqref{eq:lqg-iterations}.
		\textbf{A:} Effect of position perception noise weight $\gamma$ (between $0$ and $5$) with fixed gaze noise level $\sigma_e=0.1$ and saccade time step $n_{s}=50$, i.e., at 0.1s.
		\textbf{B:} Effect of gaze noise level $\sigma_e$ (between 1e-10 and $0.1$) with fixed position perception noise weight $\gamma=10$ and saccade time step $n_{s}=50$.
		\textbf{C:} Effect of saccade time step $n_{s}$ (between $0$ and $100$) with fixed position perception noise weight $\gamma=10$ and gaze noise level $\sigma_e=0.1$.
		\textbf{D:} Development of the internal target estimate probability density function over time, with position perception noise weight $\gamma=10$, gaze noise level $\sigma_e=0.1$, and the saccade occurring after 0.05s.
	}
	\label{fig:E-LQG_effects}
	\Description{Fully described in caption and text.}
\end{figure}

The effect of the parameters $\gamma$, $\sigma_{e}$, and $n_{s}$ on both the internal target estimate and the resulting (expected) end-effector position time series is shown in Figure~\ref{fig:E-LQG_effects} A-C, with darker lines corresponding to larger parameter values. 
The parameter $\gamma$ denotes the scaling weight of the Gaussian noise term added to the visually observed %
positions, which is multiplied by the respective distance to the eye fixation point. %
The constant magnitude of the Gaussian noise added to the eye fixation position is denoted by $\sigma_{e}$, and $n_{s}$ denotes the time step at which the saccade occurs. %

Starting with an eye fixation of the initial position (which is known at the beginning of the movement and thus correctly estimated), a higher position perception noise weight $\gamma$ leads to a slower update of the internal target estimate from initial position to true target position (red dash-dotted lines in plot A). Interestingly, this does not delay the end-effector movement, but rather results in a faster movement towards the target %
(green dotted lines in plot A).
A possible explanation for this phenomenon could be the that moving the end-effector early after the saccade towards the internal target estimate can improve the internal estimate of the end-effector position, since the eyes fixate the target center after the saccade. 
This means that the variance of the end-effector position observations linearly decreases as the distance between end-effector and target decreases.
Since the terminal costs create an incentive to keep the end-effector at the target center with zero velocity and acceleration in the final time step $N$, it is thus important to obtain reliable estimates of both the target position (which is quite accurate after the saccade has taken place, see next paragraph) and the own end-effector position early in time, to avoid expensive last millisecond corrections.
A large scaling parameter $\gamma$ intensifies this problem, as a smaller distance between end-effector and target is necessary to achieve the same amount of visual observation noise.
Thus, a larger position perception noise weight $\gamma$ results in an earlier movement towards the target.
However, this effect only holds for moderate values ($\gamma\leq5$); if the visual observation noise $\gamma$ becomes too large, more time is required to obtain a reliable internal target estimate, resulting in a more tentative (i.e., slower) movement towards the estimated target position (not shown). 

Similar effects can be observed for the gaze noise level $\sigma_{e}$, that is, the (constant) magnitude of Gaussian noise that is added to the observation of the eye fixation.
As soon as the saccade towards the target has taken place (after 0.1s in the shown example), 
the target estimate is significantly improved, as it can be estimated from both visual input and the eye fixation observation. %
Thus, $\sigma_{e}$ %
mainly determines the convergence rate of the internal target estimate \textit{after} the saccade (red dash-dotted lines in plot B). %
Moreover, a larger $\sigma_{e}$ incentivizes a (slightly) faster movement towards the internal target estimate, in order to further improve this estimate and thus reliably keep the end-effector inside the actual target at the end of the movement, when terminal costs incur. %

The effect of the saccade time step $n_{s}$ on both the target estimate and the end-effector position profile is shown in plot C. %
An increase in $n_{s}$ delays the movement towards the target, as world-centered information on the target position, which become available to the controller at time $n_{s}$ via observation of the eye fixation point, considerably improve the internal target estimate, i.e., it is worth waiting a little longer. %

In plot D, %
the development of the internal target estimate over time is depicted. %
Starting with the prior target estimate $T_{0}$ (dashed line), the mean of the normal distributions shifts towards the true target $T$ (solid line) as more sensory input becomes available.
Note that the largest improvements occur at the beginning and after the saccade, i.e., after 0.05s.
The variance first increases\footnote{Note that the initial variance estimate of the target component equals zero, i.e., the LQG model initially %
assumes that the target is at the initial position $T_{0}$ with probability 1, which is why we skip the density function at time $0$ and plot the density function starting at the second time step, i.e., after 0.002s.} and then slowly decreases towards zero. %

\subsection{Results of Parameter Fitting}

The E-LQG model uses the following 9 parameters:

\begin{itemize}
	\item the (terminal) velocity cost weight $\omega_{v}$,
	\item the (terminal) force cost weight $\omega_{f}$,
	\item the effort cost weight $\omega_{r}$,
	\item the signal-dependent control noise level $\sigma_{u}$,
	\item the velocity perception noise level $\sigma_{v}$ and the force\footnote{Recall that the controlled system is assumed to have unit mass, i.e., the applied force is equivalent to the acceleration.} perception noise level $\sigma_{f}$,
	\item the gaze noise level $\sigma_{e}$,
	\item the position perception noise weight $\gamma$, and
	\item the saccade time step $n_{s}$.
\end{itemize}

We again identify the optimal parameter vector for each combination of participant, task condition, and movement direction, using the stochastic parameter fitting procedure from Section~\ref{sec:param-fitting}.
Qualitatively and quantitatively, the results do not change much from the LQG model.
We thus focus only on parameters which exhibit noticeable differences.

\begin{figure}
	\centering
	\subfloat{\includegraphics[width=0.5\linewidth]{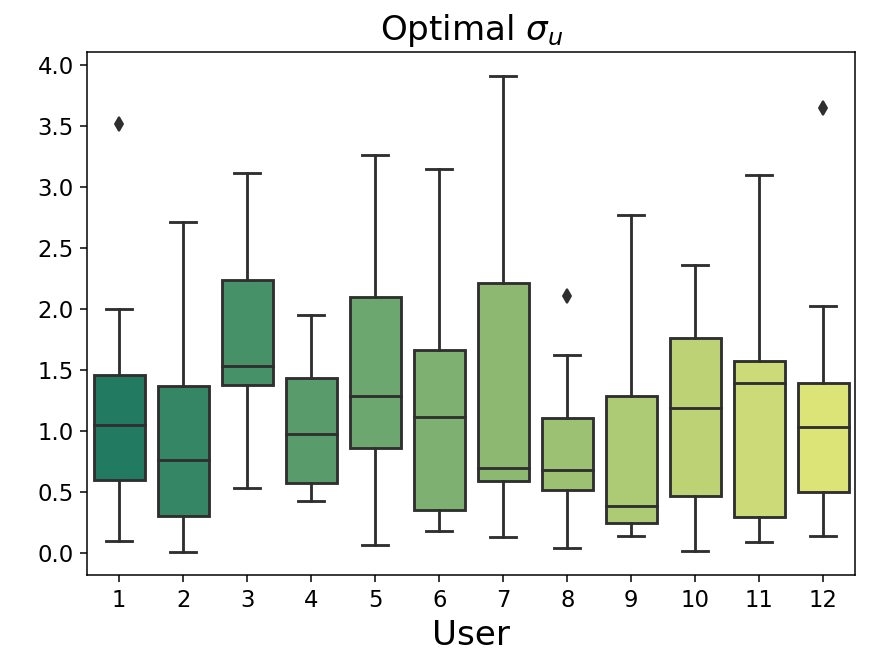}}
	\subfloat{\includegraphics[width=0.5\linewidth]{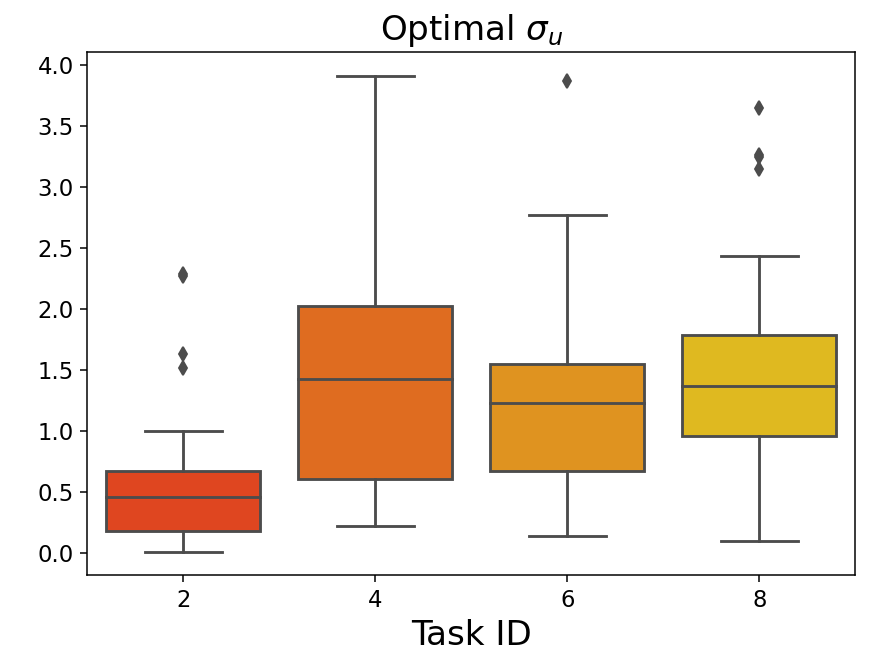}}
	\caption{Control noise parameter $\sigma_{u}$ of the E-LQG, optimized for the mean trajectories of all participants, tasks, and directions, grouped by participants (left) and by ID (right).
	}~\label{fig:E-LQG_opt}
	\Description{Fully described in caption and text.}
\end{figure}

As shown in Figure~\ref{fig:E-LQG_opt}, the optimal control noise level $\sigma_{u}$ exhibits clear differences between individual users and task IDs, similar to the LQG case. 
However, the optimal values are considerably lower 
(the mean value amounts to $1.18$ (E-LQG) and $2.52$ (LQG)).
In Section~\ref{sec:model-comparison}, we show that the E-LQG model replicates the observed user trajectories similarly well compared to the LQG model.
This suggests that the extended observation model of E-LQG allows to replicate the typical variance profiles with lower (i.e., more plausible) signal-dependent control noise levels.

The decrease of observation noise with higher ID, which was clearly noticeable for the single observation noise parameter $\sigma_{s}$ in the LQG model, occurs only for the velocity perception noise level $\sigma_{v}$ (see Figure~\ref{fig:E-LQG_opt_2a} in the Appendix; note the logarithmic scale in both plots). 
Instead, the magnitude of the noises %
$\sigma_{f}$ and $\gamma$ %
do not exhibit a clear dependency on the ID, %
and vary between individual participants. 
This suggests that the observation model of E-LQG makes it more meaningful than the LQG model. %
The optimal parameter values of %
$\sigma_{e}$ and %
$n_{s}$ increase with ID (at least for ID $\geq4$), and also exhibit characteristic differences between individual users (see Figure~\ref{fig:E-LQG_opt_2b} in the Appendix).

\section{The Intermittent Control Model (IC)}\label{sec:IC}

In the remaining part of this paper, we will compare the presented models against each other, both qualitatively and quantitatively.
In particular, we analyze and discuss the ability of the stochastic models LQG and E-LQG to predict not only individual trajectories, but entire trajectory distributions.
Since 2OL-Eq, which we use as a baseline for the deterministic optimal control models, is not capable of predicting movement variability, we need another, stochastic baseline model. %
We decided to use a model from Intermittent Control (IC) theory, which recently has been proposed by Martin et al.%
~\cite{Martin21}.
In the following, we will give a short overview of the similarities and differences between IC and OFC. %

In both IC and classical OFC models, internal models of the interaction loop are used to find controls that are optimal with respect to a certain cost function.
As a major difference, OFC continuously integrates the stream of obtained sensory input signals to account for unexpected disturbances and correct internal state estimates accordingly, whereas IC only intermittently makes use of these observations%
~\cite{Gawthrop11, Park20, Martin21, Do21}.
In particular, IC allows to include a minimum open-loop interval between two successive events, in which no feedback is available to the controller%
~\cite{Martin21}.
From a neuroscientific perspective, this is consistent with the theory of \textit{psychological refractory periods}, which assumes the existence of short periods of time after a visual stimulus has been processed, in which the controller cannot react to further changes in the environment %
~\cite{Telford31, Gawthrop11}.

More precisely, in IC, an open-loop control based on an internal representation of the system dynamics (the so-called \textit{hold}) %
is applied until the difference between predicted and observed state %
exceeds some predefined threshold, i.e., until the unaccounted disturbances become too large. In this case, an event is triggered, which updates the internal model based on a new sample from the continuously perceived stream of observations. %
Afterwards, the open-loop control that is optimal for the updated internal model is applied until the next event is triggered, and so on, resulting in an \textit{intermittent} control. %
IC models can thus be regarded as a hybrid of open-loop and closed-loop models.

We decided to use the IC model from~\cite{Martin21}, since it is based on similar assumptions (optimal control with respect to accuracy, stability, and effort costs, subject to the constraints imposed by the system dynamics), and also has been applied to mouse movements, using the same Pointing Dynamics Dataset and a parameter identification process similar to ours.
Moreover, the IC model is also able to replicate movement variability in terms of phase space probability distributions~\cite{Martin21}.
It thus constitutes a suitable baseline for the considered stochastic OFC models. %

However, the variability is generated completely differently in the two approaches.
In LQG/E-LQG, between-trial variability arises from noise terms that are explicitly modeled in the Human-Computer System Dynamics (e.g., signal-dependent control noise, or observation noise), which allows to analytically compute the expected mean and covariance matrices (see Section~\ref{sec:LQG}).
In contrast, both the control and observations dynamics of the IC model from~\cite{Martin21} are assumed deterministic.
While motor and/or observation noise could be included in principle, the IC model is based on the LQR, i.e., it is not capable of taking into account the expected variance due to such noise terms when computing the optimal control strategy.

In contrast, in the IC model, motor variability is only due to a \textit{multiple-model approach}, i.e., multiple movements are generated by using different parameter vectors, which are randomly drawn from a bank of identified parameter vectors.
This is a major difference from the presented SOFC models, where we have identified only \textit{one parameter vector} for each user, task condition, and direction, such that the resulting trajectory distribution captures both average user behavior and between-trial variability. %

Further differences between the IC model and the LQG/E-LQG include the system dynamics (in the IC model, the same fourth-order dynamics are used, but with slightly different time constants $\tau_{1}=\tau_{2}=0.05$), and the observation model, which for the IC only yields (unperturbed) positional information. %

\subsection{Technical Details}

The IC simulation trajectories, which we will use as a baseline for the LQG and the E-LQG models in the following, were generated by Martin et al.~\cite{Martin21}. %
For each combination of participant and task condition, they performed 200 simulations of 20 subsequent ``slices'' (i.e., combinations of rightward and leftward movements), by randomly drawing from a bank of 20 identified parameter vectors.
Since we analyze unidirectional movements in this paper, we split each of these slices at the respective target switch time, resulting in a total of 4000 simulated IC movements for each participant, task condition, and direction. %

For the comparison between the IC simulation data and observed user data, we then clip the trajectories of \textit{both} datasets to some $N\in\N$, since the lengths of the IC trajectories 
were chosen to match the lengths of the respective dataset slices (in contrast to LQG/E-LQG, where the optimal trajectory distribution sequence necessarily yields sample trajectories of pre-defined, equal length).
We define the maximum IC trajectory length $N_{\text{IC}}$ using the same outlier criteria (both with respect to trajectory length and position values at each time step) as for the maximum user data trajectory length $N_{\text{USER}}$ (see Section~\ref{sec:dataset}), and then cut both distribution sequences to length $N=\min(N_{\text{IC}}, N_{\text{USER}})$.\footnote{Note that, given a participant, task condition, and direction, the length $N$ used to compute a measure of similarity between simulation and user trajectories, such as MWD, %
might thus be slightly lower for IC than for the other models, where $N$ was set to $N_{\text{USER}}$.}
We also remove the reaction times from all IC simulation trajectories, using the same procedure as for the Pointing Dynamics Dataset (see Section~\ref{sec:dataset}).

Finally, we compute the sample mean and covariance matrices of the resulting set of IC trajectories on a frame by frame basis, resulting in one trajectory distribution sequence for each participant, task condition, and direction.

\section{Comparison Between Models}\label{sec:model-comparison}

In the following, we provide a detailed comparison of the six presented models, both qualitatively and quantitatively.

\subsection{Qualitative Comparison}\label{sec:qualitative-comparison}

\begin{figure}
	\centering
	\includegraphics[width=\linewidth]{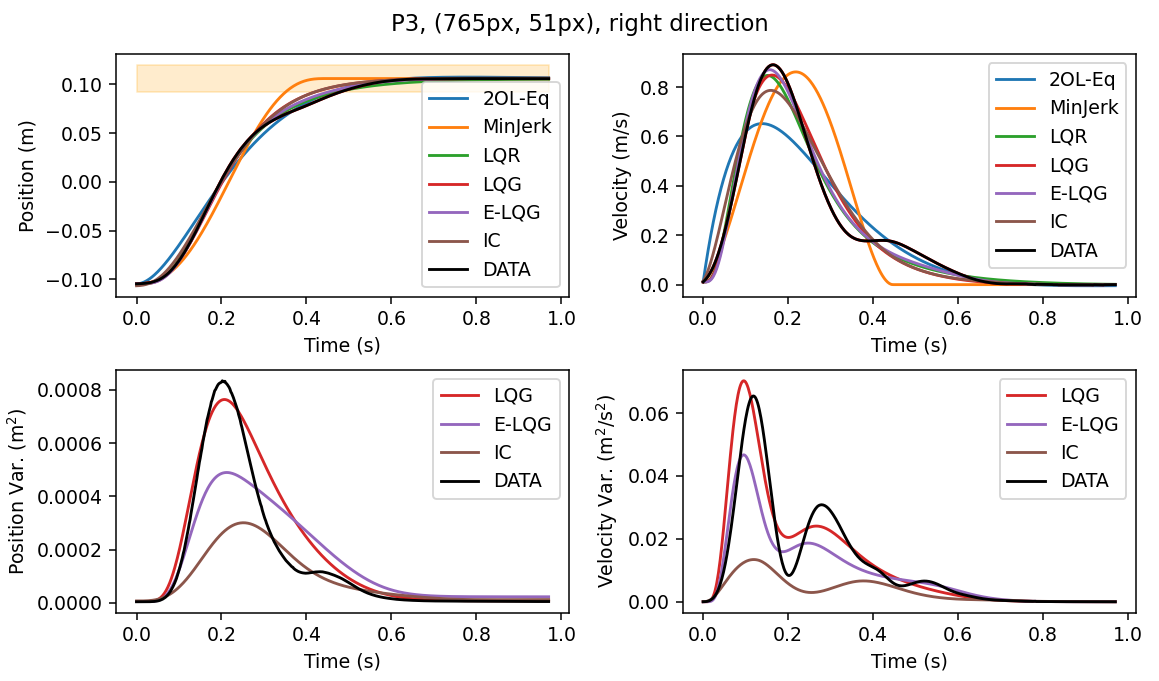}
	\caption{
		Comparison between all considered models in terms of mean and variance of both position and velocity, using the same task condition (participant 3, ID 4 (765px distance, 51px width), rightward movement) as in the previous sections.%
	}
	\label{fig:MODELS_data_ID4}%
	\Description[Mean and variance of position and velocity time series for all considered models]{Mean and variance of position and velocity time series for all considered models, together with user data from the Pointing Dynamics Dataset (Participant 3, ID 4 (765px distance, 51px width), right direction; variances are only shown for LQG, E-LQG, IC, and user data).}
\end{figure}

A comparison of all simulation trajectories for the regarded user and task condition can be found in Figure~\ref{fig:MODELS_data_ID4}, where the mean position and velocity is shown for all considered models, and the variance in position and velocity is shown for all stochastic models.

\subsubsection*{Deterministic Models}
The deterministic models can only predict average behavior (top row).
The 2OL-Eq trajectory (blue lines) has a too large velocity at the beginning of the movement, which is a direct consequence of the high initial acceleration, as discussed in Section~\ref{sec:dynamics}.
The trajectory of the MinJerk model (orange lines) exhibits a perfectly bell-shaped velocity profile, with peak velocity very close to that observed in the user trajectory (black lines). 
However, MinJerk cannot explain the required corrective submovements towards the end of the movement. 
Instead, it assumes that the target is reached after the first ballistic movement, i.e., the surge. 
For trajectories with clearly visible submovements, this results in a considerable worse overall fit of both the position and the velocity time series.
In addition, the duration of this surge does not emerge from MinJerk, but needs to be explicitly fitted to the desired user trajectory.
In contrast, the LQR model (green lines) approximates the mean trajectory well in terms of position and velocity, although %
the corrective movement is not pronounced. 

\subsubsection*{Stochastic Models}
Both the LQG and the E-LQG models do not only model average behavior well (top row), but also account for the between-trial variance observed in both position and velocity profiles (bottom row).
The variance profiles of the LQG model (red lines) are similar to those of the user data (black lines). 
For some trials (as the one shown in Figure~\ref{fig:MODELS_data_ID4}), the E-LQG trajectory distribution sequence (purple lines) fits slightly worse in terms of positional and velocity variance; in particular, the peaks of both variance profiles are considerably lower for the E-LQG compared to user data.
The IC simulation trajectories (brown lines) exhibit an even lower variance in terms of both position and velocity. Moreover, the peak velocity of IC is lower compared to user data and all other considered models (except for 2OL-Eq). %

\subsubsection*{Corrective Submovements}
Corrective submovements are not replicated well by any of the six models.
MinJerk is extended by a constant position value after the surge and thus naturally cannot account for granular corrections of the end-effector position, which are visible in the velocity time series of the user data, starting around 0.36s. %
The remaining models slowly reduce the velocity towards the end of a movement.
However, clear submovements, i.e., additional peaks in the velocity profile (around 0.42s %
in the user data), are not visible.

\subsection{Quantitative Comparison}\label{sec:quantitative-comparison}
We start with a comparison of how well each model is able to predict average user behavior, i.e., how close their simulated trajectories resemble the mean trajectories computed from the Pointing Dynamics Dataset.
Although the parameter fitting of the deterministic models (2OL-Eq, MinJerk, and LQR) was performed with respect to positional SSE only, we also evaluate the SSE with respect to mean velocity and acceleration. %
In addition, we measure the positional \textit{Maximum Error} between model and user trajectories, i.e.,
\begin{align}
	\max_{n=0,\dots,N}\vert p_{n}^{\Lambda}-p_{n}^{\text{USER}}\vert,
\end{align}
and analogously the Maximum Error in velocity and acceleration.

In addition to the $2-$Wasserstein distance, 
we also consider the mean \textit{KL divergence}~\cite{Kullback51} over time to further evaluate the performance of the stochastic models (LQG, E-LQG, and IC). %
Moreover, we compare the mean trajectories of the stochastic models with respect to SSE and Maximum Error, albeit these models were optimized to resemble the entire variability of observed user behavior.

\begin{figure}
	\centering
	\subfloat[SSE - Position]{\includegraphics[width=0.33\linewidth]{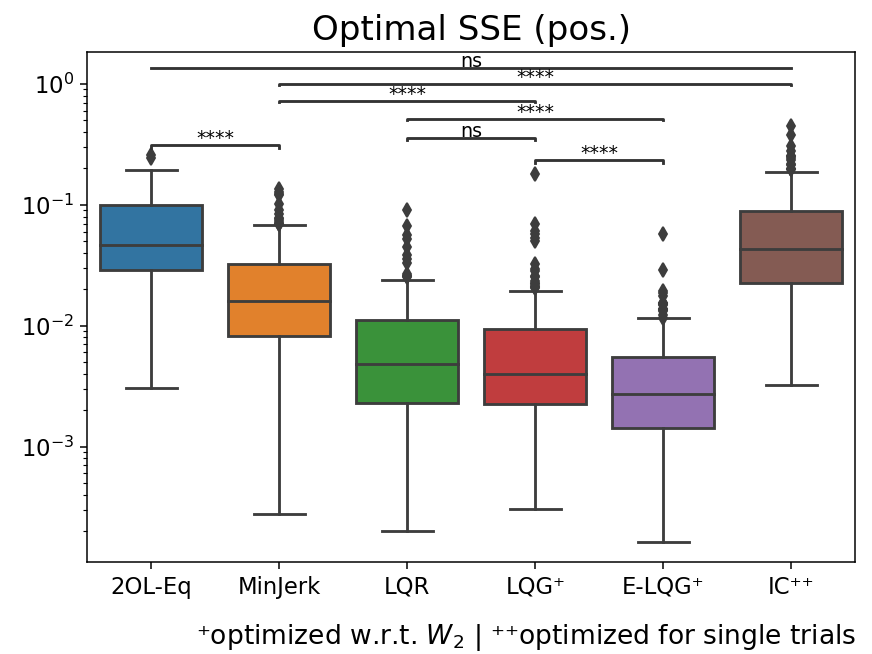}}
	\subfloat[SSE - Velocity]{\includegraphics[width=0.33\linewidth]{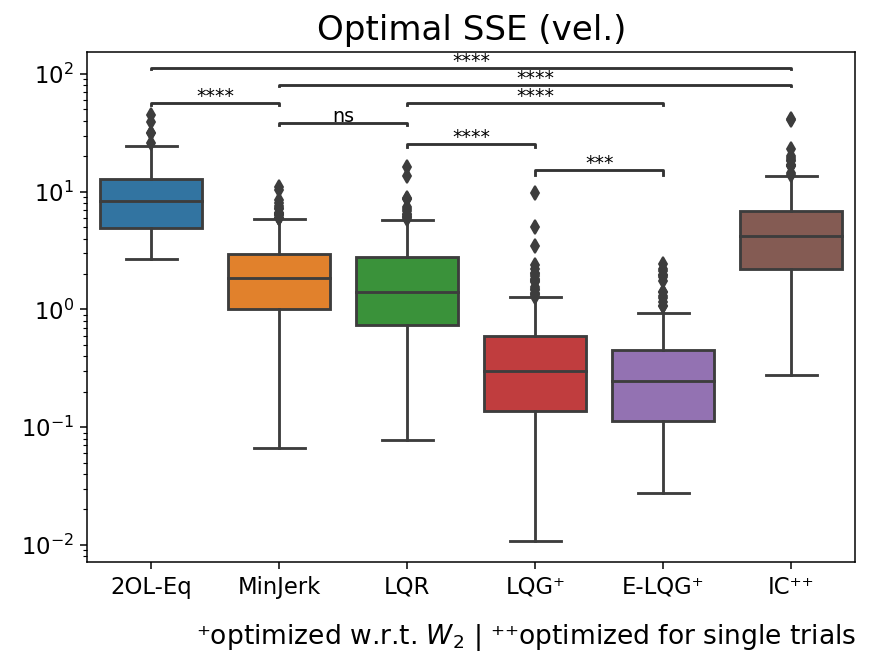}}
	\subfloat[SSE - Acceleration]{\includegraphics[width=0.33\linewidth]{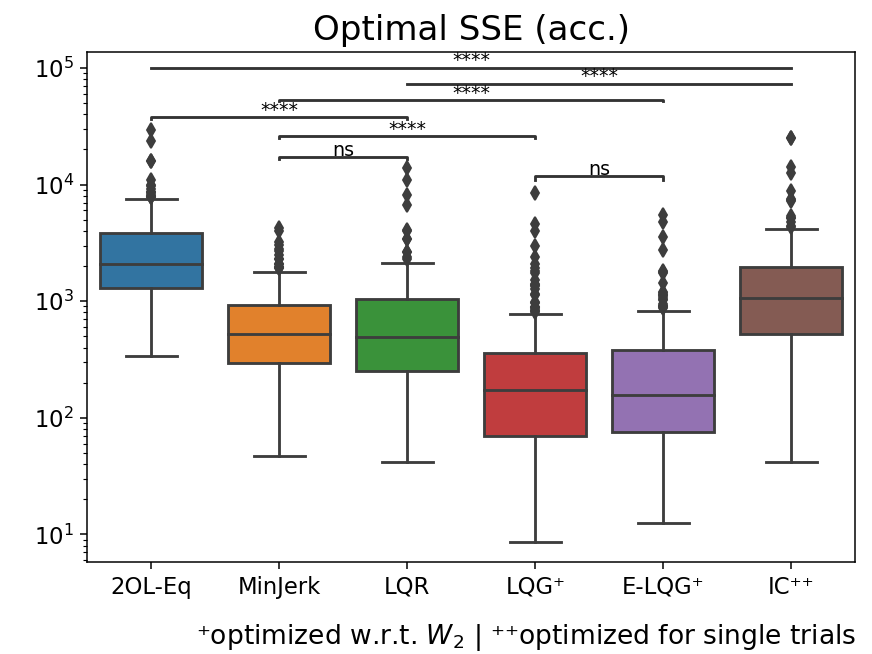}}
	\\
	\subfloat[Maximum Error - Position]{\includegraphics[width=0.33\linewidth]{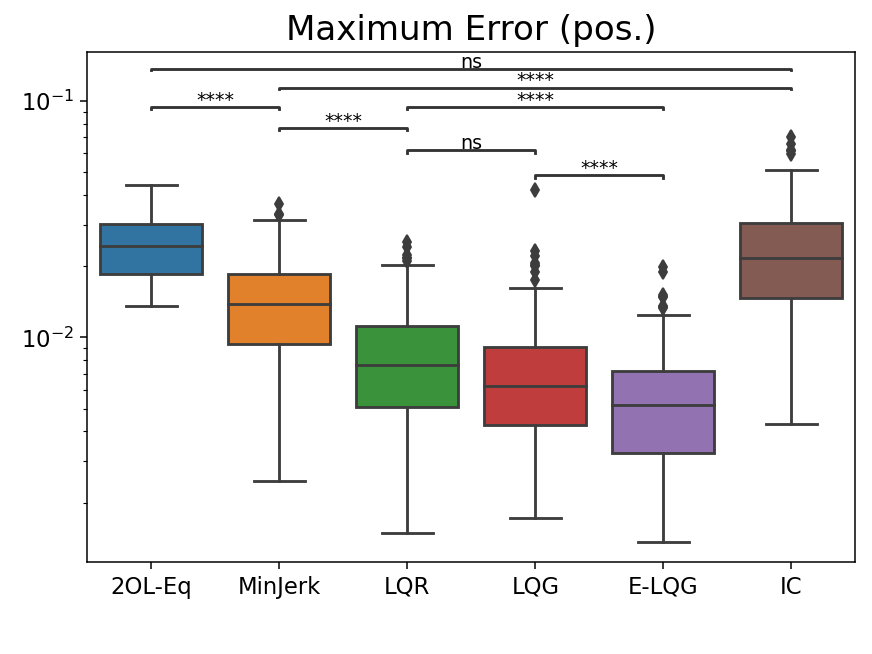}}
	\subfloat[Maximum Error - Velocity]{\includegraphics[width=0.33\linewidth]{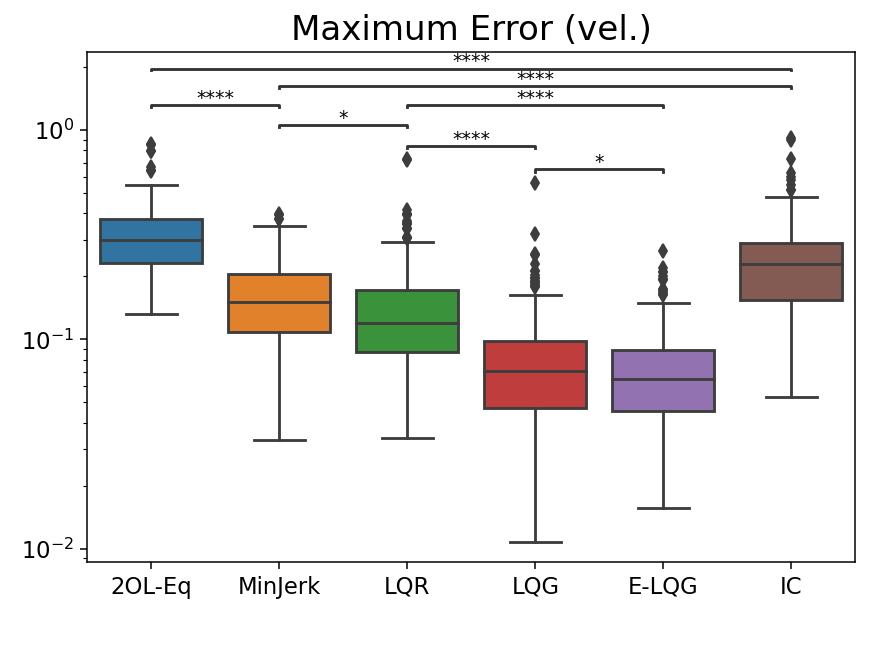}}
	\subfloat[Maximum Error - Acceleration]{\includegraphics[width=0.33\linewidth]{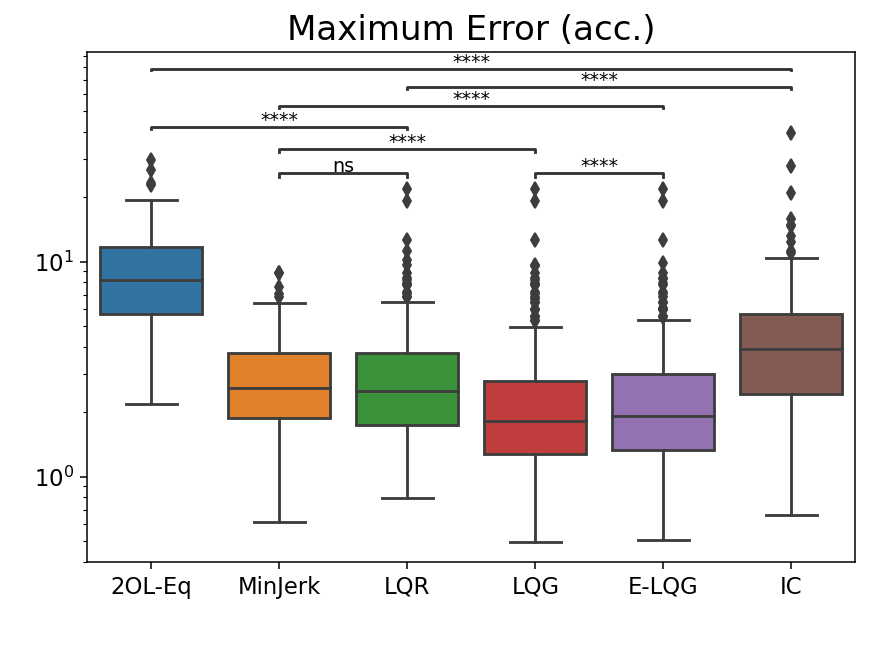}}
	\\
	\subfloat[MWD - Position \& Velocity]{\includegraphics[width=0.33\linewidth]{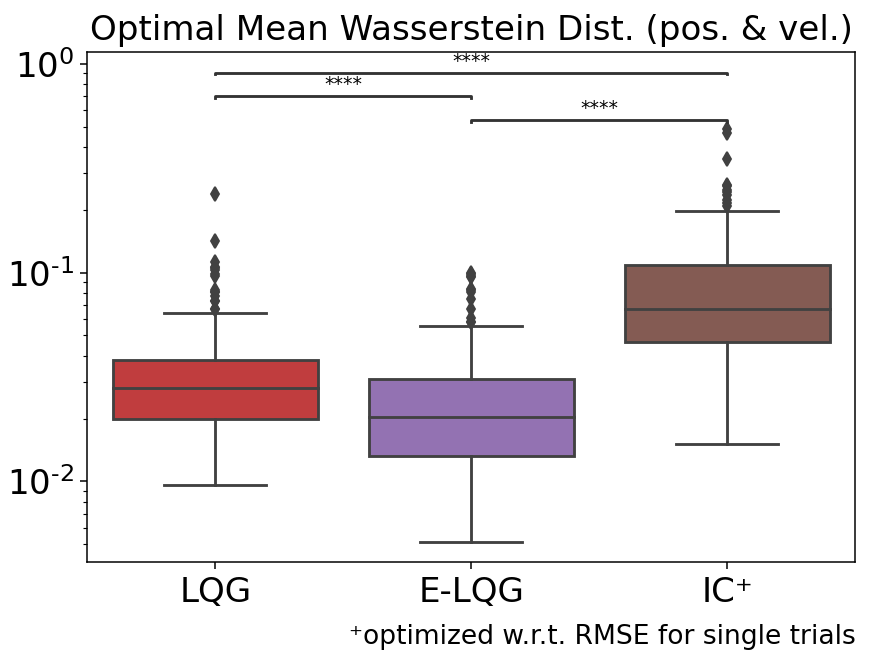}}
	\hspace*{0.2cm}
	\subfloat[MKL - Position \& Velocity]{\includegraphics[width=0.33\linewidth]{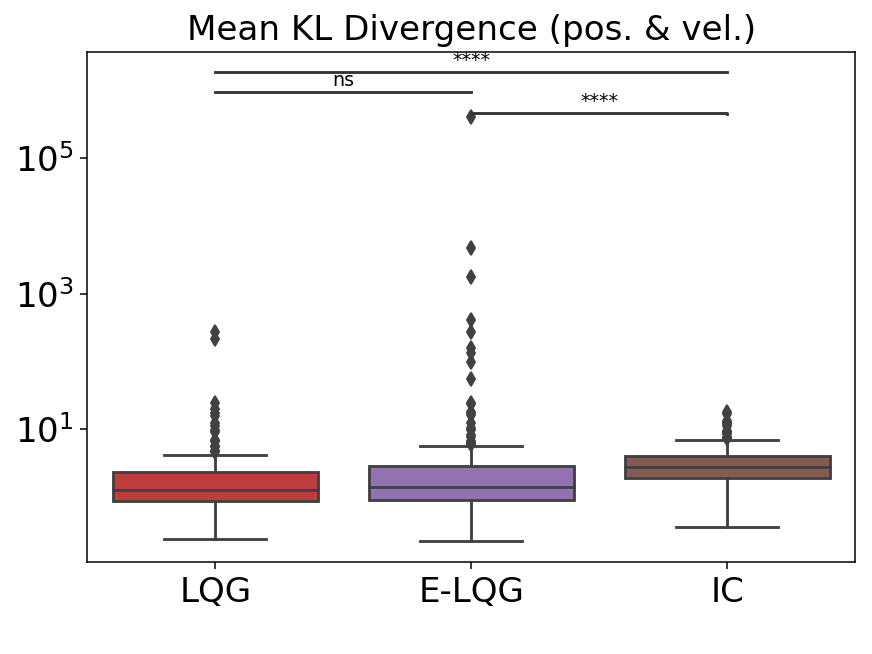}}
	\caption{\textbf{(a)-(f):} SSE and Maximum Error values of all considered models, regarding the mean position, velocity, and acceleration time series of all participants and all tasks. \\
		\textbf{(g)-(h):} Mean Wasserstein distance and mean KL divergence of LQG, E-LQG, and IC, using sequences of Gaussian distributions over end-effector position and velocity, for all participants and all tasks. \\
		The asterisks indicate whether the difference between two bars is significant according to Wilcoxon Signed Rank tests with Bonferroni corrections (*: $p\leq0.05$, **: $p\leq0.01$, ***: $p\leq0.001$, ****: $p\leq0.0001$; n.s.: $p>0.05$). 
		Note the logarithmic scale in each plot.
	}~\label{fig:ALL_quantitative}
	\Description{Fully described in caption and text.}
\end{figure}

Figure~\ref{fig:ALL_quantitative} displays the quality of the fit for all six models on a logarithmic scale.

Kolmogorov-Smirnov tests showed that the distributions of positional SSE for all six models do not fit the assumption of normality (all values $p<0.0001$; the same holds for all other considered measures). 
Thus, we carried out a Friedman Test (i.e., a non-parametric test equivalent to a repeated measures one-way ANOVA), using Bonferroni corrections.
The main factor included in the analysis was which model was used: 2OL-Eq, MinJerk, LQR, LQG, E-LQG, and IC.
The test indicated that the SSE between the six models was significantly different ($\chi^{2}(2) =~$$638$, $p<0.0001$, $n=192$). 

Additional Wilcoxon Signed Rank tests with Bonferroni corrections 
showed that the positional SSE of the E-LQG is significantly smaller than that of both LQG and LQR (E-LQG vs. LQG: $Z=-4.9$; E-LQG vs. LQR: $Z=-5.4$; $p<0.0001$), while there were no significant differences between the latter two.
However, the positional SSE of both LQG and LQR is significantly lower when compared to the MinJerk model (LQG vs. MinJerk: $Z=-9.9$; LQR vs. MinJerk: $Z=-8.4$; $p<0.0001$), which in turn shows smaller values than 2OL-Eq ($Z=-9$, $p<0.0001$).
The fit of the mean IC trajectories with respect to positional SSE is comparable to 2OL-Eq (non-significant differences), with MinJerk ($Z=-9$, $p<0.0001$) fitting the mean position profiles significantly better. %

The findings are analogous for the maximum deviations of the simulated position time series from the data (Friedman Test, $\chi^{2}(2)=~$$685.4$, $p<0.0001$, $n=192$), with Wilcoxon Signed Rank tests showing that the E-LQG model approximates user trajectories significantly better than both the LQR ($Z=-6.9$, $p<0.0001$) and the LQG ($Z=-5.4$, $p<0.0001$) models.
Moreover, they showed that the LQR model approximates user trajectories significantly better than MinJerk ($Z=-9.3$, $p<0.0001$), and that MinJerk approximates user trajectories significantly better than both the 2OL-Eq ($Z=-10.5$, $p<0.0001$) and IC ($Z=-8.9$, $p<0.0001$) models. %

Regarding velocity, a Friedman Test indicated that SSE ($\chi^{2}(2) =~$$794.4$, $p<0.0001$, $n=192$) and Maximum Error ($\chi^{2}(2) = ~$$761.9$, $p<0.0001$, $n=192$) were significantly different between the six models, with Wilcoxon Signed Rank tests showing that E-LQG approximates velocities significantly better than LQG ($Z=-4.3$, $p<0.001$ for SSE and $Z=-3.1$, $p<0.05$ for Maximum Error), LQG better than LQR (SSE: $Z=-11.7$, Maximum Error: $Z=-11.3$; $p<0.0001$), LQR comparable to MinJerk (LQR is slightly better in terms of Maximum Error, with $Z=-2.7$ and $p<0.05$, while the differences in SSE are non-significant), MinJerk better than IC (SSE: $Z=-8.4$, Maximum Error: $Z=-7.7$; $p<0.0001$), and IC better than 2OL-Eq (SSE: $Z=-9.7$, Maximum Error: $Z=-9.9$; $p<0.0001$).
The clear lead of LQG and E-LQG over LQR in this respect was to be expected, as LQG and E-LQG were optimized to minimize the Wasserstein distance with respect to \textit{both} position and velocity.

Regarding acceleration, a Friedman Test indicated that SSE ($\chi^{2}(2) =~$$723.6$, $p<0.0001$, $n=192$) and Maximum Error ($\chi^{2}(2) =~$$539.5$, $p<0.0001$, $n=192$) were also significantly different between the six models, with Wilcoxon Signed Rank tests showing that acceleration is approximated significantly better by both LQG and E-LQG than by LQR or MinJerk (E-LQG vs. LQR -- SSE: $Z=-12$, Maximum Error: $Z=-10$; LQG vs. LQR -- SSE: $Z=-12$, Maximum Error: $Z=-11.6$; E-LQG vs. MinJerk -- SSE: $Z=-8.4$, Maximum Error: $Z=-4.7$; LQG vs. MinJerk -- SSE: $Z=-8$, Maximum Error: $Z=-4.7$; $p<0.0001$), MinJerk and LQR show no significant differences in acceleration, both LQR and MinJerk approximate acceleration significantly better than IC (LQR vs. IC -- SSE: $Z=-11.6$, Maximum Error: $Z=-11.2$; MinJerk vs. IC -- SSE: $Z=-8$, Maximum Error: $Z=-8.9$; $p<0.0001$), and IC approximates acceleration significantly better than 2OL-Eq (SSE: $Z=-8.1$, Maximum Error: $Z=-9.3$; $p<0.0001$).
In terms of SSE, no significant differences between LQG and E-LQG were found, whereas in terms of Maximum Error, LQG approximates user acceleration profiles significantly better than E-LQG ($Z=-7.1$, $p<0.0001$).

In summary, the OFC model LQR achieves similar or better fits than both the dynamics model 2OL-Eq and the open-loop model MinJerk on all accounts, while being able to incorporate both control dynamics and objectives that are assumed to be optimized, given a specific interaction task.
In addition, the duration of the initial ballistic movement (i.e., the surge), emerges naturally from the model and does not need be known in advance (in contrast to MinJerk).
The mean trajectories predicted by the stochastic extensions LQG and E-LQG approximate average user behavior significantly better than the LQR model in terms of velocity and acceleration, which could also result from taking the velocity explicitly into account in the LQG/E-LQG parameter optimization.
Intermittent Control strategies on the other hand show a sub-optimal fit of mean user trajectories.
A probable reason is that parameters were fitted to match particular user trajectories instead of trial-independent trajectory distributions~\cite{Martin21}. %

Regarding the Mean Wasserstein distance (MWD) and the Mean KL divergence (MKL) (see Figure~\ref{fig:ALL_quantitative}(g)-(h)), both the LQG and the E-LQG trajectories show a significantly better approximation of user behavior than the IC trajectories according to Wilcoxon Signed Rank tests (LQG vs. IC -- MWD: $Z=-12$, MKL: $Z=-7.9$; E-LQG vs. IC -- MWD: $Z=-12$, MKL: $Z=-5.3$; $p<0.0001$).
Between the LQG model and its extension E-LQG, significant differences in favor of the E-LQG were only found in terms of MWD ($Z=-10.1$, $p<0.0001$).

These results suggest that the LQG and its extension E-LQG capture both average user behavior and the specific variance profiles observed in position and velocity time series better than all other considered models.

\section{Discussion and Future Work}\label{sec:discussion-future}

In the following, we provide a discussion of the above results.
We also discuss the applicability of the proposed framework to HCI tasks other than mouse pointing.

\subsection{Deterministic Models}

As shown in Section~\ref{sec:dynamics}, the optimal damping ratio $\zeta$ of 2OL-Eq is always lower than 1 and exhibits a relatively small between-user variance. This implies that 2OL-Eq considers all regarded user movements as underdamped, which, to the best of our knowledge, has not yet been shown.
While this could provide an indication that users rather tend to over- than undershoot for spatially-constrained 1D mouse movement, this is not consistent with our findings from all considered optimal control models.
We thus conclude that the second-order dynamics of 2OL-Eq are not sufficient to capture the complex human behavior that is already apparent in 1D end-effector trajectories.
In other words, interpreting the mouse cursor as a mass attached to one edge of the screen via a spring and a damper might suggest an underdamped system; however, it is the interpretation itself that seems to be inappropriate. %
This can also be seen from the left-skewed velocity profiles caused by (unrealistic) instantaneous peak acceleration, whereas typical user trajectories rather exhibit bell-shaped velocity profiles, as it is captured by the remaining models.

The user-specific values of the stiffness $k$ in 2OL-Eq, which indicates how fast the end-effector is moved towards the target, are closely related to those identified for the surge duration parameter $N_{MJ}$ from MinJerk.
This indicates that different parameters in different models can play a similar role in explaining user behavior when fit through our parameter fitting process, even for dynamics-only (2OL-Eq) and kinematics-only (MinJerk) models.

The MinJerk model is able to predict typical velocity and acceleration profiles.
However, in its standard form, it only covers the first ballistic movement and does not account for any corrective movements. 
Our proposed variant, which is constantly extended by its last position, thus faces a trade-off between (i) fitting the perfectly bell-shaped velocity profile to that observed in user trajectories during the surge (which, however, is often truncated), and (ii) exhibiting a non-zero velocity during the subsequent correction phase.
A tempting approach to resolve this issue is an iterative-submovement version of MinJerk, which is composed of the minimum jerk trajectories for a number of identified path segments~\cite{Viviani95}.
Indeed, such a concatenation has been shown to account for the characteristics of typically observed trajectories, e.g., in case of handwriting~\cite{Edelman04}, gesture typing~\cite{Quinn18}, or arm movements~\cite{FlashHenis91, Lee97}.
At its core, however, it requires the manual definition of when a submovement starts and terminates.
Even more critically, the kinematic properties of the end-effector (i.e., the position, velocity, and acceleration) need to be known at the beginning and at the end of each path segment. %
Thus, several \textit{via-points} need to be placed along the path, none of which can be inferred from the task description.
This contradicts the minimum intervention principle~\cite{Todorov02}, which suggests that only the ``deviations'' that interfere with task performance are being corrected.
Moreover, several user studies have shown that there is no fixed via-point users aim at when being asked to repetitively navigate around a setup of obstacles~\cite{Liu07}. %
In summary, such an extension might be appropriate to replicate kinematic characteristics of human movements to a certain extent.
However, it cannot resolve the fundamental unsuitability of the MinJerk model for explaining how \textit{and why} motion is generated, as the underlying movement dynamics are not modeled at all.

Instead, we propose to follow the argumentation of Liu and Todorov~\cite{Liu07}, %
which have empirically shown that (apart from the physical constraints induced by the environment, i.e., the Human-Computer System Dynamics in Figure~\ref{fig:genmodel_extended}) there are no internalized ``hard'' constraints users would comply \textit{at any cost}, %
but only a (non-trivial) trade-off between several objectives.
For example, users asked to reach a small target \textit{as accurate as possible} and \textit{within a certain time limit}, %
which might be very difficult to achieve together, were shown to relax both requirements to some extent~\cite{Liu07}, rather than trying to keep exactly on schedule.
This is consistent with 
the \textit{cost combination hypothesis}~\cite{Berret11}, which argues that the flexibility in coordinating motor behavior is due to the optimization with respect to a \textit{combined} objective function, e.g., including both accuracy and effort costs.
An extension of these theories is the ``reward is enough'' hypothesis~\cite{Silver21}, which claims that different forms of intelligent behavior (e.g., learning, social intelligence, or generalization) can be directly deduced from maximizing an internalized reward function (which can be considered equivalent to minimizing a cost function).
All of these findings support the idea that human movement arises as the result of an internal optimization, with objectives that can be directly deduced from the cognitive system and the task-specific instructions, and are thus fully compatible with the proposed optimal control framework for Human-Computer Interaction.

Combining the assumption of such a task-dependent optimal control with a simplified muscle model that yields overall fourth-order system dynamics, as we have done with LQR\footnote{Note that the LQR model (as well as all other presented models) depends on the task under consideration, since the target position needs to be included in state space, and the several objectives (accuracy, stability, and effort costs) were derived from the task description.}, %
allows to replicate average mouse pointing trajectories both qualitatively and quantitatively. %
In particular, the LQR model yields a significantly better fit of average user trajectories than both 2OL-Eq and MinJerk (in terms of velocity and acceleration, LQR and MinJerk show comparable fits).
However, we found it necessary to penalize both the remaining distance to target as well as its time derivatives (velocity and acceleration, which equals to applied force under the assumption of unit mass) in every time step.
In contrast, the LQR model with state costs only applied at the final time step does not replicate typical mouse pointing movements (for ID $\geq4$), as it predicts a smooth movement towards the target with symmetric, bell-shaped velocity profile and thus cannot account for the large correction phase typically observed towards the end of the movement (see Figure~\ref{fig:LQR_data_ID4}). %

\subsection{Stochastic Models}

The stochastic extension of the LQR model -- the LQG model --
is naturally capable of modeling and explaining between-trial variability observed from experimental data.
It includes both signal-dependent control noise and constant observation noise.
In contrast to the LQR model, the state cost do not need to incur in every time step, but only at the final time.
We also introduced an extension of the LQG model, called E-LQG, which incorporates %
both visual input %
and observations of the eye fixation point, %
assuming accurate saccades between the initial and the target position, %
thus being more plausible from a visuomotor perspective.

Both mean and variance of typical mouse trajectories can be replicated fairly well by the proposed LQG and E-LQG models.
In terms of SSE, E-LQG approximates mean user trajectories as good as or significantly better than all other models, including LQR. %
In terms of the mean Wasserstein distance, E-LQG shows significant improvements over LQG, suggesting that the extended observation model indeed captures characteristic end-effector variance profiles of mouse pointing tasks considerably better than the saccade-free observation model proposed by Todorov~\cite{Todorov05}. %

For the considered 1D reciprocal mouse pointing movements, we found that the assigned noise levels considerably vary between individual participants, with some of the variability better explained by large signal-dependent control noise and some better explained by higher observation noise.
In contrast, the identified optimal cost weights depend on task ID rather than the user.
We hypothesize that the trade-off between end-point accuracy (determined by a large distance weight) and stability (determined by large velocity and force weights) explicitly given by the experimental instructions was interpreted differently for each task condition, leading to different optimal control strategies depending on the task difficulty.

The good fit of the LQG and the E-LQG models in terms of variance is partially attributed %
to relatively high signal-dependent noise levels.
For the LQG model, the average optimal value of $\sigma_{u}$ amounts to $2.52$, that is, an (unbiased) deviation from the desired control value $u_{n}$ with a magnitude of $252\%$ of this value can be expected at each time step $n\in\{0,\dots,N\}$.
For the E-LQG model, the fitted signal-dependent control noise levels are reduced by approximately $50\%$ compared to the LQG model, i.e., more of the variability that is observed from user data can be explained by the extended observation model instead of attributing it to large control noise.
The control noise levels are still relatively high ($1.18$ on average, %
i.e., the desired control and the effectively applied control differ by $118\%$ of the desired control value on average).

In contrast, the literature suggests signal-dependent control noise levels between $10\%$ and $25\%$, based on several empirical findings~\cite{Clamann69, HarrisWolpert98, vanBeers03}. 
We believe that this mismatch does not render the LQG inappropriate in explaining human movements during interaction on a continuous level.
Instead, the large amount of variability observed in the Pointing Dynamics Dataset can at least partially be attributed to \textit{temporal noise}, i.e., different cognitive reaction and/or motor activation times between individual trials.
This is consistent with previous findings, which suggest that both signal-dependent and temporal noise (as well as constant noise) are required to explain the characteristic variance profiles observed in two-dimensional goal-directed arm movements~\cite{vanBeers03}.
In this paper, we have cut off the reaction times of each trial as accurately as possible. 
However, it is difficult to reliably identify reaction times using properties of the (one-dimensional) end-effector trajectories only.
Moreover, the reciprocal nature of the considered bi-directional movements further enhances the variance of the movement onset times, as these become susceptible to learning effects as well as fatigue and a temporary lack of attention (among others), whose overall effect is unclear.
These effects are not accounted for in the assumption that the internal forward model is identical to the actual Human-Computer System Dynamics.
In particular when using more complex dynamics, it might be too restrictive.
Using an inaccurate internal model might act as an additional source of variability (often referred to as \textit{dynamic uncertainty}~\cite{Bian20}).
In the future, it will be interesting to consider different internal models, extend the LQG model by temporal delays of uncertain length (preliminary attempts to include \textit{fixed} reaction times can, e.g., be found in~\cite{Todorov98_thesis}), and to combine it with RL-based methods that allow to model optimal learning behavior~\cite{Bian20}. %

Another point that requires a more thorough discussion is the choice of parameters that are included in the parameter fitting process. For each considered model, we decided to optimize all parameters that we suspected to differ between users or task conditions, and which could not be directly inferred from literature (in contrast to, e.g., the time constants in LQR, LQG, and E-LQG). However, for some parameters (e.g., the force perception noise level $\sigma_f$) the identified values for each participant span several magnitudes, suggesting rather minor effects on the resulting movement trajectories. In this case, it might be reasonable to conduct a second parameter identification, where these parameters are set constant.
Similarly, parameters corresponding to task-independent user strategies or inherent body characteristics could be identified once per user instead of differentiating between task conditions.

It is also important to note that the fit of a given model can be expected to improve with the number of optimized parameters (i.e., degrees of freedom of the model). This implies a trade-off between model simplicity (low number of interpretable parameters) and goodness of fit (ability to ``explain'' observed data), which can be captured by several measures, including the \textit{Akaike information criterion} and the \textit{Bayesian information criterion}~\cite{Stoica04}.
However, these criteria are not directly applicable to our model comparison, since the considered models not only differ in terms of parameters, but most importantly with respect to their scope, which is also reflected in the used reward function, system dynamics, and observation model. For example, LQG can account for movement variability even with fixed (non-zero) noise levels, while LQR naturally cannot.
Instead of comparing the proposed models only based on a single quantitative value, their scope and the phenomena that can be predicted should also be taken into account (also see the three-stage process proposed at the end of Section~\ref{sec:benefits-advice}).

Compared to a recently proposed intermittent control model (IC)~\cite{Martin21}, simulation trajectories of the LQG and the E-LQG models exhibit a considerably better fit in terms of all considered metrics (SSE, Maximum Error, MWD, and MKL).
Qualitatively, both OFC models predict the positional variance at the beginning of the movement and during the correction phase, as well as the velocity variance profiles, more accurate than the IC.
However, it is important to note that the IC simulations are based on a multiple-model approach, i.e., an individual parameter vector is identified for every single trial.
In particular, the variability of the IC only results from random sampling from this set of identified parameter vectors during run-time. %
In contrast, the LQG/E-LQG parameters were explicitly fitted to match the user position and velocity \textit{distributions}, incorporating all trials of a given user, direction, and task condition.
In other words, the variability of LQG/E-LQG is intrinsic to the considered stochastic optimal control problem.
While the between-trial variability of IC %
has shown to match the trajectory variance of the Pointing Dynamics Dataset relatively well in terms of phase plane densities~\cite{Martin21}, it seems that the time sequences of state distributions are replicated considerably better by both the LQG and the E-LQG models.
This suggests that a well-defined model of the sources of variability, such as the signal-dependent control noise and observation noise in the (E-)LQG models, is necessary to replicate the characteristic development of position and velocity variance over time.

A promising next step would thus be to develop an intermittent control model that includes both signal-dependent control noise and observation noise, and investigate its ability to account for the characteristic variance profiles observed for mouse pointing.
A rigorous analysis of the individual components of the new IC model, as we have done for several OFC models in this paper, would allow to examine whether a combination of %
OFC and IC theory (i.e., continuously perceived observations, motor and observation noise, and intermittency of control) may account for typical phenomena such as reaction times, bell-shaped velocity profiles, and characteristic variance profiles. %

One major limitation of the presented OFC models that has not yet been discussed is the need to determine the total movement duration in advance. %
While this is true for most optimal motor control models, including the minimum torque-change~\cite{Uno89} and the minimum end-point variance~\cite{HarrisWolpert98} models, a few attempts have been made towards a model that predicts movement time instead of requiring it.
These approaches include the open-loop constrained minimum-time model~\cite{Tanaka06}, which can be solved analytically in case of linear dynamics, Markov decision processes using state- and action-space discretizations and being solved via dynamic programming~\cite{Liu07}, as well as models that explicitly assign an optimal ``cost of time'', either based on prior assumptions on the human sensorimotor system~\cite{Shadmehr10} or computed via an inverse optimal control approach~\cite{Berret16}.

Another promising framework is \textit{infinite-horizon OFC}~\cite{Phillis85, Qian13}, which is based on the same assumptions as the finite-horizon framework presented in this paper, the main difference being that the total movement duration does not need to be specified in advance.
Instead, the controller is assumed to apply the optimal, time-independent steady-state strategy for \textit{both} the transient movement and the subsequent posture maintenance phase (the so-called \textit{steady-state-control hypothesis}). %
While infinite-horizon OFC constitutes an interesting alternative that allows for a more in-depth analysis of the speed-accuracy trade-off using Fitts' Law type studies, it might be inappropriate for tasks without a clear posture maintenance phase, or where it is unclear whether the controller should consider such a phase during planning, e.g., in fast, repetitive movements as those from the dataset considered in this paper~\cite{Qian13}. %
Its applicability to HCI thus needs to be explored in future work, possibly using a similar approach as in this work.

\subsection{Application to Other HCI Tasks}

In this paper, we analyzed the applicability of optimal feedback control models %
to 1D pointing tasks.
In this section, we discuss how these models can be applied to other common tasks in Human-Computer Interaction, to highlight the generalizability and limitations of these models.
The main limitations of the LQR/LQG approach are their restriction to linear dynamical systems and quadratic costs.
If one of these properties does not hold for a particular task, nonlinear optimal feedback control approaches need to be applied.

While the extension of LQR/LQG to via-point tasks\footnote{In \textit{via-point tasks}, multiple targets (the via-points) need to be reached in a pre-defined order.
Usually, no timing of when each via-point needs to be reached is prescribed.
Via-point tasks can be used to model pointing to several targets in a row.
They have been used to model handwriting or drawing, where the via-points are chosen such that a certain letter or shape is created.} with fixed passage times is straightforward (see~\cite{Todorov98_thesis}), tasks where only the order of targets is specified, but not the specific times they are reached, cannot be directly covered by the proposed models. 
This is mainly due to the assumption of quadratic state costs, which, in combination with linear dynamics, does not allow to penalize the distance to the next via-point depending on the already reached via-points.
One possible way to model via-point tasks with free timing is to integrate the LQR/LQG models into an outer optimization loop (similar to the parameter fitting process introduced in this paper), which, e.g., identifies the minimum passage times such that every via-point is reached~\cite{Todorov98_thesis}.

Following \textit{moving targets} is commonly called pursuit tracking.
Moving targets occur, e.g., in computer games.
If the movement of the target can be modeled %
by a linear differential equation (which includes straight lines, curves, and ellipses), including moving targets in the LQR/LQG models is straightforward.
Simply add the state of the target(s) (e.g., position) to the state space, and extend the system dynamics matrix~$A$ to model the dynamics of the target movement.
The target dynamics can even depend on the end-effector trajectory.
For example, it is possible to model a target that tries to evade the pointer. 
The main restriction is imposed by the linear dynamics, requiring that the target position also evolves linearly.
Pre-defined target trajectories that cannot be described by a linear differential equation are more difficult to implement (in fact, they either need to be ``hard-coded'', i.e., each state needs to be augmented by the complete discrete-time sequence of target positions, which significantly increases the computational effort due to the curse of dimensionality, or approximated by linear dynamics).

\textit{Path following}, or tracing or drawing tasks, are considerably harder to model, as they usually impose spatial-only constraints and leave the temporal profile, i.e., the movement kinematics, up to the user.
In particular, tracing can be considered the limiting case of via-point tasks (with non-determined passage times) with the distance between two via-points approaching zero.
For this reason, the same issues as for via-point tasks occur. 
More precisely, since the cost matrices need to be specified before movement onset, %
the time steps at which the end-effector should reach the desired path/via-point positions need to be known in advance. %
Thus, it is currently unclear whether and how LQG could be used to model path following.

The same holds for \textit{steering tasks} (i.e., tasks with \textit{constrained motion}), where the end-effector needs to be moved from an initial position to an end-point as quickly as possible, while keeping it inside a tunnel of possibly varying width. The most prominent examples include command selection via a hierarchical drop-down menu, parameter sliders, and scroll-bars%
~\cite{Accot97, Accot99, Zhai04b}.
While in the LQR/LQG models, a composite cost function that penalizes both the distance to target and the distance to the center of the tunnel perpendicular to the movement direction would create an incentive to move the end-effector towards the target while keeping it inside the tunnel, this intuitive approach has two major limitations.

First, the boundary constraints are implemented ``softly'' in the sense that the costs for being shortly outside the tunnel are only infinitesimally larger than the costs for being shortly inside. 
This follows directly from the LQR/LQG assumption of costs that are quadratic in the system state and thus necessarily continuous.\footnote{For the same reason, in the considered pointing task, it is not possible to only apply costs when being outside the target, as such costs would necessarily be discontinuous at the target boundary.}

Second, to account for the tunnel constraint, the quadratic costs %
require some reference position that exhibits minimum costs along the direction perpendicular to the movement direction. 
The most obvious choice for this minimum would be the center of the tunnel, as this corresponds to an unbiased penalization of deviations in either direction. 
However, empirical user studies suggest that users do not necessarily aim to follow the central path within the tunnel~\cite{Nancel17, Bailly16}. %
Instead, they deliberately make use of the respective tunnel widths by adjusting their movements, e.g., to achieve higher speeds by ``cutting of the corner''~\cite{Pastel06}.
Regardless of the specific reference trajectory, the usage of costs that penalize the distance to \textit{any} fixed movement trajectory that is not explicitly apparent from the task description (as it is the case for steering tasks) contradicts the \textit{minimum intervention principle}~\cite{Todorov02}, which suggests that only task-relevant deviations are being corrected.

The first issue does not necessarily impose a severe restriction to modeling plausible user behavior, since users also tend to associate a certain internal cost to fulfilling the boundary constraints, i.e., they consider staying inside the tunnel boundaries as one goal among many, rather than viewing it as an inevitable ``hard'' constraint~\cite{Liu07}.
The second issue, however, might constitute a serious limitation and possibly prevent a reasonable application of LQR/LQG to constrained movement tasks. %

Regarding \textit{free-hand inking} tasks (e.g., to write a certain word or draw a specific sketch), it is not clear how an appropriate cost function that includes all relevant information from the task description should look like. 
In addition, capturing high-level characteristics such as  user-specific stroke styles or connections between individual characters might be difficult to model. %
However, the case of gesture-based keyboard typing has recently been successfully modeled as a via-point task with minimum jerk trajectories between two subsequent via-points~\cite{Quinn18}.
It is important to keep in mind that the scope of the proposed methods clearly is to model pointing movements, while more creative tasks would require some high-level cognition process that instantiates and coordinates multiple subprocesses~\cite{Yule13}.
While we do not want to rule out the possibility that LQG can be adapted to the modeling of handwriting or drawing, further research in this regard is certainly needed.

Note that each of the tasks discussed above can be solved using several input methods, e.g., mouse-, pen-, or touch-based input. %
Accurate modeling of the complete interaction loop, as depicted in Figure~\ref{fig:genmodel_extended}, thus requires to take into account device-specific properties such as a pointing transfer function or internal dynamics~\cite{Casiez08, Casiez11}.
Similarly, the Human Body Dynamics can be modeled with arbitrary granularity. For example, the fourth-order dynamics with simplified muscle activations used in this paper could be replaced by complex (non-linear) biomechanical models, e.g., those implemented in state-of-the-art physics engines such as MuJoCo or OpenSim~\cite{Fischer21, Hetzel21, Dembia21}.

In general, non-linearities in the body dynamics, input devices, or interface dynamics cannot be modeled accurately using LQR/LQG.
However, as long as the movements are not too big, a small signals approach can be applied and a linear approximation around an operating point can be found.
To take several operating points into account, it is possible to iteratively linearize non-linear dynamics~\cite{Li04, Todorov05b, Li07}.
Further investigation into the suitability of this approach for the different dynamics of human-computer interaction is definitely needed and constitutes a promising direction for future research. %

Finally, the above discussed limitations regarding the applicability to general HCI tasks only refer to the linear finite-horizon LQR/LQG case.
The infinite-horizon LQR/LQG formulation~\cite{Qian13} is less suitable for many HCI tasks, as it does not allow to take into account multiple, time-dependent objectives during optimization, which, e.g., is inevitable for via-point tasks that need to be reached in a given order, or moving targets. %
However, the general class of optimal control models of Human-Computer Interaction, as discussed in Section~\ref{sec:oc-framework}, is much larger and consists of a variety of modeling approaches and solution methods, including Direct and Indirect Collocation~\cite{Betts10}, Model-Free and Model-Based Reinforcement Learning~\cite{Haith13, Sutton18}, (Semi-)Supervised Learning~\cite{Russel02, Nakada18}, Model-Predictive Control~\cite{Camacho13}, and mixtures of these~\cite{Lee19, Bian20, Berret21}, each of which has its own requirements on the problem, advantages, and disadvantages.
While some HCI tasks might be too complex to solve using the linear methods presented in this paper, the general OFC framework offers exciting opportunities to model, simulate, explore, and eventually improve the interaction between humans and computers, using a mathematically profound description. %

\subsection{Practical Benefits and Advice for HCI Researchers}\label{sec:benefits-advice}

Building on the above discussion on the applicability and generalization in the context of Human-Computer Interaction, we clarify the concrete benefits of our proposed framework and methods to HCI researchers, using the example of the so-called \textit{Bubble Lens}.

Previously, the \textit{Bubble Lens} method~\cite{Mott14} has been proposed as one of the few target acquisition techniques that explicitly takes into account kinematic movement profiles. The main idea of this method is to automatically magnify the desired target area as soon as the first corrective submovement has been detected (``kinematic triggering'').
While this technique has shown to significantly outperform the standard \textit{Bubble Cursor}~\cite{Grossman05} (the fastest pointing method at this time), the authors did not account for the fact that users might adapt their behavior once the magnification has been observed. Moreover, the criteria of when to trigger the magnification have been chosen manually, based on effectiveness and practicability.
Using our proposed optimal control framework of interaction, it would be possible to analyze the effects of temporary magnification on visual input, internal estimates, predictions and subgoals, and the resulting movement trajectory (including ergonomic quantities such as muscle energy consumption or fatigue). This allows to gain a deeper understanding of \textit{why} this technique outperforms existing methods. Finally, our unifying framework can be used to optimize the technique's remaining parameters (e.g., trigger time and duration of the lens, or visual properties such as smooth transitions), consider specific body characteristics of individual user groups, and perform simulation-based comparisons with existing methods, thus considerably improving comparability between different approaches. %

As a general advice for HCI researchers, we recommend the following procedure when using our proposed optimal control framework of Human-Computer Interaction:
\begin{enumerate}
	\item Make a \textbf{preliminary selection} of model(s) based on the phenomena of interest.
	
	Not every model is suitable for every purpose. 
	For example, the deterministic models 2OL-Eq, MinJerk, and LQR can only predict (optimal) average movement behavior, while the stochastic models LQG, E-LQG, and IC predict entire trajectory distributions. 
	Closed-loop models can explain how humans respond to unexpected perturbations during movement.
	If modeling muscle activations and fatigue is of interest, then torque- or muscle-driven models of the human arm and hand are much more appropriate than kinematic models such as 2OL-Eq and MinJerk.
	Finally, the extended observation model included in E-LQG provides an opportunity to analyze gaze using the saccade time step parameter.
	\item\label{itm:model-choice} Select a \textbf{specific model} based on qualitative and quantitative criteria. 
	
	The used metrics and decision criteria crucially depend on the overarching goal of the analysis. For example, replication of user data requires a more quantitative evaluation based on some metric that incorporates all relevant aspects of the observed trajectories, whereas a comparison of models regarding their explanatory and predictive power should rather be based on qualitative results, e.g., whether well-established phenomena such as bell-shaped velocity profiles, corrective submovements, or specific eye-movement coordination patterns can be inferred. In either case, the evaluation and visualization tools from our \href{https://github.com/fl0fischer/OFC4HCI}{OFC4HCI toolbox} may be helpful.
	
	Note that this stage requires choosing one (or multiple) reasonable parameter sets for each model to be compared, e.g., from the literature. Alternatively, the model parameters that "best explain" observed user data can be identified within an (outer) optimization loop, using the method presented in Section~\ref{sec:param-fitting}.
	In general, we suggest to include all parameters that are suspected to differ between independent variables (e.g., the user ID or the task condition) or which cannot be reasonably inferred from literature in the parameter fitting process.
	\item (Optional:) Fine-tune the specific \textbf{model parameters}. 
	
	If the analysis of parameters from stage~\ref{itm:model-choice} suggests that some parameters could be set constant, as they do not depend considerably on the user ID or task condition, another iteration of the parameter identification process could be performed.
	In this case, %
	information criteria such as the \textit{Akaike information criterion}~\cite{Stoica04}, which account for the trade-off between the goodness of fit and the simplicity of a model, might be considered.
	
\end{enumerate}

\section{Conclusion}\label{sec:conclusion}

In this paper, we have provided an introduction to the concepts of optimal feedback control for an HCI audience and have presented a %
generic parameter fitting process that can be used to identify system and strategy parameters of any given control model.
Using the example of mouse pointing, we have shown that both a non-trivial dynamic model of the human-computer interaction loop, which includes signal-dependent control noise, and continuously perceived noisy feedback are necessary to explain user behavior both qualitatively and quantitatively. 
These optimal control models show a significantly better fit to the considered user trajectories than pure dynamics models such as 2OL-Eq or pure kinematic models such as MinJerk. 

The \textit{optimal control framework for Human-Computer Interaction} that we have proposed is \textit{versatile}, as it can be used to model interaction with different interfaces using various input devices, and \textit{comprehensive}, as it allows to model the complete interaction loop, including body, input device, and interface dynamics, as well as feedback properties, each depending on the task and/or the user under consideration.
While the basic assumptions of LQG (linear dynamics, quadratic costs, Gaussian noise) are relatively restrictive, we have shown its ability to replicate typical mouse movements both in terms of average behavior and between-trial variability.
More importantly, we have demonstrated how the proposed framework can be used to identify characteristic differences in movement behavior between participants or task conditions. The degree to which these differences can be interpreted naturally depends on the model complexity.
In particular, aggregated Human-Computer System Dynamics as used in the presented case of one-dimensional mouse pointing do not allow to simulate motion of the human body per se, but only predict movement in end-effector space (i.e., mouse cursor trajectories). For an in-depth analysis of the intrinsic characteristics and strategies of the human biomechanical and cognitive system, independent of the used interaction technique, more detailed and separate submodels of both the Human Body Dynamics and the Interface Dynamics would be required.
 
We have also discussed the applicability of the framework to several other HCI tasks, as well as possible extensions (e.g., regarding non-linear body and interface dynamics) that remain as future work.
As a more general advice for HCI researchers, we recommend first making a preliminary selection of models based on the phenomena of interest, then selecting a specific model based on qualitative and quantitative criteria, and finally fine-tuning the model parameters.

We hope that this paper, along with our \href{https://github.com/fl0fischer/OFC4HCI}{OFC4HCI toolbox}, provides an easy-to-understand overview of how recent methods and concepts from optimal control theory can be applied to HCI using the example of mouse pointing, and encourages HCI researchers to use them in their own studies and simulations. Optimal feedback control provides a concise and mathematically exact explanation of movement in interaction with computers that we hope will be useful not only for HCI research, but also for teaching HCI and ultimately for interface design.  %

\begin{acks}
We would like to thank Lars Grüne for his very helpful advice and comments on a preliminary version of this paper.
\end{acks}

\balance{}

\bibliographystyle{ACM-Reference-Format}
\bibliography{bib}

\clearpage

\appendix

\renewcommand\thefigure{\thesection.\arabic{figure}}
\setcounter{figure}{0}

\renewcommand\thetable{\thesection.\arabic{table}}
\setcounter{table}{0}  

\section{LQR equations}\label{sec:appendix-LQR}

The proposed LQR model can be described as the discrete-time linear-quadratic optimal control problem with finite horizon $N\in\N$
\begin{subequations}\label{eq:appendix-LQR}
	\begin{small}
		\begin{align}\label{eq:appendix-LQR-objective}
			\begin{gathered}
				\textsl{Minimize} \quad J_{N}^{\text{(LQR)}}(x,u)= \sum_{n=0}^{N}x_{n}^{\top}Q_{n}x_{n} + \sum_{n=0}^{N-1} u_{n}^{\top}R_{n}u_{n} \\
				\textsl{with respect to } u= (u_{n})_{n\in\{0,\dots,N-1\}}\subset \R, %
			\end{gathered} 
		\end{align}
	\end{small}
	where $x= (x_{n})_{n\in\{0,\dots,N\}} \subset \R^{5}$ with $x_{n} = (p_n, v_n, f_n, g_n, T)^{\top}$ satisfies
	\begin{align}\label{eq:discrete-control_3}
		\begin{gathered}
			x_{n+1}=A x_{n} + B u_{n}, \quad n\in\{0,\dots,N-1\}, \\
			x_{0}=\bar{x}_{0},
		\end{gathered}
	\end{align}
	given some $\bar{x}_{0}\in\R^{5}$,
	with sampling time $h>0$ and system dynamics matrices
	\begin{align}
		A = \begin{pmatrix}
			1	& h	& 0 & 0 & 0 \\
			0	& 1	& h & 0 & 0 \\
			0	& 0	& 1 - \frac{h}{\tau_{2}}  & \frac{h}{\tau_{2}} & 0 \\
			0   & 0	& 0	& 1 - \frac{h}{\tau_{1}} & 0 \\
			0	& 0	& 0	& 0	& 1
		\end{pmatrix}, \quad B = \begin{pmatrix}
		0 \\
		0 \\
		0 \\
		\frac{h}{\tau_{1}} \\
		0
		\end{pmatrix}
	\end{align}
\end{subequations}
corresponding to the combination of a simplified second-order muscle model with time constants $\tau_{1}$, $\tau_{2}>0$ and a double integrator.
\\
The state cost matrices are defined by
\begin{align}\label{eq:appendix-LQR-cost-matrices_2}
	Q_{n}= 
	\begin{pmatrix}
		1 & 0 & 0 & 0 & -1 \\
		0 & \omega_{v} & 0 & 0 & 0 \\
		0 & 0 & \omega_{f} & 0 & 0 \\
		0 & 0 & 0 & 0 & 0 \\
		-1 & 0 & 0 & 0 & 1
	\end{pmatrix}\in\R^{5\times 5}, \quad n\in\{0,\dots,N\},
\end{align}
which implies
\begin{align}
	x_{n}^{\top}Q_{n}x_{n}=D_{n}^{2} +  \omega_{v} v_{n}^{2} + \omega_{f} f_{n}^{2},
\end{align}
i.e., the distance $D_{n}=\vert T - p_{n} \vert$ between mouse and target position as well as the end-effector velocity $v_{n}$ and force $f_{n}$ are quadratically penalized at every time step $n\in\{0,\dots,N\}$.
In our case of one-dimensional pointing tasks, the control cost matrices are scalar and given by
\begin{align}\label{eq:2OL-LQR-jerk-weights_2}
	R_{n}=\frac{\omega_r}{N-1} \in\R, \quad \omega_r>0, \quad n\in\{0,\dots,N-1\},
\end{align}
which yields the quadratic cost terms \begin{align}
	u_{n}^{\top}R_{n}u_{n}=\frac{\omega_{r}}{N-1}u_{n}^{2}.
\end{align}
\\
\\[0.2cm]
It can be shown that the unique solution $u^{*}=(u_{n}^{*})_{n\in\{0,\dots,N-1\}}$ to the optimization problem \eqref{eq:appendix-LQR} is given by
\begin{align}\label{eq:appendix-LQR-feedback}
	\begin{gathered}
		u_{n}^{*} = \pi(x_{n}) = -L_{n} x_{n}, \quad n\in\{0,\dots,N-1\}, \\
		L_{n}=(R_{n}+B^{\top}\mathcal{S}_{n+1}B)^{-1}B^{\top}\mathcal{S}_{n+1}A, %
	\end{gathered} \nonumber \\
	n\in\{0,\dots,N-1\},
\end{align}
where the symmetric matrices $\mathcal{S}_{n}\in\R^{5\times 5}$ can be determined by solving the \textbf{Discrete Riccati Equations}
\begin{subequations}\label{eq:modified-discrete-riccati}
	\begin{align}
		\begin{gathered}
			\mathcal{S}_{n}= Q_{n} + %
			A^{\top}\mathcal{S}_{n+1}A %
			- A^{\top}\mathcal{S}_{n+1}B(R_{n} + B^{\top}\mathcal{S}_{n+1}B)^{-1}B^{\top}\mathcal{S}_{n+1}A
		\end{gathered}
	\end{align}
	for $n\in\{0,\dots,N-1\}$ backward in time with initial value 
	\begin{align}
		\mathcal{S}_{N}=Q_{N}.
	\end{align}
\end{subequations}

\vfill

\section{Supplementary Material}

	\aboverulesep = 0mm
	\belowrulesep = 0mm

	\begin{table}[!htb]%
	\centering
	\begin{tabular}{|c|c||c|c|c|}
		\toprule
		\rule{0pt}{10pt}\noindent
		\textbf{Model} & \textbf{Parameter} & \textbf{Minimum} & \textbf{Maximum} & \textbf{Type} \\
		\midrule
		\multirow{2}{*}{2OL-Eq} & $k$ & 0 & 500 & Continuous \\
		\rule{0pt}{10pt}\noindent
		& $d$ & 0 & 500 & Continuous \\
		\hline
		\rule{0pt}{10pt}\noindent
		MinJerk & $N_{MJ}$ & 0 & $N$ & Continuous (Relaxed) \\
		\hline
		\multirow{3}{*}{LQR} & $\omega_{r}$ & 2e-9 & 20 & Continuous \\
		\rule{0pt}{10pt}\noindent
		& $\omega_{v}$ & 0 & 10e-2 & Continuous \\
		\rule{0pt}{10pt}\noindent
		& $\omega_{f}$ & 0 & 10e-4 & Continuous \\
		\hline
		\multirow{3}{*}{LQG / E-LQG} & $\omega_{r}$ & 4e-18 & 7e-3 & Continuous \\
		\rule{0pt}{10pt}\noindent
		& $\omega_{v}$ & 0 & 10 & Continuous \\
		\rule{0pt}{10pt}\noindent
		& $\omega_{f}$ & 0 & 10 & Continuous \\
		\hline
		\multirow{2}{*}{LQG} 
		& $\sigma_{u}$ & 10e-10 & 5 & Continuous \\
		\rule{0pt}{10pt}\noindent
		& $\sigma_{s}$ & 0 & 5 & Continuous \\
		\hline
		\multirow{6}{*}{E-LQG} 
		\rule{0pt}{10pt}\noindent
		& $\sigma_{u}$ & 10e-10 & 5 & Continuous \\
		\rule{0pt}{10pt}\noindent
		& $\sigma_{v}$ & 0 & 10 & Continuous \\
		\rule{0pt}{10pt}\noindent
		& $\sigma_{f}$ & 0 & 50 & Continuous \\
		\rule{0pt}{10pt}\noindent
		& $\sigma_{e}$ & 0 & 5 & Continuous \\
		\rule{0pt}{10pt}\noindent
		& $\gamma$ & 4e-18 & 100 & Continuous \\
		\rule{0pt}{10pt}\noindent
		& $n_{s}$ & 0 & $N$ & Continuous (Relaxed) \\
		\bottomrule
	\end{tabular}
	\caption{\label{tab:param_bounds}Boundaries of all model parameters used for parameter fitting.}
\end{table}

\begin{figure}[!htb]%
	\resizebox{0.8\linewidth}{!}{
		\begin{tikzpicture}[auto, thick, node distance=1.5cm, >=triangle 45,inner sep=2mm]
			\draw
			node [input] (input1) {} 
			node [input,right of=input1] (plhold){} 
			node [sum, right of=input1,align=left,node distance=2cm] (suma1) {}
			
			node [block, right of=suma1,node distance=2cm] (inte1) {\inte}
			node [split, right of=inte1] (split1) {}
			node [block, right of=split1] (inte2) {\inte}
			node [split, right of=inte2] (split2) {}
			node [output,right of=split2] (output1) {} 
			node [gain, below of=inte1,node distance=0.85cm] (gaind) {$d$}
			node [gain, below of=inte2,node distance=1.35cm] (gaink) {$k$}
			node [input,left of=gaind, node distance=2.65cm] (plhold3) {}
			;
			\draw[->](input1) --  node[near start] {$u(t)$}
			node[pos=0.9]{$+$}(suma1);
			\draw[->](suma1) -- node[near start]{$\ddot{y}(t)$} (inte1);
			\draw[-](inte1) -- node {$\dot{y}(t)$} (split1);
			\draw[->](split1) -- node{} (inte2);
			\draw[->](split1) |- node{} (gaind);
			\draw[->](inte2) -- node {} (output1);
			\draw[-](inte2) -- node {$y(t)$} (split2);
			\draw[->](split2) |- node{} (gaink);
			\draw[->](gaind) -| node[pos=0.9,right]{$-$} (suma1);
			\draw[-](gaink) -|  (plhold3);
			\draw[->](plhold3) -- node[pos=0.9,left]{$-$} (suma1);
			\draw[->](split2) -- node{} (output1);
		\end{tikzpicture}
		
	}
	\caption{Control-flow diagram of the second-order lag~\eqref{eq:2ol-continuous}.~\label{fig:2OL_controlflow}}
	\Description{Fully described in caption and text.}
\end{figure}
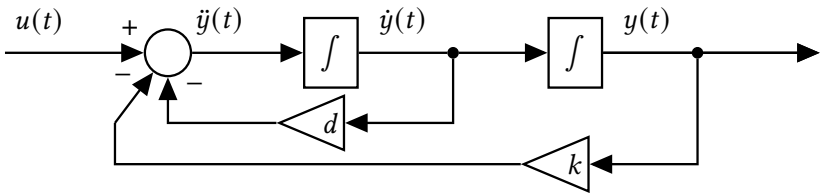

\begin{figure}[!htb]
	\centering
	\includegraphics[width=0.8\linewidth]{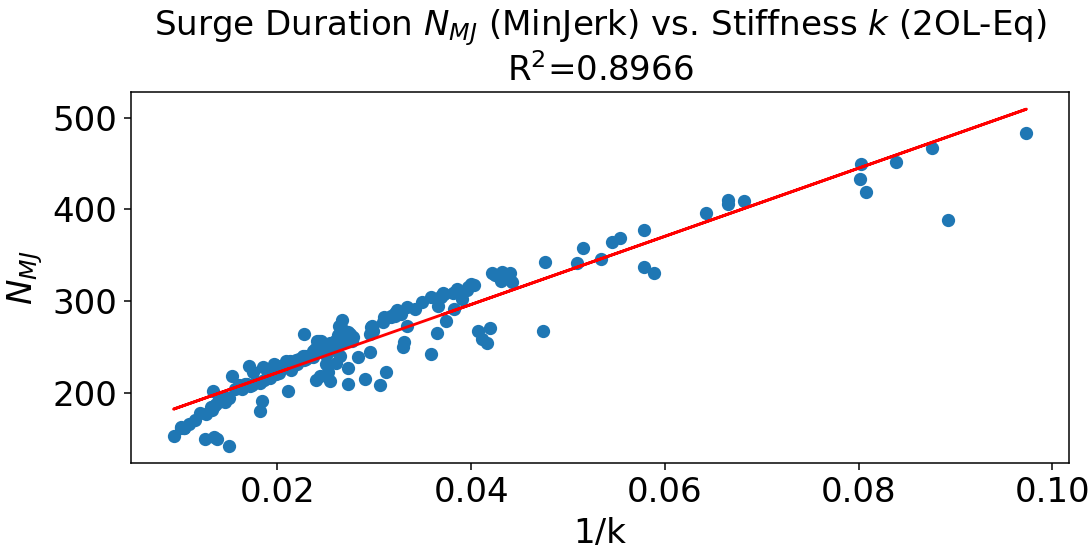}
	\caption{The relationship between the inverse of the stiffness parameter $k$ in 2OL-Eq and the surge duration parameter $N_{MJ}$ in MinJerk is captured well by a linear function. %
	}~\label{fig:MinJerk_2OL-Eq_params}
	\Description{MinJerk surge duration parameter $N_{MJ}$ plotted against inverse of 2OL-Eq stiffness parameter $k$ together with its linear approximation, resulting in a coefficient of determination of $R^{2}=0.8966$.}
\end{figure}

\begin{figure}[!htb]
	\centering

	\subfloat{\includegraphics[width=0.33\linewidth]{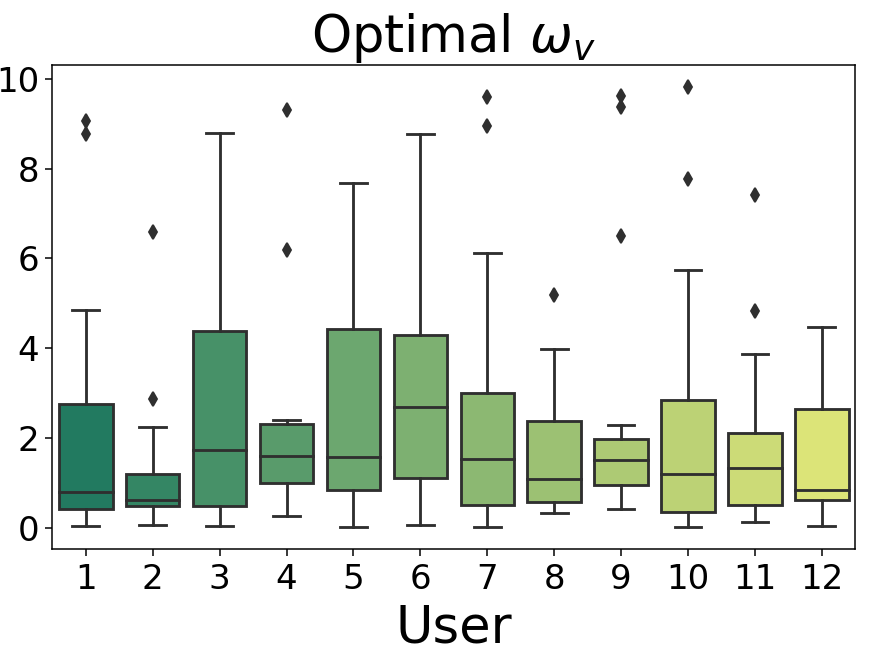}}
	\subfloat{\includegraphics[width=0.33\linewidth]{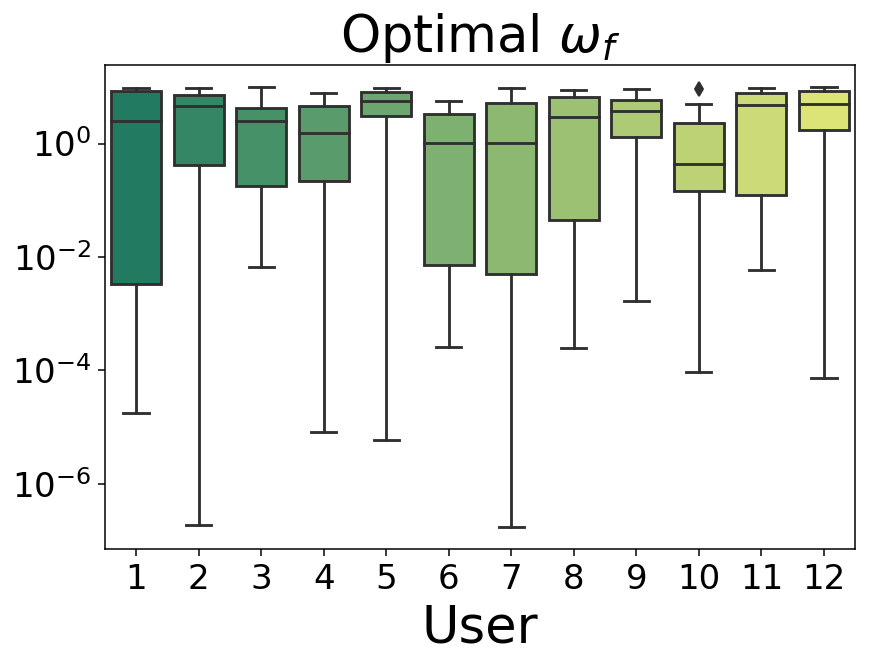}}
	\subfloat{\includegraphics[width=0.33\linewidth]{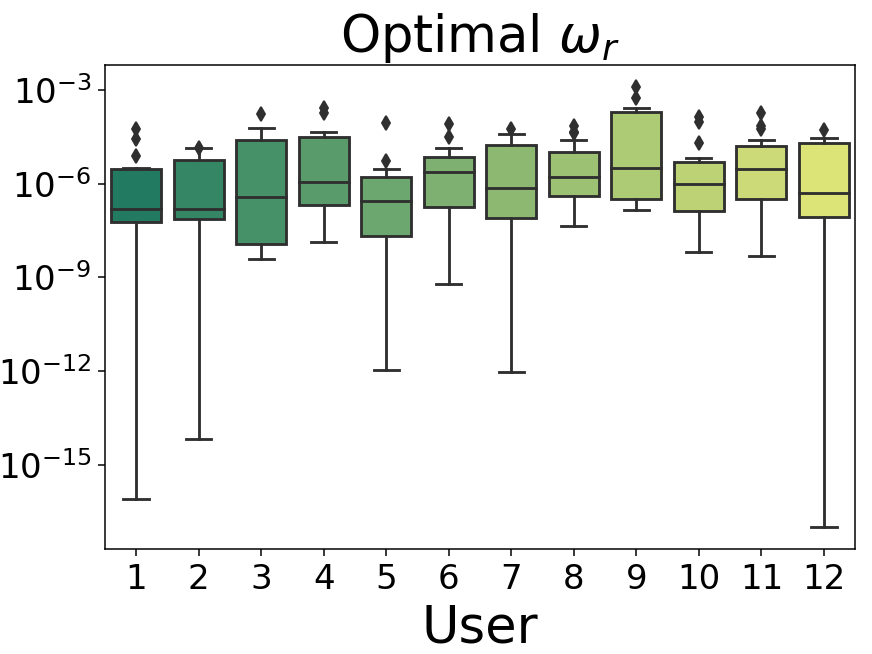}}
	\caption{Cost weight parameters of the LQG, optimized for the trajectory sets of all participants, tasks, and directions, grouped by participants. %
		Note that the optimal values of $\omega_{f}$ and $\omega_{r}$ are plotted on a logarithmic scale. %
	}~\label{fig:LQG_opt_2}
	\Description{Fully described in caption and text.}
\end{figure}

\begin{figure}[!htb]
	\centering
	\subfloat{\includegraphics[width=0.45\linewidth]{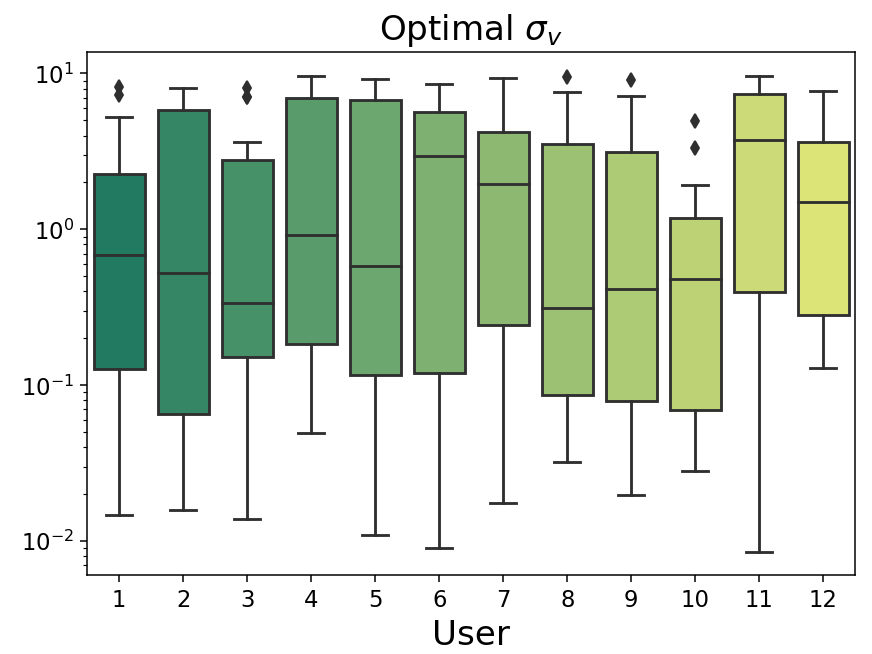}}
	\subfloat{\includegraphics[width=0.45\linewidth]{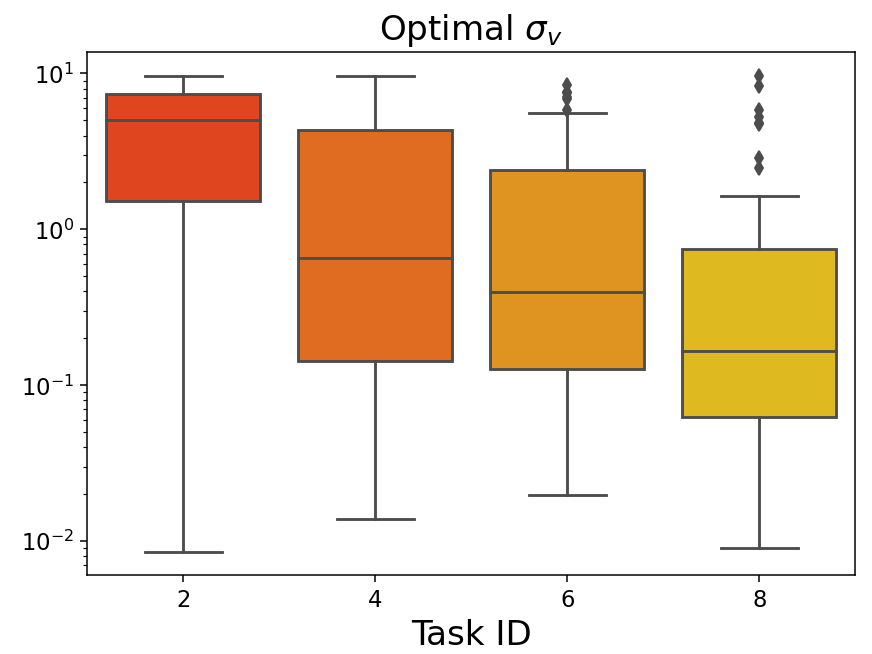}}
	\\
	\subfloat{\includegraphics[width=0.45\linewidth]{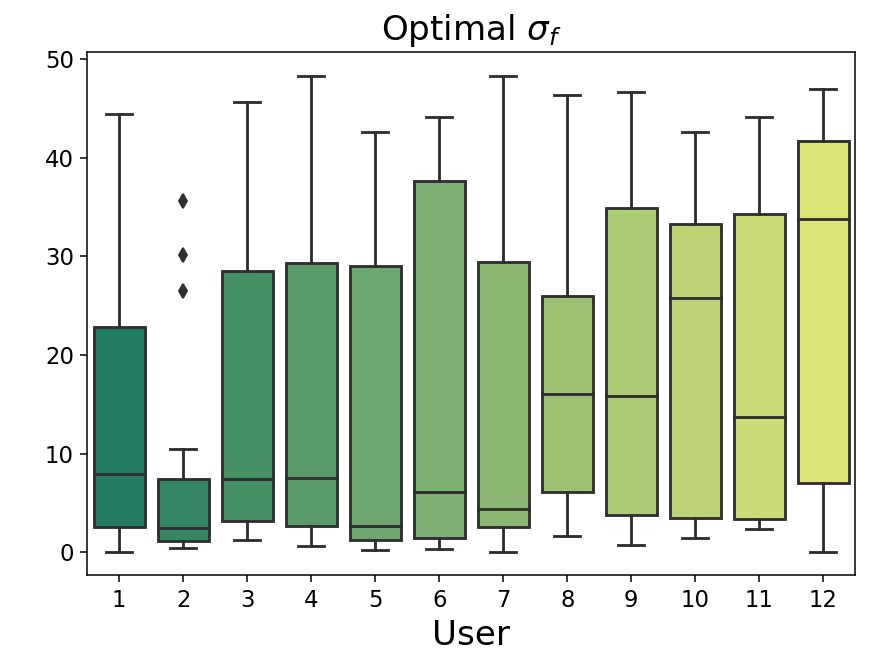}}
	\subfloat{\includegraphics[width=0.45\linewidth]{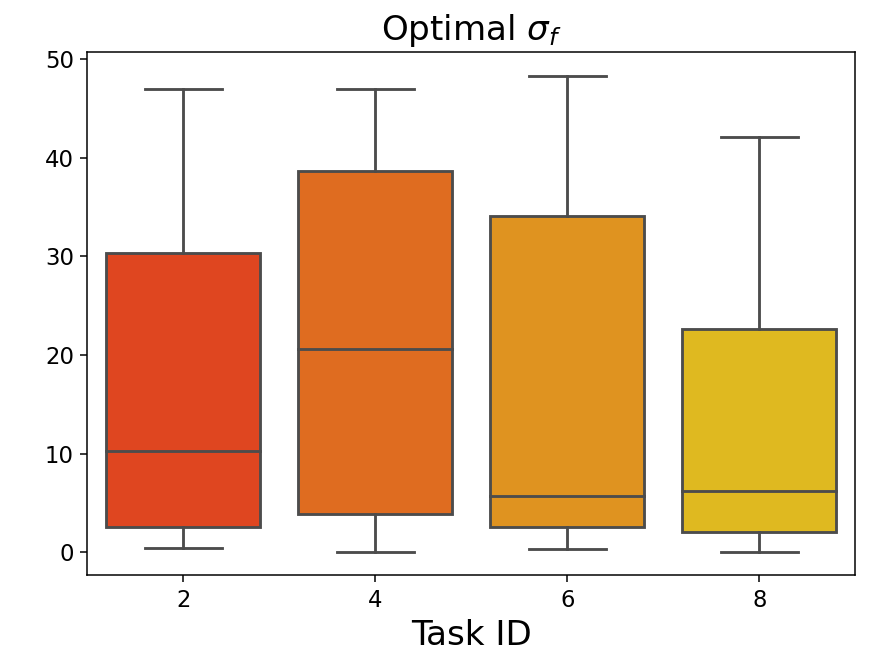}}
	\\ 
	\subfloat{\includegraphics[width=0.45\linewidth]{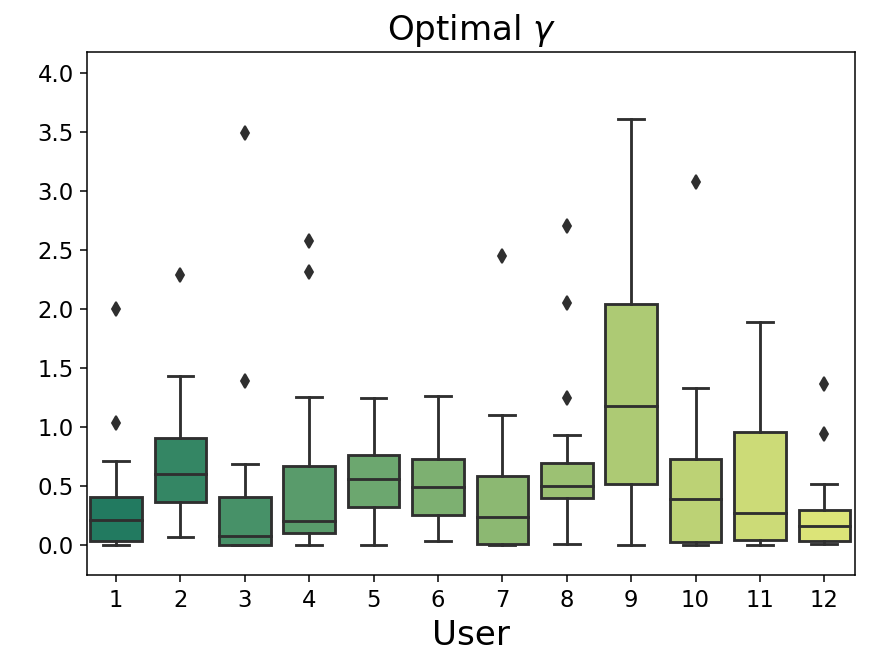}}
	\subfloat{\includegraphics[width=0.45\linewidth]{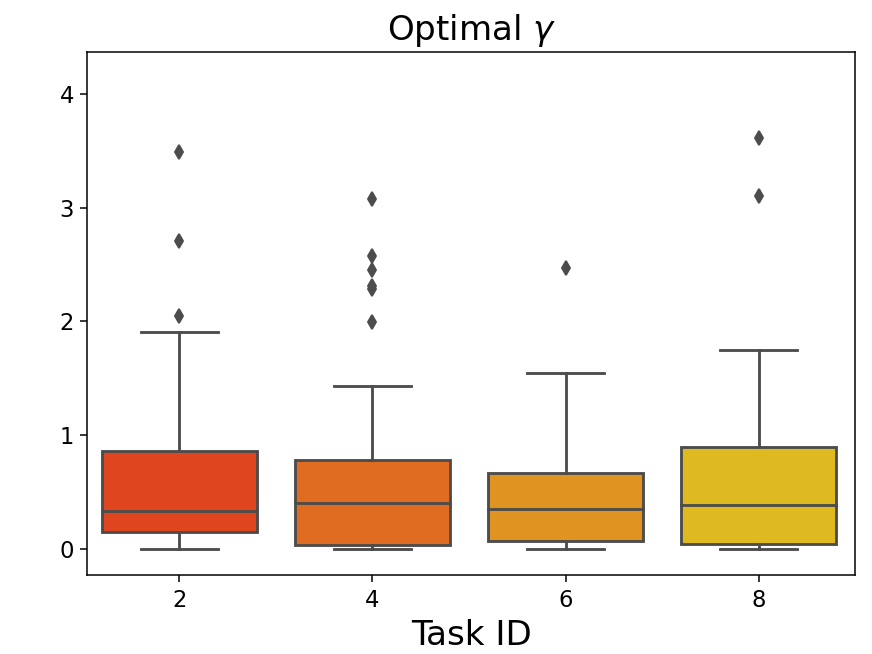}}
	\caption{Visual observation noise parameters of the E-LQG, optimized for the mean trajectories of all participants, tasks, and directions, grouped by participants (left) and by ID (right).
		For better visibility, both plots for the position perception noise weight $\gamma$ do not include the 2 largest outliers with values $19.7$ and $55.4$. %
		Note that the optimal values of $\sigma_v$ are plotted on a logarithmic scale.
	}~\label{fig:E-LQG_opt_2a}
	\Description{Fully described in caption and text.}
\end{figure}
\clearpage
\begin{figure}%
	\centering
	\subfloat{\includegraphics[width=0.45\linewidth]{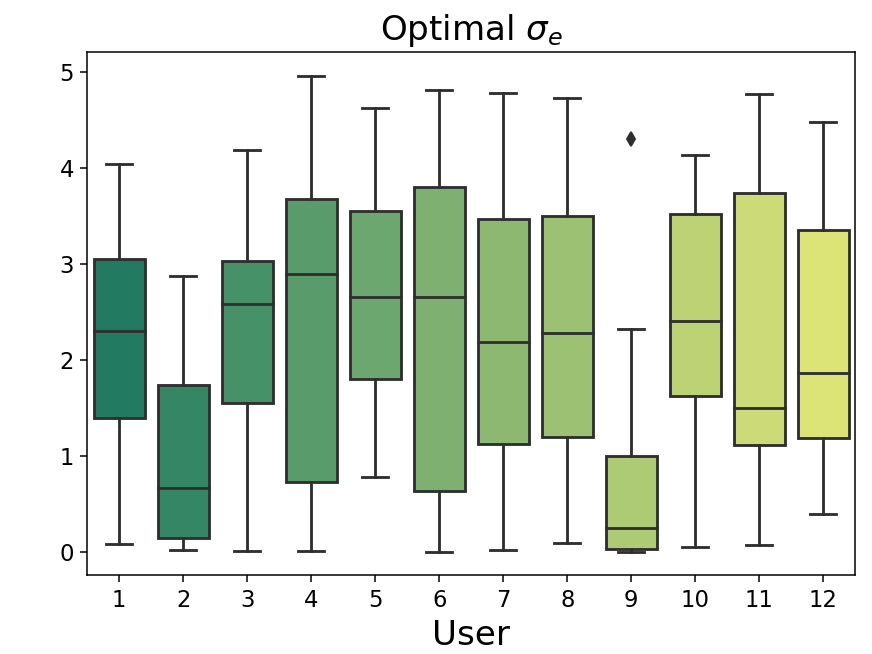}}
	\subfloat{\includegraphics[width=0.45\linewidth]{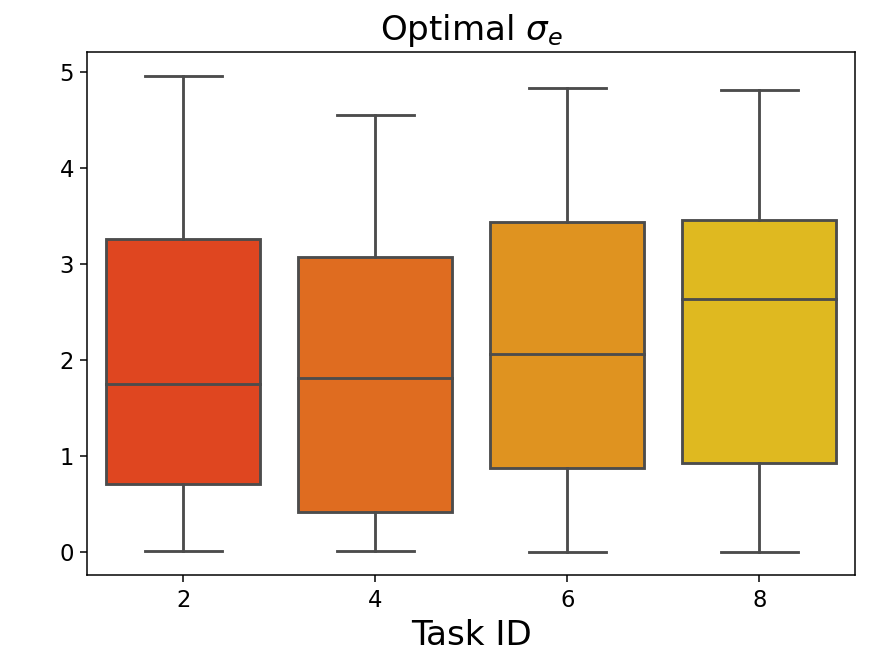}}
	\\
	\subfloat{\includegraphics[width=0.45\linewidth]{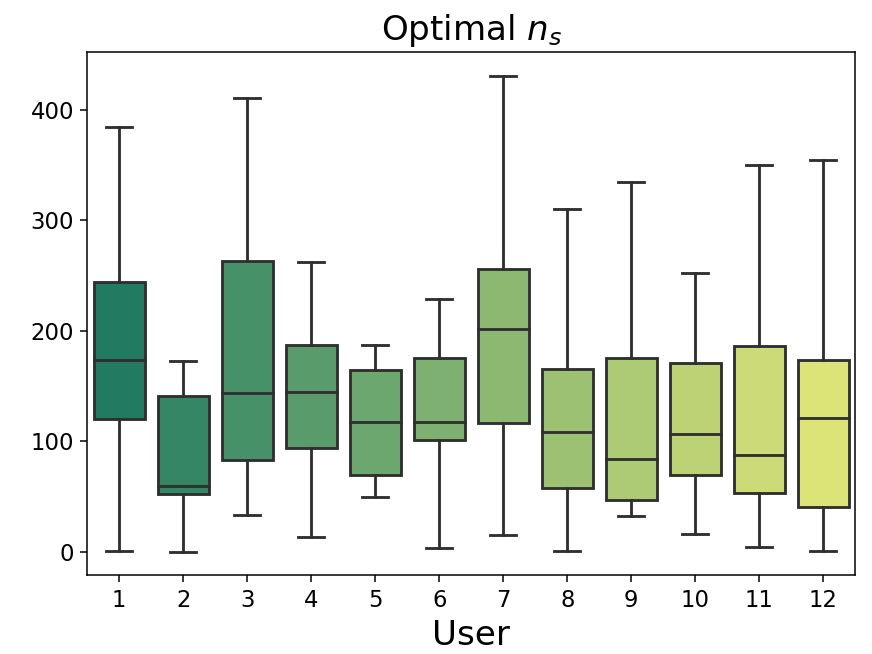}}
	\subfloat{\includegraphics[width=0.45\linewidth]{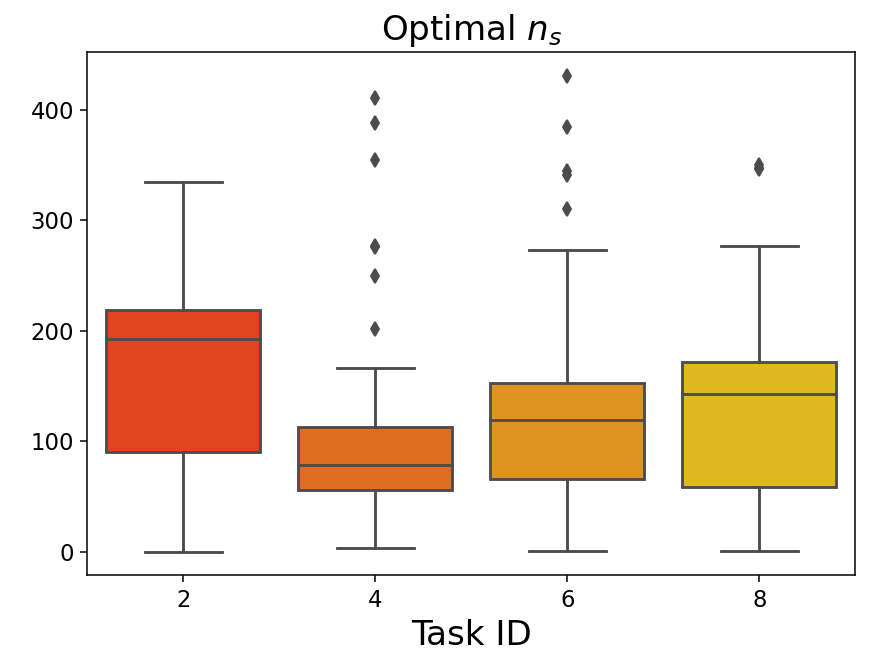}}
	\caption{The gaze noise level $\sigma_{e}$ and the saccade time step $n_{s}$ of the E-LQG, optimized for the mean trajectories of all participants, tasks, and directions, grouped by participants (left) and by ID (right).
	}~\label{fig:E-LQG_opt_2b}
	\Description{Fully described in caption and text.}
\end{figure}

\end{document}